\documentclass[aps,prl,preprint,nofootinbib,superscriptaddress,nobibnotes]{revtex4-2}

\usepackage[T1]{fontenc}
\usepackage{amsmath,amssymb,amsfonts}
\usepackage{graphicx}
\usepackage{ulem}
\usepackage{booktabs}
\usepackage{xcolor}
\usepackage{hyperref}
\usepackage{cleveref}
\usepackage{subcaption}
\usepackage{tabularx}
\usepackage{tikz}
\usetikzlibrary{decorations.pathmorphing}
\makeatletter
\def\frontmatter@abstractfont{\small\linespread{0.97}\selectfont}
\def\frontmatter@affiliationfont{\footnotesize\itshape\linespread{0.97}\selectfont}
\makeatother
\crefname{section}{Section}{Sections}
\Crefname{section}{Section}{Sections}

\crefname{subsection}{Section}{Sections}
\Crefname{subsection}{Section}{Sections}

\crefname{subsubsection}{Section}{Sections}
\Crefname{subsubsection}{Section}{Sections}

\crefname{appendix}{Appendix}{Appendices}
\Crefname{appendix}{Appendix}{Appendices}

\crefname{equation}{Equation}{Equations}
\Crefname{equation}{Equation}{Equations}

\crefname{figure}{Figure}{Figures}
\Crefname{figure}{Figure}{Figures}

\crefname{table}{Table}{Tables}
\Crefname{table}{Table}{Tables}

\newcommand{\M}{\mathcal{M}}
\newcommand{\N}{\mathcal{N}}
\newcommand{\Ltot}{\mathcal{L}_{\rm tot}}
\newcommand{\LE}{\mathcal{L}_{\rm E}}

\newcommand{\LW}{\mathcal{L}_{W}}
\newcommand{\LD}{\mathcal{L}_{\det}}
\newcommand{\LP}{\mathcal{L}_{\rm P}}
\newcommand{\LKR}{\mathcal{L}_{K{\rm rep}}}
\newcommand{\rhoSchw}{\rho_{\rm Schw}}

\newcommand{\diag}{\operatorname{diag}}
\newcommand{\dd}{\mathrm{d}}
\newcommand{\R}{\mathbb{R}}

\begin{document}

\title{\texorpdfstring{Black Hole Black Boxes: \\ Numerical Black Hole Metrics via AInstein Neural Networks}{Black Hole Black Boxes: Numerical Schwarzschild Metrics via AInstein Neural Networks}}

\author{Edward Hirst}
\email{ehirst@unicamp.br}
\affiliation{Instituto de Matemática, Estatística e Computação Científica (IMECC) da Universidade Estadual de Campinas (UNICAMP), 13083-859, Brasil}

\author{Tancredi Schettini Gherardini}
\email{tsg@math.uni-bonn.de}
\affiliation{Mathematisches Institut (MI), University of Bonn, Bonn, Germany}
\affiliation{Max Planck Institute for Mathematics, Bonn, Germany}

\author{Alexander G. Stapleton}
\email{a.g.stapleton@qmul.ac.uk}
\affiliation{Centre for Theoretical Physics and Astronomy, Queen Mary University of London, London E1 4NS, United Kingdom}

\par

\date{\today}

\begin{abstract}
The AInstein architecture introduced an unsupervised neural method for solving the Riemannian Einstein equations on arbitrary manifolds. 
This Physics Informed Neural Network approach (PINN) is extended here to Lorentzian signature, validated by recovering the maximally extended Schwarzschild geometry, and tested as novel search method for arbitrary black hole solutions. 
The topology is built into the architecture by treating $S^{2}$ globally through its standard embedding, such that the network learns an ambient metric on the manifold $\mathbb{R}^{2} \times \mathbb{R}^{3}$, where Penrose coordinates are chosen for $\mathbb{R}^2$ and the metric on $S^{2}$ is obtained by pullback. 
The architecture is first trained with the objective of recovering the Schwarzschild metric via losses encoding the vacuum Einstein equation, a quadratic Weyl scalar constraint, and the $SO(3)$ symmetry of the resultant metric; directly motivated by the Birkhoff--Jebsen theorem. 
Following this, the objective is generalised to use the Petrov speciality index, a horizon curvature anchor, and a trapped‑surface constraint, to allow search for algebraically general Petrov type I solutions, finding potentially novel general-type Lorentzian Einstein metrics with a genuinely trapped interior.
\end{abstract}

\maketitle
\clearpage
{\small\linespread{0.97}\selectfont
\tableofcontents
}

\small
\linespread{0.97}\selectfont

\newpage
%%%%%%%%%%%%%%%%%%%%%%%%%%%%%%%%%%%%%%%%%%%%%%%
\section{Introduction}
Finding explicit Einstein metrics is a central problem in geometry and theories of gravity. More specifically, in general relativity and its generalisations, one is particularly interested in the construction of pseudo-Riemannian Lorentzian metrics. Let $\mathcal M$ be a smooth manifold equipped with a local coordinate chart $(U, x^\mu)$ on an open set $U\subset \mathcal M$. Given $U$, one may locally construct a smooth, pseudo-Riemannian metric $g_{\mu\nu}$ on $U$ with an associated Ricci curvature tensor $R_{\mu\nu}$. In this notation, an Einstein metric is defined such that Ricci curvature is everywhere proportional to the metric itself, i.e.\footnote{We work with the \textit{`mostly-positive'}, or \textit{`East-Coast'} signature $(-,+,+,+)$, and use the conventional definition $R^a{}_{bcd}=\partial_c\Gamma^a_{db}-\partial_d\Gamma^a_{cb}+\Gamma^i_{bd}\Gamma^a_{ic}-\Gamma^i_{bc}\Gamma^a_{id}$ for the Riemann tensor.}
\begin{equation}
  R_{\mu\nu}=\lambda g_{\mu\nu},
  \label{eq:einstein_metric}
\end{equation}
for some constant of proportionality $\lambda$ referred to as the \textit{Einstein constant}.
In Riemannian geometry, \cref{eq:einstein_metric} says that the space has the same averaged curvature in every tangent direction, whilst in general relativity, its pseudo-Riemannian version is the vacuum Einstein equation with $\lambda$ interpreted as a cosmological constant. Although \cref{eq:einstein_metric} is concise, it represents a non-linear second-order system for the metric components. The global problem also requires the local chart components to assemble into a single smooth tensor on the whole manifold. 

In numerical relativity and stationary black-hole construction, this non-linear geometric problem is usually made tractable by turning the Einstein equations into a well-posed computational system after choosing a formulation, coordinates, gauge conditions, boundary data, and a mesh or spectral discretisation. 
The dynamical branch of the field is built around the Cauchy problem, where the spacetime is foliated into spatial hypersurfaces, initial data are chosen subject to the Hamiltonian and momentum constraints, and the data are evolved by a gauge-fixed form of Einstein's equations. 
The ADM decomposition is the classical starting point for this viewpoint \cite{ArnowittDeserMisner1962}, while standard reviews of initial data and numerical-relativity methodology emphasise that the practical problem is not only to discretise the equations, but also to control constraints, gauges, singularities, outer boundaries, and wave extraction, see \cite{Cook2000,Lehner2001} for instance. 
A major development was the conformal BSSN formulation, originating in the work of \cite{ShibataNakamura1995} and systematised by \cite{BaumgarteShapiro1998}, which greatly improved the stability of three-dimensional evolutions relative to direct ADM evolution. 
The binary-black-hole breakthrough was obtained in \cite{Pretorius:2005gq} via the first stable full evolution of a binary black-hole spacetime using generalized harmonic evolution and excision, while the moving-puncture methods of \cite{Campanelli2006,Baker2006} made long-term inspiral, merger, ringdown, and gravitational-wave extraction broadly practical.
Subsequent reviews describe this as the point at which binary black-hole merger simulations became routine tools for strong-field gravity and gravitational-wave astronomy \cite{Centrella2010}, with high-accuracy spectral evolutions and community infrastructure such as the Einstein Toolkit further maturing the field \cite{Scheel2009,Loeffler2012}. 
Stationary black-hole construction is different in character, instead of evolving initial data one typically solves a coupled elliptic boundary-value problem for the metric, with boundary conditions encoding asymptotic regions, axes, horizons, conformal boundaries, or compact directions. 
The construction of \cite{Wiseman:2002zc}, for static axisymmetric vacuum solutions and non-uniform black strings, illustrates the role of adapted ansatz choices and elliptic relaxation methods. Similarly, the Einstein--DeTurck formulation of \cite{Headrick:2009pv} gives a systematic gauge-fixed elliptic framework for static numerical relativity and Kaluza--Klein black holes. 
This approach was extended toward more general stationary vacuum black holes in Lorentzian signature in \cite{Adam:2011dn}.
This stationary toolkit is reviewed in detail by \cite{DiasSantosWay2015,Dias:2015nua} and has been applied in less symmetric and asymptotically AdS settings, for example in numerical constructions of black resonators and related AdS instabilities. 
In all of these cases, the metric is represented by finite-dimensional data tied to a coordinate grid, finite-difference stencil, finite-element mesh, or spectral basis, and geometric consistency is enforced through the discretised field equations together with carefully chosen gauge, regularity, and boundary conditions.

Machine learning provides a complementary representation: the metric is treated as a differentiable function of the coordinates by representing it with a (smooth) neural network, with geometric tensors computed by automatic differentiation and inserted directly into the objective\footnote{This is the Physics Informed Neural Network (PINN) formalism \cite{Raissi:2017zsi}.}.
This program of repurposing the efficient differentiation mechanisms of AI for differential geometry and geometric analysis has been recently pursued in \cite{Hirst:2025seh, Cortes:2026kfx} for various problems concerning spheres, as well as by pioneering works in the context of Calabi-Yau metrics \cite{Ashmore:2019wzb, Douglas:2020hpv, Larfors:2021pbb, Gerdes:2022nzr, Berglund:2022gvm}, $G_2$ metrics \cite{Douglas:2024pmn, Heyes:2026rch}, hyperbolic spaces \cite{DeLuca:2024, gherardini2026minimalsurfacesknotsneural}, modes related to black holes \cite{Kumar:2023hlu, Patel:2024wzo, Cranganore:2025vsu}, and for broader contexts\footnote{Here we note complementary applications of pure supervised learning to the black hole context \cite{He:2022fxp, Jejjala:2023zxw, Jejjala:2025hgv}.} such as minimal surfaces \cite{Hashimoto:2025zmi, Hirst:2026qwi, Hirst:2026cpm, gherardini2026minimalsurfacesknotsneural}.

AInstein\footnote{AInstein is available at: \href{https://github.com/xand-stapleton/ainstein}{\texttt{https://github.com/xand-stapleton/ainstein}}.} \cite{Hirst:2025seh} introduced this strategy for Riemannian Einstein metrics on spheres. 
In the original construction, the manifold was represented by an atlas, each chart had its own subnetwork, points in overlap regions were passed through the relevant transition maps, and an explicit overlap loss enforced the tensorial transformation law for the metric. In this way, global consistency became a learnable constraint rather than an assumption built into the architecture.

This work adapts AInstein to first find and approximate the maximally extended Schwarzschild solution, then presents a more general Lorentzian metric search. The main changes are a Lorentzian metric parametrisation, compactified Penrose-domain sampling, coordinate-invariant scalar losses, and an embedded treatment of the topologically non-trivial part of the manifold. The embedding is not only a visual convenience; it changes how global consistency is imposed. Rather than learning two independent sets of metric components and matching them with a loss term, the model learns a single ambient metric, with the two stereographic hemispheres being related analytically by pullback.

%%%%%%%%%%%%%%%%%%%%%%%%%%%%%%%%%%%%%%%%%%%%%%%
\section{Theoretical Prerequisites}
\subsection{The Schwarzschild Problem}
\label{sec:Theoretical_schw}

The Schwarzschild metric is an exact solution to Einstein's field equations that describes the spacetime geometry outside a spherically-symmetric non-rotating mass such as a black hole. In this section, the coordinate description used throughout the work is developed.

Let $\mathcal{P}$ denote the two-dimensional compactified Penrose domain \cite{Penrose1963Asymptotic}, i.e.~an open hexagonal subset of $\mathbb{R}^2$ (see below), and let $S^{2}$ denote the 2-sphere. The four-dimensional manifold is
\begin{equation}
  \M=\mathcal{P}\times S^{2},
  \label{eq:model_manifold}
\end{equation}
where it should be noted \cref{eq:model_manifold} does not assume a product metric -- the product notation only separates the causal coordinates from the angular coordinates. Although the Schwarzschild solution is spherically symmetric, the class of metrics considered later is a general symmetric four-dimensional pullback metric, not an ansatz restricted to the diagonal Schwarzschild form.

Let $(T,X)$ denote Penrose coordinates. A Penrose diagram is a conformally compactified representation of spacetime, infinite regions are mapped to finite boundaries while the light-cone structure is preserved. Penrose coordinates are useful because they place the event horizons inside a finite computational domain; the angular $S^{2}$ directions are suppressed, so each point in the two-dimensional diagram represents a sphere, and radial light rays follow diagonal lines. Let $m>0$ denote the Schwarzschild mass parameter and let $r$ denote the Schwarzschild areal radius. In Schwarzschild coordinates, the horizon at $r=2m$ is a coordinate singularity; in Kruskal or Penrose coordinates, it is a regular null surface. For any centre $c$, define the two boundary functions
\begin{equation}
  \tau_{c}^{-}(X)=|X-c|-\frac{\pi}{4},
  \qquad
  \tau_{c}^{+}(X)=-|X-c|+\frac{\pi}{4}.
  \label{eq:tau_boundaries}
\end{equation}
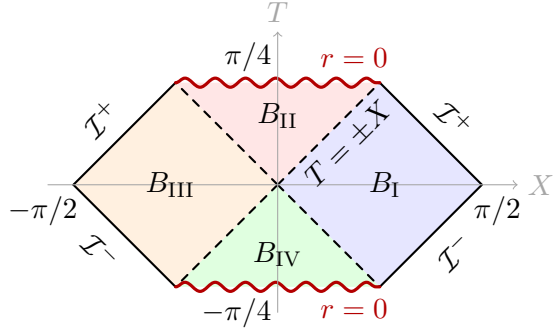
\begin{figure}[t!]
  \centering
  \begin{tikzpicture}[scale=1.35, line cap=round, line join=round]
    \coordinate (L) at (-2,0);
    \coordinate (TL) at (-1,1);
    \coordinate (TR) at (1,1);
    \coordinate (R) at (2,0);
    \coordinate (BR) at (1,-1);
    \coordinate (BL) at (-1,-1);
    \coordinate (O) at (0,0);

    \fill[blue!10] (O) -- (TR) -- (R) -- (BR) -- cycle;
    \fill[orange!12] (O) -- (TL) -- (L) -- (BL) -- cycle;
    \fill[red!10] (TL) -- (TR) -- (O) -- cycle;
    \fill[green!12] (BL) -- (O) -- (BR) -- cycle;

    \draw[thick] (L) -- (TL);
    \draw[thick] (TR) -- (R) -- (BR);
    \draw[thick] (BL) -- (L);
    \draw[thick, dashed] (O) -- (TL);
    \draw[thick, dashed] (O) -- (TR);
    \draw[thick, dashed] (O) -- (BL);
    \draw[thick, dashed] (O) -- (BR);
    \draw[very thick, red!70!black, decorate, decoration={snake, amplitude=0.6mm, segment length=4mm}] (TL) -- (TR);
    \draw[very thick, red!70!black, decorate, decoration={snake, amplitude=0.6mm, segment length=4mm}] (BL) -- (BR);

    \draw[->, gray!70] (-2.25,0) -- (2.35,0) node[right] {$X$};
    \draw[->, gray!70] (0,-1.25) -- (0,1.5) node[above] {$T$};

    \node at (1.05,0) {$B_{\rm I}$};
    \node at (-1.05,0) {$B_{\rm III}$};
    \node at (0,0.68) {$B_{\rm II}$};
    \node at (0,-0.68) {$B_{\rm IV}$};

    \node[above, red!70!black] at (0.75,1.05) {$r=0$};
    \node[below, red!70!black] at (0.75,-1.02) {$r=0$};
    \node[rotate=45, above] at (-1.58,0.50) {$\mathcal{I}^{+}$};
    \node[rotate=-45, below] at (-1.58,-0.50) {$\mathcal{I}^{-}$};
    \node[rotate=-45, above] at (1.58,0.50) {$\mathcal{I}^{+}$};
    \node[rotate=45, below] at (1.58,-0.50) {$\mathcal{I}^{-}$};
    %\node[font=\scriptsize, above right] at (0.4,0.2) {$T=\pm X$};
    \node[rotate=45, below] at (0.55,0.55) {$T=\pm X$};

    \node[below=8pt] at (-2.3,0.2) {$-\pi/2$};
    \node[below=8pt] at (2.15,0.2) {$\pi/2$};
    \node[left=12pt] at (0.35,1.25) {$\pi/4$};
    \node[left=12pt] at (0.35,-1.25) {$-\pi/4$};
  \end{tikzpicture}
  \caption{Compactified Penrose domain used in \cref{eq:penrose_regions}. The horizontal and vertical axes are the Penrose coordinates \(X\) and \(T\), respectively, with the plot drawn in units of \(\pi/4\). The right and left diamonds are the two exterior regions \(B_{\rm I}\) and \(B_{\rm III}\), while \(B_{\rm II}\) and \(B_{\rm IV}\) are the future and past black-hole interior regions. Dashed diagonal lines denote the null horizons \(T=\pm X\), and the red horizontal boundaries are the Schwarzschild singularities.}
  \label{fig:penrose_regions}
\end{figure}
\cref{fig:penrose_regions} shows the corresponding compactified domain. Using \cref{eq:tau_boundaries}, the four exterior/interior diamonds we consider are
\begin{equation}\label{eq:penrose_regions}
\begin{split}
 B_{\rm I}
 &=\{0<X<\pi/2,\ \tau_{\pi/4}^{-}<T<\tau_{\pi/4}^{+}\}, \\
 B_{\rm II}
 &=\{-\pi/4<X<\pi/4,\ |X|<T<\pi/4\}, \\
 B_{\rm III}
 &=\{-\pi/2<X<0,\ \tau_{-\pi/4}^{-}<T<\tau_{-\pi/4}^{+}\},\\
 B_{\rm IV}
 &=\{-\pi/4<X<\pi/4,\ -\pi/4<T<-|X|\}.
\end{split}
\end{equation}
The null horizons are the diagonal lines $T=\pm X$ in the central diamond, whilst the future and past singularities are the horizontal boundaries $T=\pm\pi/4$ with $|X|<\pi/4$. The remaining null sides are the conformal boundaries of the exterior regions.

In the Penrose coordinates of \cref{eq:penrose_regions}, the target Schwarzschild line element may be written as
\begin{equation}
  \dd s^{2}
  =F(T,X)(-\dd T^{2}+\dd X^{2})
   +r(T,X)^{2}\dd\Omega_{2}^{2},
  \label{eq:schw_metric}
\end{equation}
where $\dd\Omega_{2}^{2}$ is the round metric on the unit two-sphere, $r(T,X)$ is the Schwarzschild areal radius and $F(T,X)$ is the Penrose conformal factor. This conformal factor is
\begin{equation}
  F(T,X)=
  \frac{32m^{3}e^{-r(T,X)/(2m)}}
       {r(T,X)\left(\cos^{2}T-\sin^{2}X\right)^{2}} \, ,
  \label{eq:F_factor}
\end{equation}
with $r(T,X)$ defined below.
For stereographic angular coordinates $(q_{1},q_{2})$ on $S^{2}$,
\begin{equation}
  \dd\Omega_{2}^{2}
  =\frac{4}{(1+q_{1}^{2}+q_{2}^{2})^{2}}
   (\dd q_{1}^{2}+\dd q_{2}^{2}).
  \label{eq:stereo_s2}
\end{equation}
The conformal factor $F$ is finite and non-zero at the horizons; the singular behaviour of the Schwarzschild solution is instead carried by $r(T,X)\rightarrow0$ on the spacelike singularity (in addition to the conformal boundary divergence). Thus, the singular behaviour is restricted to the boundary of the Penrose domain. Let $W$ denote the real Lambert-$W$ function, and let $\mathcal{B}$ label one of the four regions in \cref{eq:penrose_regions}. The radius $r$ is computed over $B_{\rm I},\ldots,B_{\rm IV}$ by
\begin{equation}
  r=2m(1+W_{\mathcal{B}}),
  \label{eq:radius_lambert}
\end{equation}
where $\chi=\cos 2T/\cos 2X$ and
\begin{equation}
\begin{split}
  W_{\mathcal{B}}
  &=
  W\!\left[\exp\{-2\,\operatorname{arccoth}\chi-1\}\right],
   \qquad\qquad\qquad\quad \mathcal{B}=B_{\rm I},B_{\rm III},\\
  W_{\mathcal{B}}
  &=
  W\!\left[-\exp\{-2\,\operatorname{arctanh}\chi-1\}\right],
  \qquad\qquad\qquad \mathcal{B}=B_{\rm II},B_{\rm IV}.
  \label{eq:piecewise_W}
\end{split}
\end{equation}
For a detailed account on these coordinates the reader is referred to \cite{roken2026constructionapplicationpenrosediagrams}, for instance. The tensor equation to be solved remains the vacuum condition $R_{\mu\nu}=0$.

\subsection{Birkhoff's Theorem and Scalar Invariants}
\label{subsec:schwarzschild_identification}
This section collects the invariant statements that characterise the Schwarzschild and Type-I geometries. The key point here is not that any single scalar uniquely determines the metric in general, but rather the specification of increasing numbers of invariants constrains the space of possible solutions.

\subsubsection{Petrov Classes, Weyl Invariants, and Curvature Scales}
\label{par:schwarzschild_inv}

Any smooth region of a 4D classical black-hole spacetime has a local Petrov type. A classifier of the type may be defined through the complex speciality index (\cite{BakerCampanelli2000, Stephani:2003tm}). Let \(C_{abcd}\) denote the Weyl tensor and define the self-dual Weyl tensor
\begin{equation}
  \mathcal{C}_{abcd}
  :=
  C_{abcd}
  -
  i\,{}^{\star}C_{abcd}
  \qquad \text{where} \qquad
  {}^{\star}C_{abcd}
  :=
  \frac{1}{2}\epsilon_{ab}{}^{ef}C_{efcd}.
  \label{eq:self_dual_weyl_petrov}
\end{equation}
With our normalisation, the complex Weyl invariants are
\begin{equation}
  I
  :=
  \frac{1}{32}\mathcal{C}_{abcd}\mathcal{C}^{abcd}
  \qquad \text{and} \qquad
  J
  :=
  \frac{1}{384}
  \mathcal{C}_{ab}{}^{cd}
  \mathcal{C}_{cd}{}^{ef}
  \mathcal{C}_{ef}{}^{ab}.
  \label{eq:weyl_petrov_invariants}
\end{equation}
The corresponding \textit{speciality index} is
\begin{equation}
  \mathcal{S}
  :=
  \frac{27J^{2}}{I^{3}},
  \label{eq:speciality_index}
\end{equation}
following the invariant normalisation used by \cite{BakerCampanelli2000}, and is defined away from the conformally flat locus \(I=0\). In this convention, an algebraically special Kerr background has \(\mathcal{S}=1\) (\cite{BakerCampanelli2000}). Schwarzschild is the \(a=0\) Kerr limit; in the Schwarzschild principal tetrad the only non-zero Newman--Penrose Weyl scalar is real, \(\Psi_{2}=-m/r^{3}\) (see \cite{Cherubini2003}). Therefore
\begin{equation}
  I=3\Psi_{2}^{2}=\frac{3m^2}{r^6},
  \qquad
  J=-\Psi_{2}^{3}=\frac{m^{3}}{r^{9}},
  \qquad
  \mathcal{S}_{\rm Schw}=1+0i .
  \label{eq:schwarzschild_weyl_speciality}
\end{equation}
Equivalently, the algebraically special Petrov condition is measured by the discriminant
\begin{equation}
  \Delta_{\rm Petrov}
  :=
  I^{3}-27J^{2}.
  \label{eq:petrov_discriminant}
\end{equation}
Schwarzschild is Petrov type \(\mathrm{D}\), so the invariant target is \(\mathcal{S}=1\), or equivalently \(\Delta_{\rm Petrov}=0\), at every non-flat point. The implication is not reversible: \(\Delta_{\rm Petrov}=0\) is the algebraically special condition and also includes type \(\mathrm{II}\), while the more degenerate types \(\mathrm{III}\), \(\mathrm{N}\), and \(\mathrm{O}\) have \(I=J=0\) and hence a discriminant with an undefined speciality index \cite{Stephani:2003tm}. Petrov type I is associated with a non-zero discriminant. Thus the discriminant is a test for repeated principal null directions, not a complete type-\(\mathrm{D}\) classifier, yet corroborates type D and can be used to determine the solution is \textit{not} type D. The same invariant, \(\mathcal{S}=27J^{2}/I^{3}\), is also used in later local Petrov-classification diagnostics (\cite{RosatoNakanoLousto2021}).

The \textit{Kretschmann invariant} fixes the curvature scale of Schwarzschild giving partial information on the coordinate representation. We write
\begin{equation}
  K := R_{abcd}R^{abcd}
  \label{eq:kretschmann_definition}
\end{equation}
for the quadratic curvature scalar. For the Schwarzschild solution,
\begin{equation}
  K_{\rm Schw}=\frac{48m^{2}}{r^{6}},
  \label{eq:schwarzschild_kretschmann_scale}
\end{equation}
where \(r\) is the areal radius \cite{Cherubini2003}. Thus the dimensionless product
\begin{equation}
  \frac{Kr^{6}}{m^2}=48
  \label{eq:schwarzschild_kr6}
\end{equation}
is constant on the non-singular Schwarzschild region. This provides an invariant radial scale against which the learnt metric can be checked.

One may also employ a useful Schwarzschild-specific cubic curvature ratio. Let
\begin{equation}
  J_{\rm cub}
  :=
  R_{ab}{}^{cd}R_{cd}{}^{ef}R_{ef}{}^{ab}.
  \label{eq:cubic_riemann}
\end{equation}
With the Riemann-sign convention fixed throughout this work, the Schwarzschild solution has \(J_{\rm cub}=96m^{3}/r^{9}\), using \cite{Zakhary1997}.
In the Weyl-invariant normalisation of \cref{eq:weyl_petrov_invariants}, this is equivalent to \(J_{\rm cub}=96\,\mathrm{Re}(J)\) on Schwarzschild, allowing definition of the dimensionless reference scalar
\begin{equation}
  \rhoSchw
  :=
  -\frac{96\,\mathrm{Re}(J)}{|K|^{3/2}}
  =
  -\frac{J_{\rm cub}}{|K|^{3/2}}
  =
  -\frac{1}{\sqrt{12}},
  \label{eq:schwarzschild_cubic_ratio}
\end{equation}
everywhere the metric is defined, with \(|\rhoSchw|=1/\sqrt{12}\); further details are provided in \cref{app:kretschmann_schwarzschild}. 
The value follows from the chosen normalisation of the diagnostic \(\rho\); the literature input is the standard Schwarzschild Weyl and Kretschmann invariants \cite{Cherubini2003}, not the diagnostic name itself. This ratio is simply an additional scalar check of the Schwarzschild curvature pattern, independent of the (areal) radius.

\subsubsection{Killing Vectors and Spherical Symmetry}
\label{par:killing_vectors}
The analytic Schwarzschild solution is spherically symmetric. The relevant infinitesimal symmetries are the three rotations of the round two-sphere. Equivalently, the Killing algebra of the round $S^{2}$ is $\mathfrak{so}(3)$, generated by the infinitesimal rotations inherited from the ambient Euclidean space (\cite{Lee:2018Riemannian}). In the Cartesian embedding $S^{2}\subset\R^{3}$ these are the familiar angular-momentum vector fields
\begin{equation}
  J_x=-z\partial_y+y\partial_z,\qquad
  J_y=z\partial_x-x\partial_z,\qquad
  J_z=-y\partial_x+x\partial_y .
  \label{eq:cartesian_rotation_generators}
\end{equation}
To obtain the coordinate formulae used here, one may write the northern inverse stereographic map as
\begin{equation}
  x=\frac{2q_{1}}{1+|q|^{2}},
  \qquad
  y=\frac{2q_{2}}{1+|q|^{2}},
  \qquad
  z=\frac{1-|q|^{2}}{1+|q|^{2}} \, ,
  \label{eq:north_stereo_rotation_pullback}
\end{equation}
which yields the following generators in coordinates:
\begin{align}
  \xi_{z}&=-q_{2}\,\partial_{q_{1}}+q_{1}\,\partial_{q_{2}},
  \nonumber\\
  \xi_{x}&=-q_{1}q_{2}\,\partial_{q_{1}}
  -\frac{1}{2}(1-q_{1}^{2}+q_{2}^{2})\,\partial_{q_{2}},
  \nonumber\\
  \xi_{y}&=\frac{1}{2}(1+q_{1}^{2}-q_{2}^{2})\,\partial_{q_{1}}
  +q_{1}q_{2}\,\partial_{q_{2}}.
  \label{eq:s2_killing_vectors}
\end{align}
The corresponding formulae on the southern chart differ only by the expected stereographic sign changes for two of the generators. A metric is invariant under these rotations precisely when
\begin{equation}
  (\mathcal{L}_{\xi_A}g)_{\mu\nu}
  =
  \xi_A^{\rho}\partial_{\rho}g_{\mu\nu}
  +g_{\rho\nu}\partial_{\mu}\xi_A^{\rho}
  +g_{\mu\rho}\partial_{\nu}\xi_A^{\rho}
  =0,
  \qquad A=x,y,z.
  \label{eq:killing_equation_s2}
\end{equation}
The Killing equations are the infinitesimal statement of spherical symmetry. In the numerical workflow below, the same operators are used both as a training loss and as validation residuals. This choice makes the Birkhoff--Jebsen hypothesis an explicit part of the Schwarzschild targeted search rather than a purely posterior observation. Equivalently, one may construct and implement any other symmetry property in this manner to perform other targeted searches. 

The Birkhoff--Jebsen theorem states that a four-dimensional spherically symmetric vacuum solution is locally Schwarzschild; for $\lambda\neq0$ the corresponding solution is Kottler/Schwarzschild--de Sitter (\cite{Birkhoff:1923,Jebsen:1921}). Israel's theorem gives a complementary static black-hole result, under the standard assumptions of asymptotic flatness, regular horizon, vacuum equations, and horizon topology, the regular static vacuum black hole is Schwarzschild, as discussed in \cite{Israel:1967wq}. These criteria are intentionally redundant. Ricci flatness and spherical symmetry already place the metric on the local Schwarzschild branch by the Birkhoff--Jebsen theorem, while non-degeneracy excludes collapsed metric limits and the Kretschmann scale fixes the mass parameter. Since each condition is stated tensorially or through scalar curvature invariants, the combined identification is insensitive to the coordinates used to represent the metric -- it is a diffeomorphism-invariant identification of the geometry as Schwarzschild.

\subsection{Spacetime Embedding}

The $S^2$ factor is represented by two stereographic hemispherical charts. This two-chart representation is only a coordinate device. Let
\begin{equation}
  w=(T,X,q_{1},q_{2},p),
  \label{eq:intrinsic_input}
\end{equation}
be an intrinsic coordinate together with a chart label, where $p=0$ denotes the northern chart and $p=1$ denotes the southern chart. We write $q=(q_{1},q_{2})$, $|q|^{2}=q_{1}^{2}+q_{2}^{2}$, and define $s_{0}=+1$, $s_{1}=-1$, allowing the embedding to be written as
\begin{equation}
  W(w)=
  \left(T,X,\frac{2q_{1}}{1+|q|^{2}},
            \frac{2q_{2}}{1+|q|^{2}},
            s_{p}\frac{1-|q|^{2}}{1+|q|^{2}}\right) \, .
  \label{eq:embedding}
\end{equation}
\cref{eq:embedding} maps the four intrinsic coordinates in \cref{eq:intrinsic_input} to $\R^{2}\times S^{2}\subset\R^{5}$. The first two coordinates are unchanged, while the final three are the inverse stereographic projection of the angular point.

The embedding Jacobian $J^{A}{}_{\mu}=\partial W^{A}/\partial w^{\mu}$ reads
\begin{equation}
  J=
  \begin{pmatrix}
  1&0\\
  0&1
  \end{pmatrix} \oplus 
  \frac{1}{(1+q_{1}^{2}+q_{2}^{2})^{2}}
  \begin{pmatrix}
  2(1+q_{2}^{2}-q_{1}^{2}) & -4q_{1}q_{2}\\
  -4q_{1}q_{2} & 2(1+q_{1}^{2}-q_{2}^{2})\\
  -4s_{p}q_{1} & -4s_{p}q_{2}
  \end{pmatrix}.
  \label{eq:sphere_embedding_block}
\end{equation}
The pullback of the flat Euclidean metric on the last three coordinates through \cref{eq:embedding} gives \cref{eq:stereo_s2}. The sign $s_{p}$ changes which pole is represented by $q=0$, while the induced round metric is the same in both charts. This equality of the induced metric is the mathematical reason the two hemispheres can be used as one embedded angular manifold.

All tensorial calculations on $\mathcal{P}\times S^{2}$ are taken with respect to $(T,X,q_{1},q_{2}) = w$. The two sets of coordinates therefore define a single embedded sphere, rather than two a-priori independent charts (which was the case in \cite{Hirst:2025seh}).

\subsection{Trapped Surfaces}
\label{subsec:trapped_surfaces}

The invariants of \cref{subsec:schwarzschild_identification} characterise the Schwarzschild solution among vacuum geometries, but the defining property of a \textit{black hole} is causal rather than curvature-based: the presence of a horizon, and locally of \textit{trapped surfaces}. The event horizon is a global, teleological (physical) object, the boundary of the causal past of future null infinity, and cannot be evaluated pointwise on a bounded domain. The quasi-local notion of a trapped surface \cite{Penrose:1964wq, Hayward:1993wb, Ashtekar:2004cn} is therefore used, which is local and requires no symmetry.

Let $\Sigma$ be a closed spacelike two-surface with the two future-directed null normals $\ell,n$ (normalised by $g(\ell,n)=-1$), and let $q_{\mu\nu}=g_{\mu\nu}+\ell_\mu n_\nu+n_\mu\ell_\nu$ project onto $\Sigma$. The null expansions
\begin{equation}
  \theta_\ell=q^{\mu\nu}\nabla_\mu\ell_\nu,
  \qquad
  \theta_n=q^{\mu\nu}\nabla_\mu n_\nu,
  \label{eq:null_expansions}
\end{equation}
give the fractional rate of change of the area element of $\Sigma$ along the two light fronts emanating from it. In a normal region the outgoing family diverges and the ingoing family converges, $\theta_\ell>0>\theta_n$. The surface is \textit{(future) trapped} when both expansions are negative, $\theta_\ell,\theta_n<0$, so that even the nominally outgoing light is dragged inward; $\theta_\ell=0$ is the marginal (apparent-horizon) case, and $\theta_\ell,\theta_n>0$ is the time-reversed, anti-trapped (white-hole) case. By the Penrose singularity theorem (\cite{Penrose:1964wq}) a closed trapped surface, together with the null energy condition (saturated in vacuum), forces geodesic incompleteness and lies within the black-hole region; trapped surfaces are the standard local certificate of a black-hole interior.

The manifold \cref{eq:model_manifold} supplies a natural family of such surfaces: the $S^2$ fibres at fixed $(T,X)$. Their induced metric is exactly the angular block of the pullback \cref{eq:pullback_metric}, with no symmetry assumed, so each fibre has a well-defined area $A=4\pi R^{2}$ and areal radius $R$. It is convenient to encode the trapping in the single scalar
\begin{equation}
  \Xi:=g^{\mu\nu}\,\partial_\mu R\,\partial_\nu R,
  \qquad
  \operatorname{sign}\Xi=-\operatorname{sign}(\theta_\ell\theta_n),
  \label{eq:trapped_scalar}
\end{equation}
as the squared norm of the areal-radius gradient, where $\Xi>0$ (spacelike $\nabla R$) is untrapped, $\Xi=0$ is marginal, and $\Xi<0$ (timelike $\nabla R$) is trapped or anti-trapped, the two being distinguished by the common sign of the expansions. In spherical symmetry $\Xi$ reduces to the Misner--Sharp combination $1-2M_{\rm MS}/R$ of \cref{eq:misner_sharp_def} \cite{MisnerSharp1964}, which for Schwarzschild equals $1-2m/r$ and is negative precisely on the interior $r<2m$.

Two remarks matter for the non-spherically-symmetric metrics learnt in \cref{subsec:blackhole_results}. First, the Misner--Sharp \textit{mass} is only invariantly defined under spherical symmetry and is unavailable in general, whereas the trapped condition $\theta_\ell,\theta_n<0$ is symmetry-free and remains the operative definition. Second, $\Xi$ is a smooth scalar built from the metric and its first derivatives, hence a natural differentiable training signal, while the individual expansions $\theta_\ell,\theta_n$ -- which alone distinguish a black hole from its time reverse -- are reserved for the posterior certification. The identification \cref{eq:trapped_scalar} is exact for round fibres and is corrected only by their departure from roundness, which we find to be small.

%%%%%%%%%%%%%%%%%%%%%%%%%%%%%%%%%%%%%%%%%%%%%%%
\section{AInstein Embedding Neural Networks}
\label{sec:ainstein_networks}

\subsection{Warm up -- Local Lorentzian Schwarzschild}\label{sec:architecture_local}
Before considering the embedded Schwarzschild problem, we study a local two-dimensional Lorentzian simplification. In this mode the manifold is a single coordinate ball, the dimension is set to \(d=2\), and no \(S^{2}\) embedding is used. The network therefore represents the metric directly on the local coordinates \(u=(u^{0},u^{1})\). Its outputs are  three independent entries of a lower-triangular \(2\times2\) matrix \(L_{\theta}(u)\), and the Lorentzian metric is obtained from an indefinite-Cholesky parametrisation,
\begin{equation}
  g^{(\theta)}_{\mu\nu}(u)
  =
  \left(L_{\theta}(u)\eta_{(2)}L_{\theta}(u)^{\mathsf{T}}\right)_{\mu\nu},
  \qquad
  \eta_{(2)}=\diag(-1,1),
  \qquad
  \mu,\nu=0,1.
  \label{eq:local_lorentz_cholesky}
\end{equation}
This local architecture isolates the Lorentzian signature and curvature calculation from the additional global structure of \(\mathcal{P}\times S^{2}\) later considered.

The network has a sets of four 128 neuron-wide linear layers, and employs GELU activations; GELU allows derivatives to be $C^2$ smooth, as needed for metric derivatives. To mitigate confusion, it is emphasised that the network is trained such that all \textit{Schwarzschild losses} later defined in \cref{ssub:unsupervised_sch} are disabled. Only this Lorentzian Einstein condition is used for each of $\lambda \in \{-1,0,1\}$.

\subsection{Network Architecture}
\label{sec:architecture}

For the central, \textit{global}, architecture, let $\theta$ denote the trainable weights and biases of the numerical model. The neural architecture is built using an internal multilayer perceptron
\begin{equation}
  \N_{\theta}:\R^{5}\longrightarrow\R^{15}.
  \label{eq:network_map}
\end{equation}
This map in \cref{eq:network_map} takes the embedded point $W(w)=(T,X,x(q_1, q_2),y(q_1, q_2),z(q_1, q_2))$, where $(x,y,z)$ are the Cartesian sphere coordinates implicitly defined in \cref{eq:embedding}, and outputs the independent entries of a lower-triangular $5\times5$ matrix $L_{\theta}$. The use of this lower-triangular matrix enforces the symmetric property of the metric under the below combination meaning there are fewer components to learn. The ambient Lorentzian metric is
\begin{equation}
  G_{\theta}=L_{\theta}\eta L_{\theta}^{\mathsf{T}},
  \qquad
  \eta=\diag(-1,1,1,1,1).
  \label{eq:lorentz_cholesky}
\end{equation}
The four-dimensional metric used in the loss is the pullback
\begin{equation}
  g^{(\theta)}_{\mu\nu}(w)
  =
  (G_{\theta})_{AB}(W(w))
  J^{A}{}_{\mu}(w)J^{B}{}_{\nu}(w) \,
  .
  \label{eq:pullback_metric}
\end{equation}
Here $\mu,\nu=0,\ldots,3$ and $A,B=0,\ldots,4$. The biases of the architecture's output layer are initialised so that the model is initialised near the flat ambient metric $\eta$. In the limit that the standard deviation of the sample distribution tends to 0, one recovers the flat Minkowski space metric.

As mentioned, this is where the embedded sphere changes the original AInstein construction. In \cite{Hirst:2025seh}, separate chart metrics were matched by an overlap loss. Here both stereographic hemispheres are obtained by the push-forward of the same ambient metric through \cref{eq:embedding}, so their agreement is built into the parametrisation and no angular overlap loss is needed; any learnt metric is automatically globally consistent on the manifold.

\subsection{Data Generation}
\label{sec:Sampling}

To sample a general point on the $\mathbb{R}^2 \times S^2$ manifold, the Penrose sampler first draws a point in a 2-dimensional disc and then maps it to the compactified Penrose domain. This approach is more efficient than rejection sampling in the polygonal Penrose diagram, and gives direct control over the radial density of a sample.

\begin{figure}[!t]
    \centering
    \captionsetup[subfigure]{justification=centering,singlelinecheck=true}
    \begin{subfigure}{0.32\textwidth}
        \centering
        \includegraphics[width=0.98\textwidth]{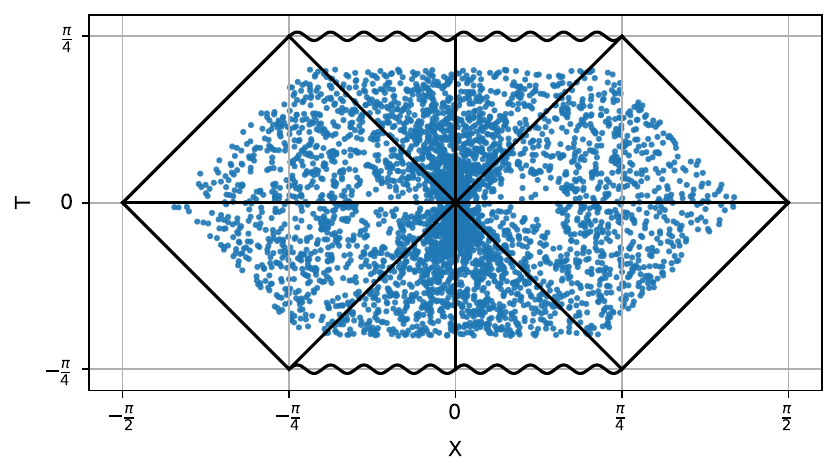}
        \caption{Penrose diagram}
    \end{subfigure}
    \begin{subfigure}{0.32\textwidth}
        \centering
        \includegraphics[width=0.98\textwidth]{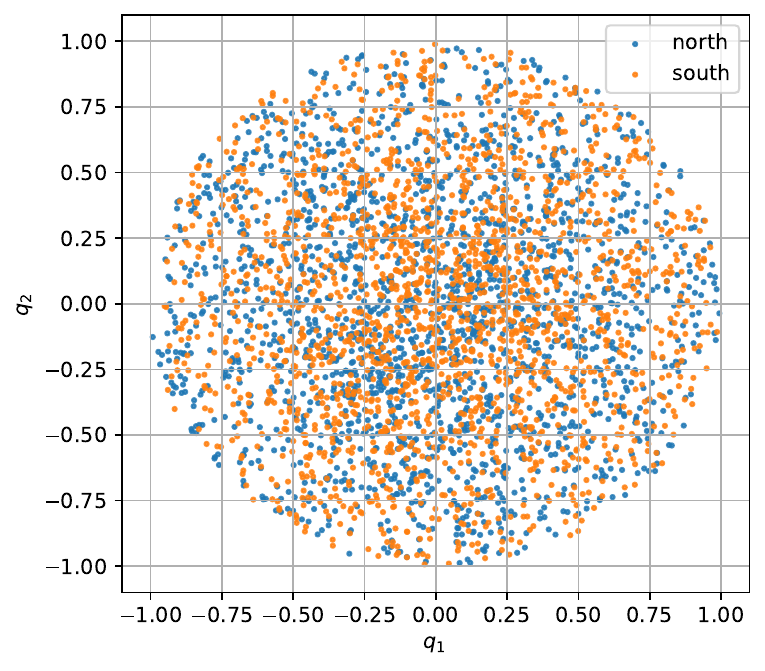}
        \caption{$S^2$ Stereographic Patches}
    \end{subfigure}
    \begin{subfigure}{0.32\textwidth}
        \centering
        \includegraphics[width=0.98\textwidth]{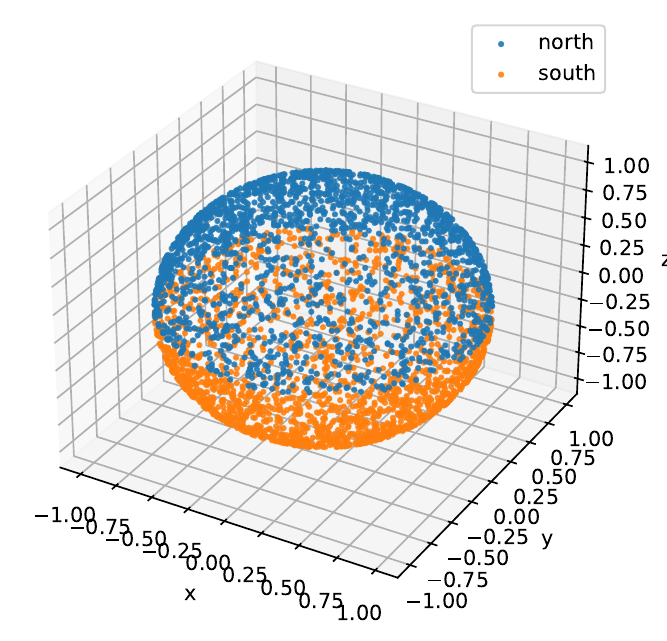}
        \caption{$S^2$ Embedded}
    \end{subfigure}
    \caption{The sampling distribution, used in training and testing, shown for 2000 points, across the (a) Penrose diagram, (b) stereographic patches of the $S^2$ sphere, and (c) the embedded $S^2$ sphere.}
    \label{fig:penrose_sampling_dist}
\end{figure}

Let $R_{\mathcal{P}}$ be the maximum radius used in the auxiliary unit disc. We draw a disc point $z=\rho u$, where $0\leq\rho\leq R_{\mathcal{P}}\leq1$, $u=(u_{T},u_{X})$ is a unit vector, and $|u|=1$. In the implementation
\begin{align}
  \rho&=R_{\mathcal{P}}\,b,
  \nonumber\\
  b&\sim{\rm Beta}(\alpha_{\mathcal{P}}^{-1},\alpha_{\mathcal{P}}),
  \qquad \arg u\sim{\rm Unif}[0,2\pi),
  \label{eq:disc_sampling}
\end{align}
where $\alpha_{\mathcal{P}}$ is the density-power hyperparameter. Define the offset distance
\begin{equation}
  \delta(\rho)=\frac{\pi}{4}(1-\rho)
  \label{eq:penrose_offset}
\end{equation}
and the three radial boundary scales
\begin{equation}
\begin{split}
  A(u)&=\frac{\pi/4-\delta}{|u_{T}|},\\
  B(u)&=\frac{\pi/2-\sqrt2\,\delta}{|u_{T}+u_{X}|},\\
  C(u)&=\frac{\pi/2-\sqrt2\,\delta}{|u_{X}-u_{T}|}.
  \label{eq:boundary_scales}
\end{split}
\end{equation}
The sampled Penrose point is obtained using the directional scale $\ell(u)$,
\begin{equation}
  (T,X)=\ell(u)\,u,
  \qquad
  \ell(u)=\min\{A(u),B(u),C(u)\},
  \label{eq:disc_to_penrose}
\end{equation}
with the limiting prescription implied by \cref{eq:boundary_scales} when a denominator vanishes. 
A representative sample of 2000 points for $R_\rho=0.85$ is shown in Figure \ref{fig:penrose_sampling_dist}, where the ill-defined nature of the singularity at the boundaries motivates this use of a $R_\rho < 1$.
At $\rho=1$, \cref{eq:disc_to_penrose} reaches the full diagram boundary; at $\rho=0$, it collapses to the centre.

\subsection{Training Objectives}
\label{sec:Training}

The first step of training a neural architecture is the specification of a training objective in the form of a loss function. In this section, the training objectives for supervised training against the known target solutions, and then four related problems are outlined. First, local two-dimensional Lorentzian Einstein runs are used as a controlled signature test. We then consider the task of constructing a simulacrum of the classic Schwarzschild black hole, before introducing two general searches for Petrov type I solutions: the first a search agnostic to solution style, and the second specifically searching for black hole trapped surfaces.

\subsubsection{Supervised Network}

The supervised model, where training instead approximates a known construction directly, is used for initialisation and benchmarking. The base network still outputs the ambient Cholesky vector, while a wrapper applies \cref{eq:pullback_metric} and returns the flattened intrinsic metric. For training samples $w_{a}$ and target metric $g^{\rm tar}_{\mu\nu}$, the default supervised loss is the componentwise relative mean-squared error
\begin{equation}
  \mathcal{L}_{\rm sup}
  =
  \frac{1}{16N}\sum_{a=1}^{N}
  \sum_{\mu,\nu}
  \frac{
  \left(g_{\mu\nu}^{(\theta)}(w_{a})-g_{\mu\nu}^{\rm tar}(w_{a})\right)^{2}}
  {\left(g_{\mu\nu}^{\rm tar}(w_{a})\right)^{2}+1}.
  \label{eq:supervised_loss}
\end{equation}
In these investigations, the target metric is either the Lorentzian identity metric on $\R^{2}\times \R^{3}$ (useful for a neutral initialisation), or the analytic Schwarzschild metric in \cref{eq:schw_metric} (useful for benchmarking the general unsupervised search performance).

\medskip

\subsubsection{Unsupervised Local Lorentzian Network}\label{ssub:unsupervised_local}

For the local two-dimensional calibration runs, for each training sample $w_{a}$, define an Einstein error tensor
\begin{equation}
  E_{\mu\nu}(w_{a})
  :=
  R_{\mu\nu}[g^{(\theta)}](w_{a})
  -\lambda g^{(\theta)}_{\mu\nu}(w_{a}), 
  \label{eq:einstein_residual}
\end{equation}
which is used to define the Einstein loss
\begin{equation}\label{eq:einstein_loss_local}
\mathcal L_{E, \text{Local}} := \frac{1}{N} \sum_{a=1}^N \bigg|E_{\mu\nu}\big(g_E^{(\theta)}\big)^{\mu\alpha}\big(g_E^{(\theta)} \big)^{\nu\beta}E_{\alpha\beta}
  \bigg|_{w_a}
\end{equation}
where $g_E$ is the \textit{Euclideanised} variant of the metric\footnote{By this we mean $g_E$ is obtained by the same vielbein as $g$, but using $\delta$ as the flat metric, instead of $\eta$.}.
The network described in \cref{sec:architecture_local} is then trained with this $\mathcal L_{E, \text{Local}}$ loss only, for \(\lambda\in\{+1,0,-1\}\).

\medskip

\subsubsection{Unsupervised Schwarzschild Network}\label{ssub:unsupervised_sch}

For the four-dimensional Schwarzschild and general Petrov type searches, the vacuum equation is $R_{\mu\nu}=0$, so $\lambda=0$ is set\footnote{Note the code also has functionality for general $\lambda$ search.}.
The default Schwarzschild Lorentzian Einstein loss is then defined as the inverse-metric contraction
\begin{equation}
  \mathcal L_{E,\text{Schw}}
  :=
  \frac{1}{N}\sum_{a=1}^{N}
  \frac{1}{\text{min}(|K|, K_\text{cap}) + \epsilon}\left|
  E_{\mu\nu}\big(g^{(\theta)}_E\big)^{\mu\alpha}\big(g^{(\theta)}_E \big)^{\nu\beta}E_{\alpha\beta}
  \right|_{w_{a}},
  \label{eq:contracted_einstein_loss_sch}
\end{equation}
where once more $g_E$ is the \textit{Euclideanised} version of $g$, producing a well-defined scalar, since all objects involved are tensorial (as opposed to using MSE). The numerical prefactor\footnote{This and other losses have $\epsilon$ terms in denominators as numerical guards. Unless otherwise specified, $\epsilon=10^{-12}$.} involves a local scalar weighting which uses $K_\text{cap} = \kappa K_\text{hor}$ to cap this factor to the horizon value $K_\text{hor}=0.75$. The numerator of \cref{eq:contracted_einstein_loss_sch} is simply the most basic statement of the Einstein constraint which is consistent between all problems (and indeed \cite{Hirst:2025seh}).

% This loss can be further improved by adding a point-wise weighting, which needs to be a coordinate scalar to yield another geometrically well-defined quantity. Such choice needs to be motivated by physical reasoning, geometric naturalness and/or numerical considerations. In our case, we considered the two following variations of the Einstein loss: \\
% ? Insert the two Einstein losses with different normalisations here. \AS{I'm on it!}
% and the loss is multiplied sample-wise by $\sqrt{|\det g|}$ which weights the loss contribution according to the local geometry, matching the traditional integral measure.

Secondly, the curvature-scale loss is imposed through the quadratic Weyl invariant \(I\), rather than directly through the full Kretschmann scalar. For a Ricci-flat metric the Riemann and Weyl tensors agree, but away from the vacuum locus the Kretschmann scalar also contains Ricci contributions. Using \(I\) therefore decouples the scale-fixing term from the Einstein residual: \(\LE\) drives the Ricci tensor to zero, while the Weyl term fixes the non-Ricci curvature scale of the Schwarzschild branch.

With the normalisation in \cref{eq:weyl_petrov_invariants}, the exact Schwarzschild solution obeys
\begin{equation}
  |I_{\rm Schw}|=\frac{3m^{2}}{r^{6}}
  \quad \implies \quad
  \sqrt{|I_{\rm Schw}|}\,r^{3}=\sqrt{3}\,m .
  \label{eq:schwarzschild_weyl_scale}
\end{equation}
For each sampled point, the implemented Weyl loss is therefore
\begin{equation}
  \LW
  :=
  \frac{1}{N}\sum_{a=1}^{N}
  \left(
  r(w_a)^{3}
  \sqrt{|I_\theta(w_a)|+\epsilon}
  -
  \sqrt{3}\,m
  \right)^{2}.
  \label{eq:weyl_invariant_loss}
\end{equation}
Here \(r(w_a)\) is the analytic areal radius of the sampled Penrose point, and \(I_\theta\) is computed from the self-dual Weyl tensor of the learnt metric. The square-root form removes the six-power radial dynamic range of the raw invariant, while still excluding the flat \(I=0\) branch.

The Birkhoff-motivated symmetry term is
\begin{equation}
  \mathcal{L}_{SO(3)} :=
  \frac{1}{3N}
  \sum_{A=x,y,z}\sum_{a=1}^{N}
  \frac{
  \left|
  (\mathcal{L}_{\xi_A}g)_{\mu\nu}
  (g_E)^{\mu\alpha}(g_E)^{\nu\beta}
  (\mathcal{L}_{\xi_A}g)_{\alpha\beta}
  \right|_{w_a}}
  {
  \left|
  g_{\mu\nu}(g_E)^{\mu\alpha}(g_E)^{\nu\beta}g_{\alpha\beta}
  \right|_{w_a}
  +\epsilon},
  \label{eq:symmetry_operator_loss}
\end{equation}
where the three operators \(\xi_A\) are the rotational fields in \cref{eq:s2_killing_vectors}, $\mathcal L_{\xi_A}$ are the usual symmetry generators, and \(g_E\) is the same Euclideanised metric used in \cref{eq:contracted_einstein_loss_sch}. In the implementation this term is evaluated by a small finite-difference pullback along each rotational flow. Its role is to impose the spherical-symmetry assumption needed for the Birkhoff--Jebsen identification, which it does by measuring the metric-contracted Killing-vector action along each basis direction relative to the metric-contracted size of \(g\). This action should be zero in each case according to \cref{eq:killing_equation_s2}. Together with the vacuum Einstein loss, this loss selects the Schwarzschild branch locally, while the Weyl curvature-scale target fixes scaling over the Penrose diagram to ensure it is fully covered and not a more trivial diffeomorphism.

Finally, in order to aid numerical stability, it is necessary to add a `determinant loss', $\LD$, which prevents the $R^2$ block becoming degenerate during training. Let $g^{(\theta)}_{(T,X)}$ denote the upper-left $2\times2$ block of the learnt metric in the Penrose directions. The determinant loss is defined by
\begin{equation}
  \LD
  :=
  \frac{1}{N}\sum_{a=1}^{N}
  \frac{1}
       {\big|\det g^{(\theta)}_{(T,X)}(w_{a})\big|+\epsilon},
  \label{eq:r2_det_loss}
\end{equation}
and prevents the degenerate attractor in which the two-dimensional Lorentzian block components flow to 0 whilst reducing the curvature terms. 

The Schwarzschild-search objective becomes
\begin{equation}
  \Ltot^{\rm Schw}(t)
  :=
  \frac{\alpha_{\rm E}(t) \mathcal L_{E,\text{Schw}} + \alpha_{W}(t)\LW + \alpha_{SO(3)}(t)\mathcal{L}_{SO(3)} + \alpha_{\det}(t)\LD}
       {\alpha_{\rm E}(t)+\alpha_{W}(t)+\alpha_{SO(3)}(t)+\alpha_{\det}(t)},
  \label{eq:total_loss}
\end{equation}
where $t$ denotes the training epoch, and the $\alpha_i(t)$ are the scheduled multipliers of the active loss terms which change the weighting of various loss components throughout training. The training samples may optionally be regenerated at every epoch using the current scheduled density powers, while the validation set is held fixed. Batches are shuffled through a dataset pipeline, and the implementation facilitates Ricci, Kretschmann, and Killing-diagnostic quantity evaluation either by the standard two-tape kernel or by the optimised forward-mode kernel; validation is computed in batches. %If a batch produces an inversion failure, it is skipped and logged; repeated failures stop the run.

Post training, on an independent test data sample the training losses are computed to evaluate performance.
In addition to the losses, the 3 invariant scalars introduced in \cref{par:schwarzschild_inv} are computed and compared to the expected values for the Schwarzschild solution as further validation of the found solution the architecture is approximating.
These include matching the speciality index $\mathcal{S} = 1 + 0i$, the Kretschmann scalar $\frac{Kr^6}{m^2} = 48$, and the cubic curvature ratio $\rho_{\text{Schw}} = -\frac{1}{\sqrt{12}}$.

The Schwarzschild-targeted experiment trains Ricci flatness, the Weyl curvature scale, spherical symmetry, and non-degeneracy, and then asks whether these independent invariant diagnostics agree with the Schwarzschild/type-\(\mathrm{D}\) values. 
By Birkhoff's theorem, if the resulting solution is spherically symmetric and vacuum, it is guaranteed to be locally Schwarzschild; these scalar checks provide additional confidence that the resulting metric is indeed a Schwarzschild solution.

\subsubsection{Unsupervised Petrov Type-I Network}
\label{subsubsec:petrov_repeller_network}
The Schwarzschild experiment above uses the quadratic Weyl loss to fix the curvature scale and then checks the Petrov speciality index and Schwarzschild cubic invariant a posteriori. A complementary search can instead use the speciality index in the loss function as a repeller from the Schwarzschild/type-\(\mathrm{D}\) value \(\mathcal{S}=1\). This is motivated by the role of \(\mathcal{S}\) as an invariant measure of deviation from algebraically special Kerr/Schwarzschild geometry \cite{BakerCampanelli2000} and by its later use in local Petrov-classification diagnostics \cite{RosatoNakanoLousto2021}. These runs do not include the analytic Schwarzschild Kretschmann loss or the \(SO(3)\) symmetry loss, so the network is not supplied with the Schwarzschild radial curvature profile.

Beginning by stating a variant of the Einstein loss, \cref{eq:contracted_einstein_loss_sch}, which is homothety\footnote{A homothety transform is simply a transformation that scales distances from a fixed point, known as the centre, by a given ratio.} invariant, the type I Einstein loss is defined as
\begin{equation}
\mathcal{L}_{E,I}
:=
\frac{1}{N}
\sum_{a=1}^{N}
\frac{1}{|I_\theta(w_a)|+\epsilon}
\left.
\left|
E_{\mu\nu}
\left(g_E^{(\theta)}\right)^{\mu\alpha}
\left(g_E^{(\theta)}\right)^{\nu\beta}
E_{\alpha\beta}
\right|
\right|_{w_a}.
\label{eq:type-i-einstein-loss}
\end{equation}

For each sample with Weyl invariants \(I(w_a)\) and \(J(w_a)\), define the speciality index as in \cref{eq:speciality_index}. The implemented Petrov-profile term fits a prescribed nonconstant profile \(\mathcal{S}_{\rm targ}(w_a)\) centred away from the algebraically special value \(1\):
\begin{equation}
  \LP^{\rm profile}
  :=
  \frac{1}{N}\sum_{a=1}^{N}
  \left|
  \frac{27J(w_a)^{2}}{I(w_a)^{3}+\epsilon_I}
  -
  \mathcal{S}_{\rm targ}(w_a)
  \right|^{2},
  \qquad
  \mathcal{S}_{\rm targ}(w_a)\not\equiv1 .
  \label{eq:petrov_repeller_loss}
\end{equation}
In the runs reported below the profile is quadrupolar in nature\footnote{The bounded quadrupolar target profile used was inspired by \cite{Quevedo:2012ttw}, with motivation from the Legendre-polynomial multipolar structure of Weyl/Zipoy-Voorhees-type static axisymmetric solutions. This choice can easily be generalised in the codebase for other searches.} and centred near the real value \(2\), so agreement with \(\mathcal{S}_{\rm targ}\) explicitly penalises collapse back to the Schwarzschild/type-\(\mathrm{D}\) value. The target profile is an arbitrary numerical training choice rather than a canonical closed-form Ricci-flat solution from the literature, and any complex value would work equivalently well. Since \(\mathcal{S}=1\) characterises the algebraically special type-\(\mathrm{D}\) value for Kerr/Schwarzschild, this term is best understood as a conservative device for searching away from that basin, not on its own a positive classifier for a particular alternative Petrov type. The retained \(I\) and \(J\) diagnostics are therefore still needed to check that the model has not collapsed to a conformally flat or otherwise degenerate case where the speciality index is ill-conditioned.

In practice, we also include a bounded curvature repeller
\begin{equation}
  \LKR
  :=
  \frac{1}{N}\sum_{a=1}^{N}
  \frac{\epsilon_K}{|K_\theta(w_a)|+\epsilon},
  \label{eq:k_repeller_loss}
\end{equation}
to discourage collapse to a conformally flat or nearly flat state. This term does not supply the Schwarzschild Kretschmann target; it only acts as a soft barrier against the \(K_\theta\simeq0\) failure mode.

The corresponding objective is thus
\begin{equation}
  \Ltot^{\rm I}(t)
  :=
  \frac{
  \alpha_{\rm E}(t) \mathcal L_{E,I}
  +\alpha_{\rm P}(t)\LP^{\rm profile}
  +\alpha_{\det}(t)\LD
  +\alpha_{K{\rm rep}}(t)\LKR}
  {\alpha_{\rm E}(t)+\alpha_{\rm P}(t)+\alpha_{\det}(t)+\alpha_{K{\rm rep}}(t)}.
  \label{eq:non_petrov_d_loss}
\end{equation}

\subsubsection{Unsupervised Petrov Type-I Black Hole Network}
\label{subsubsec:blackhole_network}

The type-I search of \cref{subsubsec:petrov_repeller_network} drives the geometry away from the algebraically special locus while keeping it numerically Ricci-flat, but it fixes neither an absolute curvature scale nor any causal structure.
The homothety-invariant loss $\mathcal L_{E,I}$ is insensitive to constant rescalings $g\mapsto\Lambda^{2}g$, so the resulting type-I vacua carry no distinguished horizon and are not, in themselves, black holes. To promote them to black holes we add two terms: a horizon curvature anchor that fixes the scale and enforces a regular finite-curvature horizon, and a trapped-surface term that installs the causal structure of \cref{subsec:trapped_surfaces}.

The anchor uses the same degree-two Weyl scalar as the Schwarzschild curvature-scale loss \cref{eq:weyl_invariant_loss}, but Gaussian-localised (sample-wise weighted by $\phi_a^{-1}$) to the horizon band $r\simeq2m$ rather than imposed everywhere,
\begin{equation}
  \mathcal L_{\rm hor}
  :=
  \frac{\sum_{a}w_a\big(r(w_a)^{3}\sqrt{|I_\theta(w_a)|+\epsilon}-\sqrt3\,m\big)^{2}}
       {\sum_{a}\phi_a},
  \qquad
  \phi_a=\exp\!\left[-\left(\frac{r(w_a)-2m}{\sigma\,2m}\right)^{2}\right].
  \label{eq:horizon_anchor_loss}
\end{equation}
As the target $\sqrt{|I_{\rm Schw}|}\,r^{3}=\sqrt3\,m$ of \cref{eq:schwarzschild_weyl_scale} is a finite constant, anchoring it at $r=2m$ simultaneously fixes the overall scale -- so that the Kretschmann invariant attains its physical value $K_{\rm hor}$ at the horizon, rather than being sent to zero by the homothety freedom -- and forbids a divergent, naked curvature there. Imposing the constraint at a single radius, instead of throughout the domain as in \cref{eq:weyl_invariant_loss}, leaves the bulk profile free to remain type-I rather than collapsing back to Schwarzschild. This term subsumes and replaces the curvature repeller \cref{eq:k_repeller_loss}.

The trapped-surface term acts on the scalar $\Xi$ of \cref{eq:trapped_scalar}, evaluated for the learnt metric by automatic differentiation of the fibre areal radius $R$. A black-hole interior requires $\Xi<0$ inside $r<2m$ and $\Xi\to0$ at the horizon; we impose this one-sidedly,
\begin{equation}
  \mathcal L_{\rm trap}
  :=
  \frac1N\sum_{a=1}^{N}
  \Big[\max\!\big(0,\ \beta(r_a)-\operatorname{sign}(r_a-2m)\,\Xi_\theta(w_a)\big)\Big]^{2},
  \qquad
  \beta(r)=\Big(\beta_0+\beta_1\frac{\max(0,2m-r)}{2m}\Big)\mathbf 1_{r<2m}.
  \label{eq:trapped_loss}
\end{equation}
The sign factor flips the target across the horizon, requiring $\Xi\le-\beta$ on the interior and $\Xi\ge0$ on the exterior. The margin $\beta(r)$ is the Schwarzschild-linearised trapped depth $\beta_1(2m-r)/2m$, which vanishes at the horizon and grows inward, plus a small floor $\beta_0>0$ that maintains a non-zero gradient just inside $r=2m$, where the linear part is weak. The one-sided $\max(0,\cdot)$ penalises only insufficient trapping and leaves $\Xi$ free to be more negative, so that the type-I radial profile is not pinned to the Schwarzschild one. Since $\Xi$ is invariant under time reversal, $\mathcal L_{\rm trap}$ enforces $\Xi<0$ but not the black-hole orientation $\theta_\ell,\theta_n<0$ specifically; the orientation is checked a posteriori.

The black-hole search objective is therefore
\begin{equation}
  \Ltot^{\rm BH}(t)
  :=
  \frac{\alpha_{\rm E}(t)\mathcal L_{E,I}
        +\alpha_{\rm P}(t)\LP^{\rm profile}
        +\alpha_{\rm hor}(t)\mathcal L_{\rm hor}
        +\alpha_{\rm trap}(t)\mathcal L_{\rm trap}
        +\alpha_{\det}(t)\LD}
       {\alpha_{\rm E}(t)+\alpha_{\rm P}(t)+\alpha_{\rm hor}(t)+\alpha_{\rm trap}(t)+\alpha_{\det}(t)}.
  \label{eq:blackhole_loss}
\end{equation}
Relative to the type-I objective \cref{eq:non_petrov_d_loss}, this drops the curvature repeller and adds $\mathcal L_{\rm hor}$ and $\mathcal L_{\rm trap}$; the trapping weight $\alpha_{\rm trap}$ is switched on only after a warm-up during which the vacuum, type-I and horizon terms are first established.

%%%%%%%%%%%%%%%%%%%%%%%%%%%%%%%%%%%%%%%%%%%%%%%
\section{Results}

This section reports the results from each numerical investigation in the search of Einstein metrics using the architectures described in \cref{sec:architecture}. All runs hyperparameters are stated in \cref{tab:best-einstein-loss}.

\subsection{Local Lorentzian Searches}
\label{subsec:local_lorentzian_results}

As a local calibration before the four-dimensional searches, we train two-dimensional Lorentzian Einstein metrics on a single ball patch (contractible, topologically trivial) for Einstein constants $\lambda\in\{+1,0,-1\}$. These runs use the same Lorentzian metric parametrisation and the Einstein residual in \cref{eq:einstein_residual}, with no other terms in the loss function. 

\cref{tab:local_losses} reports the global test losses for the three Einstein constants, together with the corresponding mean values of $\det(g)$ over the sampled points. The losses are small in all three cases, with the Ricci-flat case giving the lowest Einstein loss. %snce init is minkowski which satisfies this property
The determinant values remain negative and away from zero, indicating that the learnt metrics retain Lorentzian signature rather than collapsing towards a degenerate solution.

\begin{table}[!t]
\centering
\begin{tabular}{lcccccc}
\toprule
Measure
& \multicolumn{2}{c}{$\lambda=+1$}
& \multicolumn{2}{c}{$\lambda=0$}
& \multicolumn{2}{c}{$\lambda=-1$} \\
\cmidrule(lr){2-3}\cmidrule(lr){4-5}\cmidrule(lr){6-7}
& Mean & Std. & Mean & Std. & Mean & Std. \\
\midrule
Einstein loss
& $1.89\times10^{-4}$ & $8.76\times10^{-5}$
& $2.65\times10^{-8}$ & $2.79\times10^{-8}$
& $1.07\times10^{-4}$ & $6.30\times10^{-5}$ \\
$\det(g)$
& $-0.460$ & $0.022$
& $-1.402$ & $0.033$
& $-0.491$ & $0.027$ \\
\bottomrule
\end{tabular}
\caption{Global test losses averaged over 10 runs for neural-network approximations of two-dimensional Lorentzian Einstein metrics on single ball patches, together with the mean metric determinant over the sampled points. Values are reported to three significant figures, with standard deviations computed across the 10 runs.}
\label{tab:local_losses}
\end{table}

\begin{figure}[!t]
    \centering
    \captionsetup[subfigure]{justification=centering,singlelinecheck=true}
    \begin{subfigure}{0.32\textwidth}
        \centering
        \includegraphics[width=0.98\textwidth]{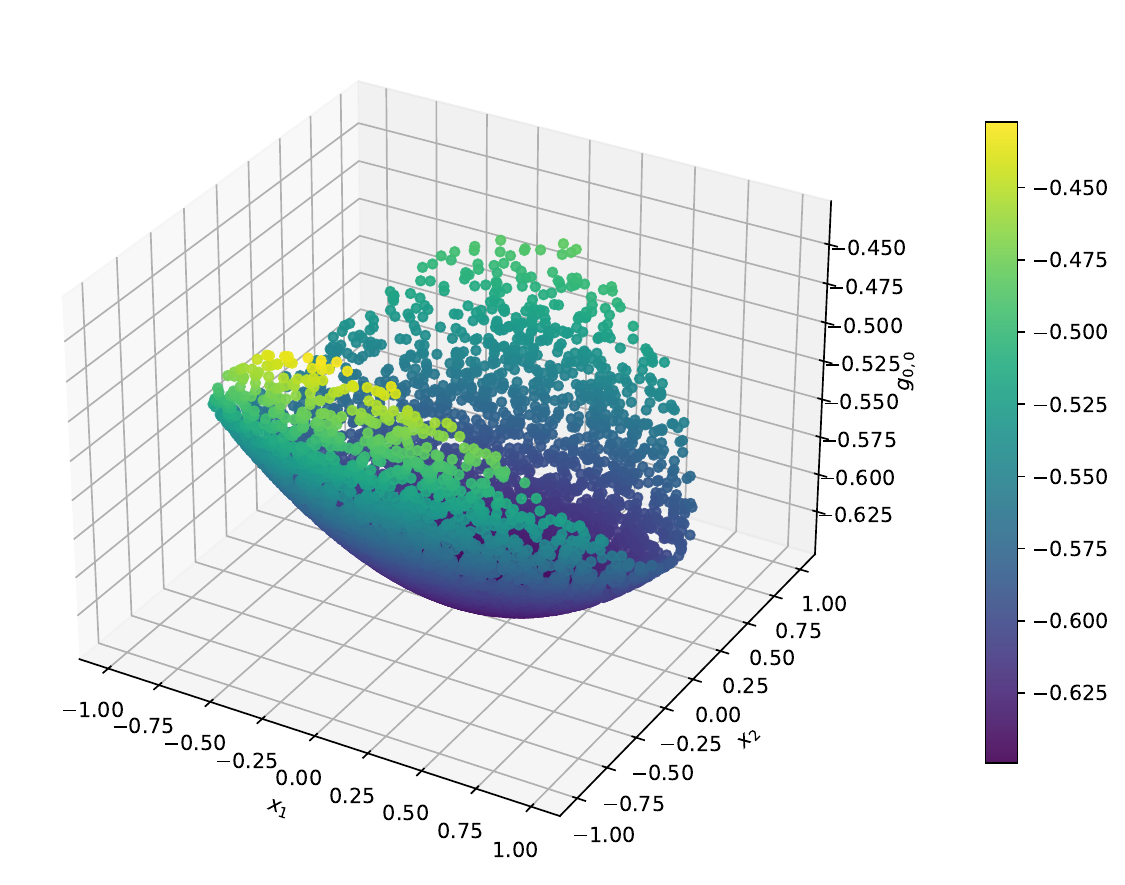}
        \caption{$g_{00}$ ($\lambda=+1$)}
    \end{subfigure}
    \begin{subfigure}{0.32\textwidth}
        \centering
        \includegraphics[width=0.98\textwidth]{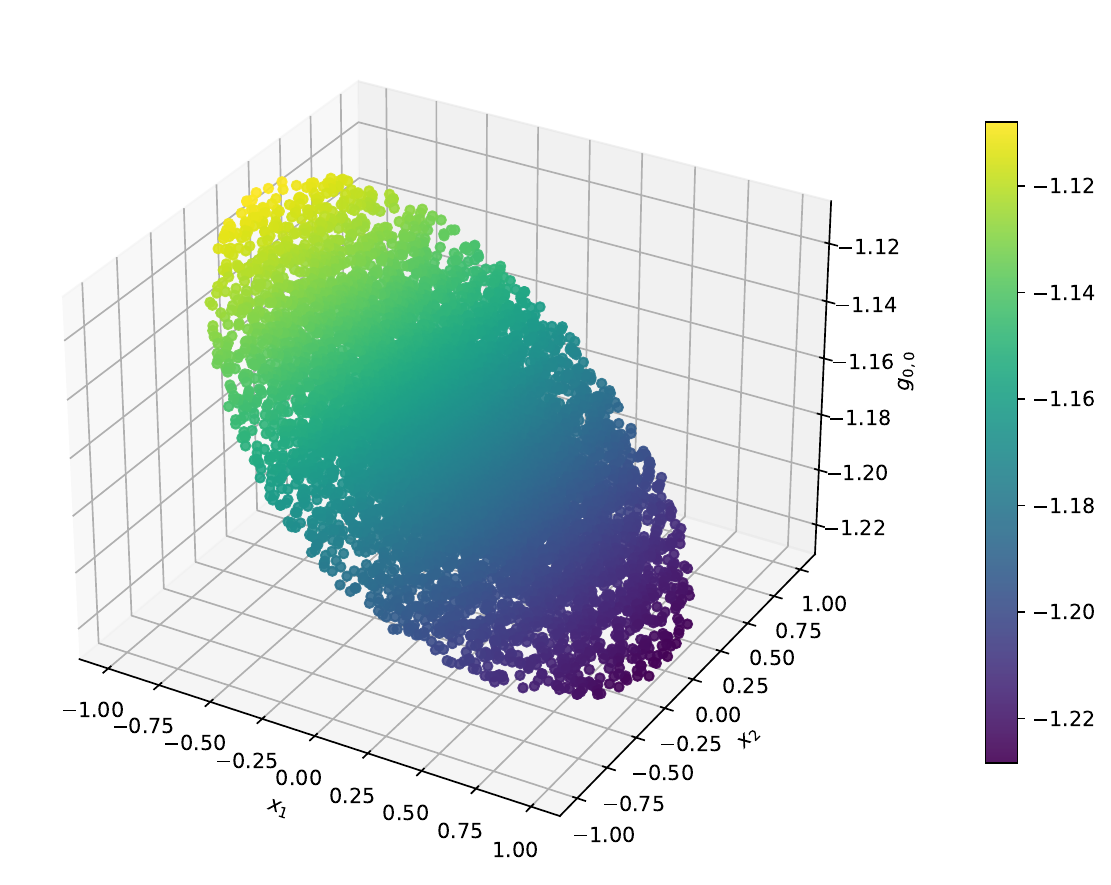}
        \caption{$g_{00}$ ($\lambda=0$)}
    \end{subfigure}
    \begin{subfigure}{0.32\textwidth}
        \centering
        \includegraphics[width=0.98\textwidth]{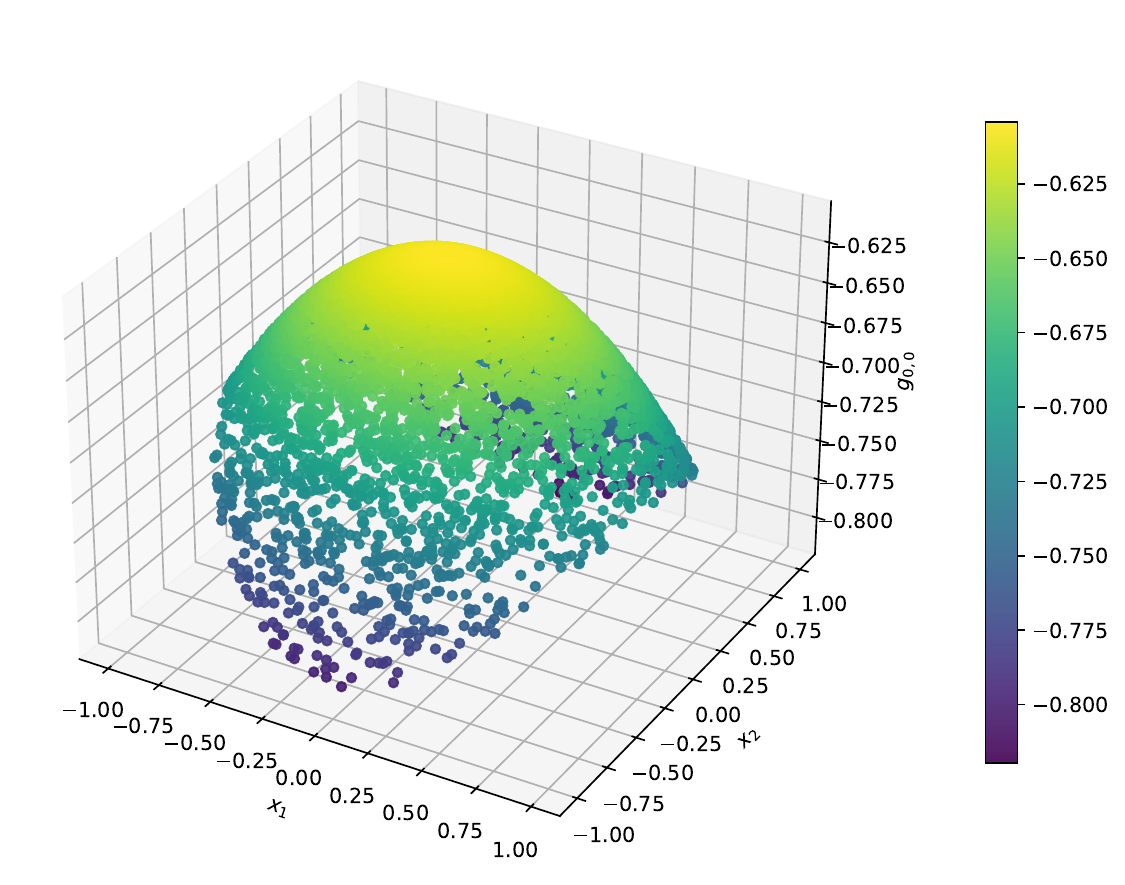}
        \caption{$g_{00}$ ($\lambda=-1$)}
    \end{subfigure}\\
    \begin{subfigure}{0.32\textwidth}
        \centering
        \includegraphics[width=0.98\textwidth]{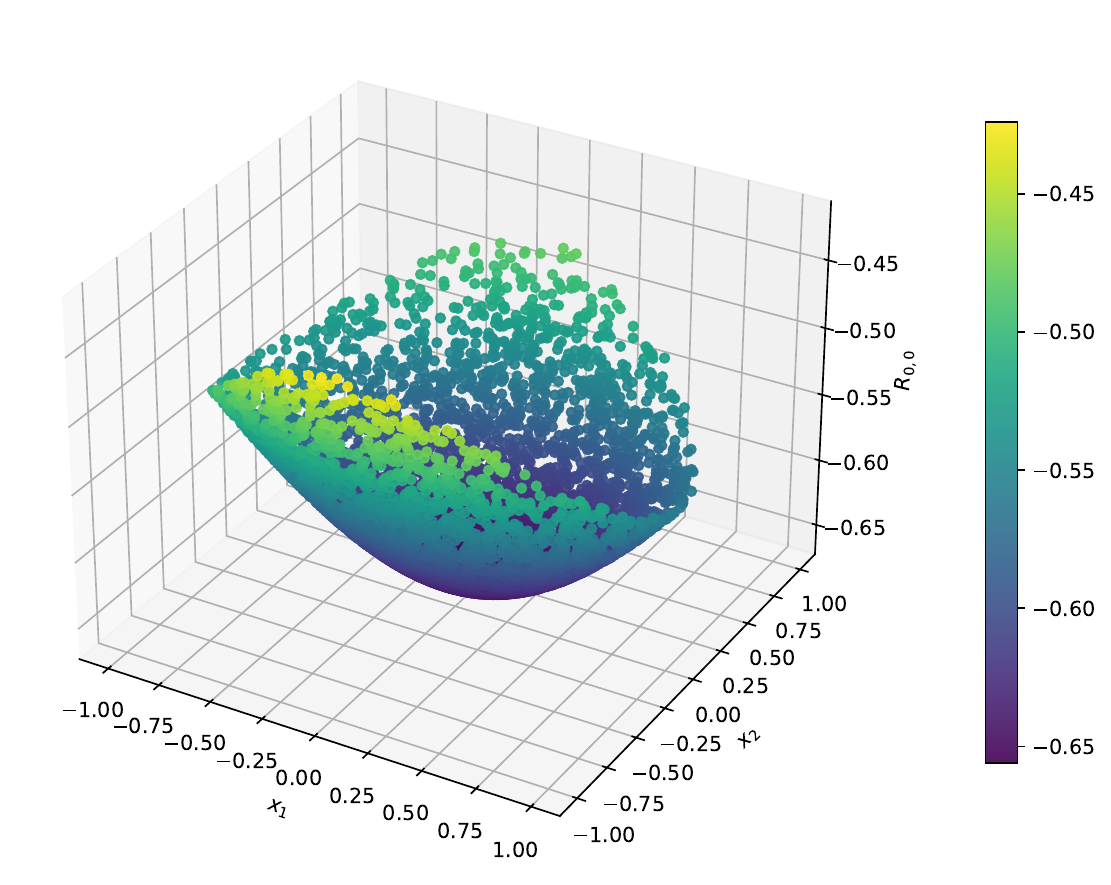}
        \caption{$R_{00}$ ($\lambda=+1$)}
    \end{subfigure}
    \begin{subfigure}{0.32\textwidth}
        \centering
        \includegraphics[width=0.98\textwidth]{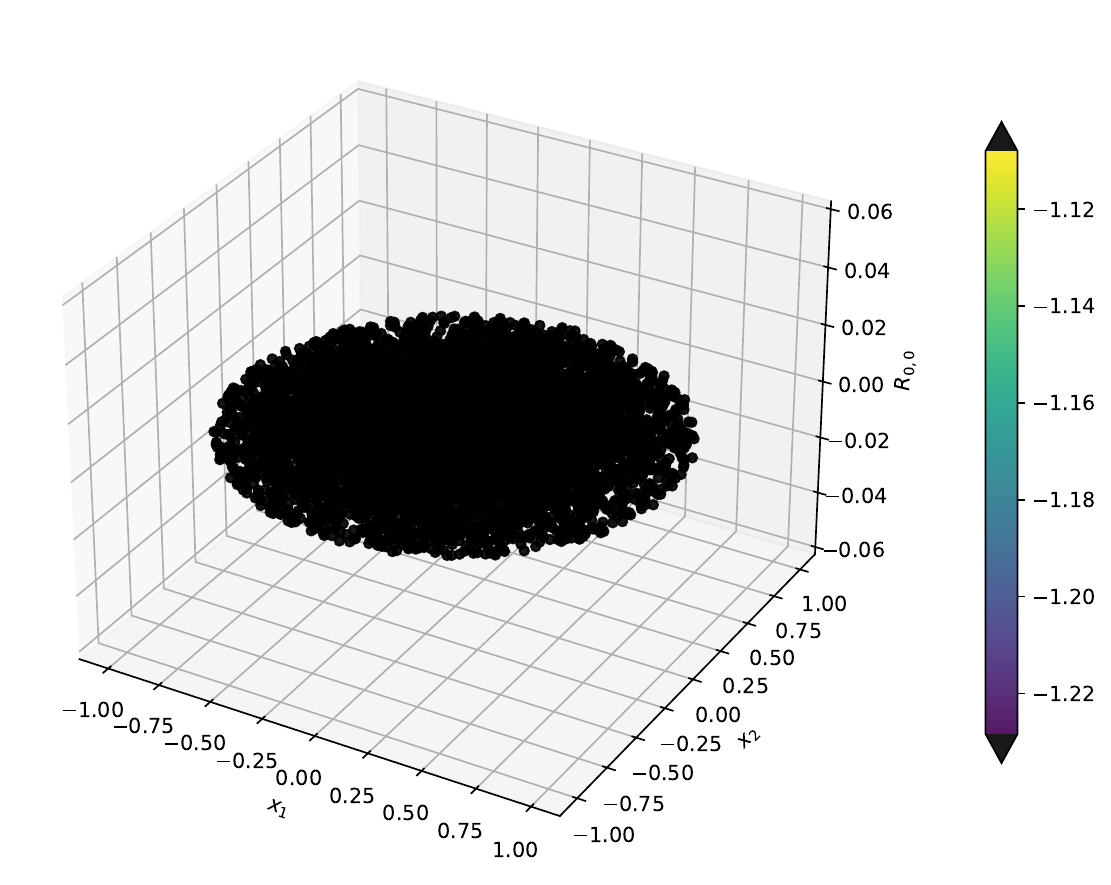}
        \caption{$R_{00}$ ($\lambda=0$)}
    \end{subfigure}
    \begin{subfigure}{0.32\textwidth}
        \centering
        \includegraphics[width=0.98\textwidth]{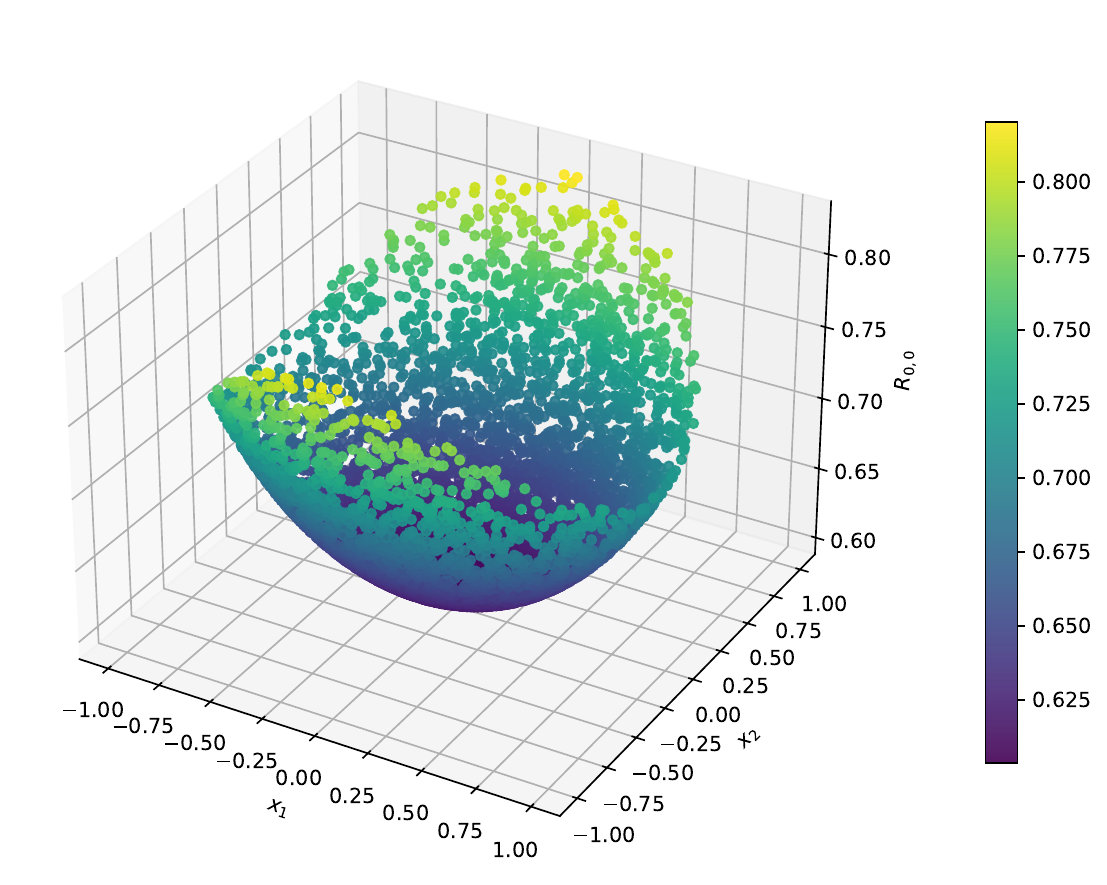}
        \caption{$R_{00}$ ($\lambda=-1$)}
    \end{subfigure}
    \caption{Visualisations of the $(0,0)$ components of the learnt two-dimensional Lorentzian metrics and their respective Ricci tensors on a single ball patch. These metrics solve the Einstein equation with Einstein constants $\lambda\in\{+1,0,-1\}$.}
    \label{fig:vis_2d1p}
\end{figure}

The component plots in \cref{fig:vis_2d1p} show representative learnt metric and Ricci components for each sign of $\lambda$. In two dimensions the Einstein condition fixes the Ricci tensor component-wise once the metric and $\lambda$ are specified, so the agreement between the small global losses, nonzero negative determinants, and the plotted components provides a compact sanity check for the Lorentzian local training setup used by the higher-dimensional experiments.

\subsection{Schwarzschild Search}

Having corroborated the most important term among the PINN's loss components, the next natural step is to test our method for the prototypical Ricci-flat solution: the Schwarzschild black hole. 
As discussed in \cref{sec:Theoretical_schw}, the choice of Penrose coordinates for this experiment provides a finite computational domain and a unified description of the whole spacetime (except the centre point of the black hole singularity).

In order to have a reference benchmark for results obtained through unsupervised training, we first performed a series of runs in supervised mode. As described at the beginning of \cref{sec:Training}, the supervised objective simply consists of approximating the target metric, which in this case is the Schwarzschild black hole written in Penrose coordinates. 
Since infinities cannot be represented computationally\footnote{For a fixed architecture there is a fixed limit of expressibility within the function class the architecture approximates. As infinities are approached this leads to numerical instabilities and overflows which need to be mitigated.}, points very close to the boundary of the Penrose diagram, which importantly include the both spatial infinity and singularity (where curvature blows up), are excluded from the experiments. The results presented in rest of the paper were obtained by sampling according to the strategy described in \cref{sec:Sampling} with $\rho = 0.85$. This value provides a substantial covering of the Penrose diagram, spanning a region from well beyond the event horizon up to multiple event horizon radii outside of the black hole.

For this benchmark $10$ independent supervised training runs were performed, which all saturated at a supervised training losses of $\sim 10^{-3}$.
For each supervised trained model an independent test dataset was used to evaluate the performance, as presented in the respective columns of \cref{tab:schwarzschild_10_seed_summary}.
The test evaluation metrics were the same physical losses used in the unsupervised runs (none of which are used in supervised training), and the relevant invariants for the Schwarzschild solution, as introduced in \cref{par:schwarzschild_inv}. 
These provide the benchmarks to assess the testing metrics of the unsupervised runs, whereas the supervised runs have exact knowledge of the analytic Schwarzschild solution to train against, and represents the limits of approximation with the allocated computational resources.
The unsupervised runs to compare against will be performing a general search with the Birkhoff-Jebsen aligned training objectives of \cref{ssub:unsupervised_sch}.

\begin{table*}[t]
  \centering
  \begin{tabular}{lccccc}
    \toprule
    Measures & Expected value
    & \multicolumn{2}{c}{Supervised Models}
    & \multicolumn{2}{c}{Unsupervised Models} \\
    \cmidrule(lr){3-4}\cmidrule(lr){5-6}
    & & Mean & Std. & Mean & Std. \\
    \midrule
    \multicolumn{6}{l}{\textit{Loss diagnostics}} \\
    Einstein loss & \(0\) & \(3.294\times10^{-2}\) & \(1.210\times10^{-2}\) & \(1.696\times10^{-4}\) & \(1.678\times10^{-5}\) \\
    Killing loss & \(0\) & \(7.814\times10^{-6}\) & \(1.914\times10^{-6}\) & \(9.322\times10^{-7}\) & \(1.579\times10^{-7}\) \\
    Weyl loss & \(0\) & \(2.118\times10^{-2}\) & \(8.165\times10^{-3}\) & \(7.191\times10^{-5}\) & \(1.328\times10^{-5}\) \\
    \midrule
    \multicolumn{6}{l}{\textit{Invariants}} \\
    \(\max(\det(g))\) & \(<0\) & \(-9044\) & \(173.8\) & \(-7857\) & \(156.4\) \\
    \(\operatorname{Re}(\mathcal{S})\) & \(1\) & \(0.9857\) & \(1.278\times10^{-2}\) & \(0.9999\) & \(2.409\times10^{-5}\) \\
    \(\operatorname{Im}(\mathcal{S})\) & \(0\) & \(-9.588\times10^{-6}\) & \(5.034\times10^{-5}\) & \(8.821\times10^{-7}\) & \(5.364\times10^{-6}\) \\
    \(K r^{6}/m^{2}\) & \(48\) & \(50.67\) & \(0.9672\) & \(47.90\) & \(4.742\times10^{-2}\) \\
    \(\rho\) & \(-1/\sqrt{12} \approx -0.2886(8)\) & \(-0.2773\) & \(4.608\times10^{-3}\) & \(-0.2886\) & \(9.668\times10^{-6}\) \\
    \bottomrule
  \end{tabular}
  \caption{Schwarzschild-search losses and invariants averaged over 10 random seeds for supervised and unsupervised models. The loss diagnostics are separated from the posterior invariant checks; the expected values are the Schwarzschild targets in the normalisations used in the text.}
  \label{tab:schwarzschild_10_seed_summary}
\end{table*}

Proceeding to the unsupervised experiment. These runs were trained subject to the losses presented in \cref{ssub:unsupervised_sch}, i.e. the optimiser is not given the analytic Schwarzschild metric components as pointwise targets, but instead learns a Lorentzian metric by minimising geometric residuals computed from its own output by automatic differentiation. 
The active constraints are Ricci flatness, the Weyl curvature-scale target, the $SO(3)$ symmetry residual, and the determinant barrier preventing degeneration of the Penrose $(T,X)$ block. These losses encode the invariant conditions needed to select the Schwarzschild branch, rather than an exact fitting of the metric component-by-component. 

Each of the 10 Schwarzschild-targeted runs were evaluated on an independent test sample of 2000 points.
The average test losses (with standard deviation), and invariant values, are given in the respective columns of \cref{tab:schwarzschild_10_seed_summary}.
Each of the test losses are orders of magnitude below the benchmark supervised training, validating that this training process has indeed found a Schwarzschild solution in each case, in alignment with the Birkhoff-Jebsen conditions.
As independent validation, each of the invariants computed for the trained models on the independent test data are also orders of magnitude closer to the target values for the Schwarzschild solution, despite the training process providing no information about them. 
Additionally, the low value of the maximum determinant across the manifold validates the solutions are consistently Lorentzian.
With both the strong loss results, and strong independent invariant corroboration, there is solid confidence in concluding the AInstein architecture can find and model the Schwarzschild solution.

\begin{figure*}[p]
  \centering
  \captionsetup[subfigure]{justification=centering,singlelinecheck=true}
  \begin{subfigure}[t]{0.48\textwidth}
    \centering
    \includegraphics[width=0.8\textwidth]{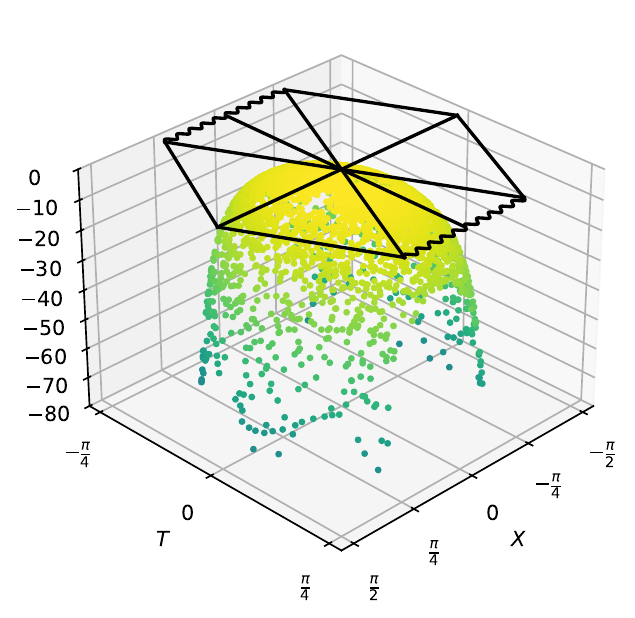}
    \caption{\(g_{00}\) Analytic}
  \end{subfigure}
  \hfill
  \begin{subfigure}[t]{0.48\textwidth}
    \centering
    \includegraphics[width=0.8\textwidth]{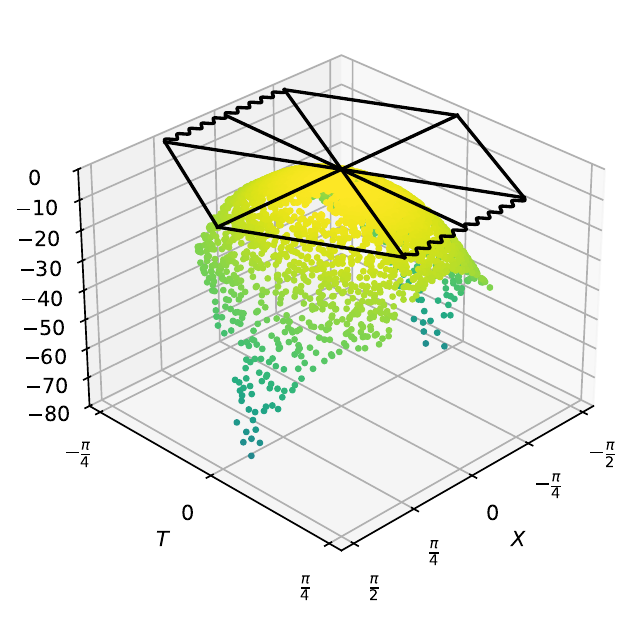}
    \caption{\(g_{00}\) Learnt}
  \end{subfigure}\\[0.35em]
  \begin{subfigure}[t]{0.48\textwidth}
    \centering
    \includegraphics[width=0.8\textwidth]{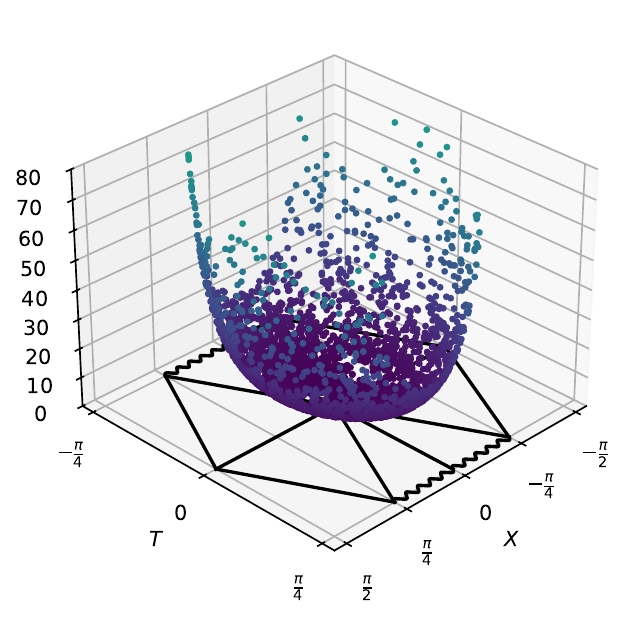}
    \caption{\(g_{11}\) Analytic}
  \end{subfigure}
  \hfill
  \begin{subfigure}[t]{0.48\textwidth}
    \centering
    \includegraphics[width=0.8\textwidth]{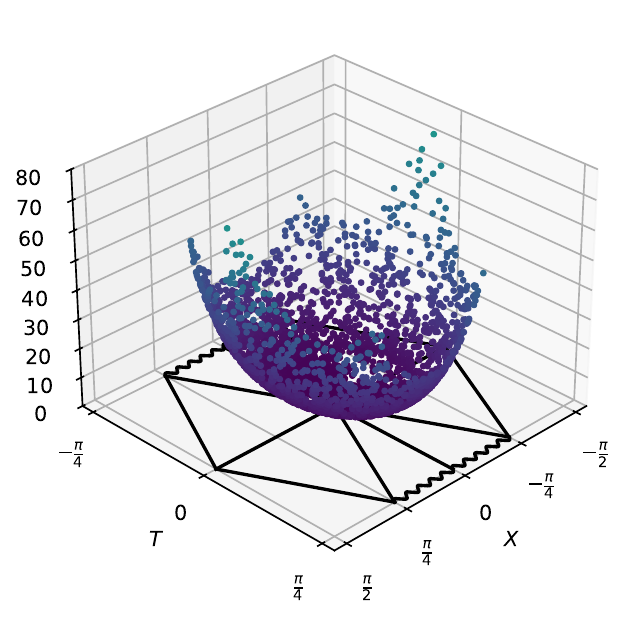}
    \caption{\(g_{11}\) Learnt}
  \end{subfigure}
  \caption{Diagonal metric components of the Penrose spacetime submanifold, for the analytic Schwarzschild solution and a representative trained solution, over the random sample of 2000 points.}
  \label{fig:schwarzschild_3d}
\end{figure*}

For further visual corroboration, \cref{fig:schwarzschild_3d} shows a learned \(g_{00}\) component for a representative run together with the corresponding Ricci component \(R_{00}\), as well as the analytic values of each from the known Schwarzschild solution for comparison.  
This metric component has the expected large variation across the compactified domain, while the Ricci component remains small on the same samples, validating the Ricci-flat behaviour of the found Schwarzschild solution for this representative component. 
The remaining components of the trained metric and respective Ricci tensor for a representative Schwarzschild run are shown in \cref{fig:schwarzschild_components_4x4} of \cref{app:component_plots}, which further corroborates the Ricci-flat nature of the found solution; additional plots on $S^2$ domain are included in the associated \href{https://github.com/xand-stapleton/ainstein}{GitHub} repository.

% \begin{figure*}[t]
%   \centering
%   \begin{subfigure}[t]{0.49\textwidth}
%     \centering
%     \includegraphics[width=\textwidth]{figures/schwarzschild/g_00_old.pdf}
%     \caption{Metric component $g_{00}$.\\ Summary stats (min, mean, max) = $(-77.3, -14.8, -5.66)$, standard deviation = $10.7$.}
%   \end{subfigure}
%   \hfill
%   \begin{subfigure}[t]{0.49\textwidth}
%     \centering
%     \includegraphics[width=\textwidth]{figures/schwarzschild/ricci_00_old.pdf}
%     \caption{Ricci component $R_{00}$.\\ Summary stats: (min, mean, max) = $(-0.291, -0.021, 0.119)$, standard deviation = $0.057$.}
%   \end{subfigure}
%   \caption{Component-level validation for the Schwarzschild run. The left panel shows the learned metric component \(g_{00}\), while the right panel shows the corresponding Ricci tensor component \(R_{00}\) on the validation samples.}
%   \label{fig:schwarzschild_g00_ricci00}
% \end{figure*}

\begin{figure*}[t]
  \centering
  \begin{subfigure}[t]{0.32\textwidth}
    \centering
    \includegraphics[width=\textwidth]{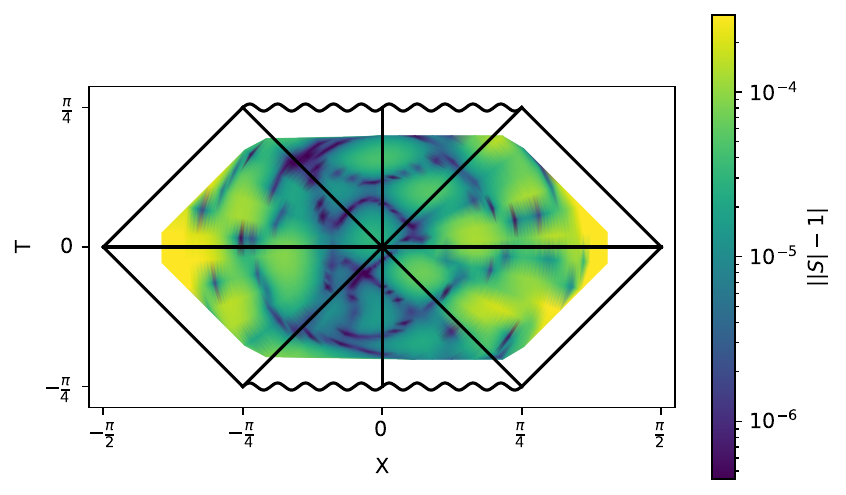}
    \caption{\(\big||\mathcal{S}|-1\big|\) residual. \\ Statistics: \(\min=1.658\times10^{-8}\), \(\mathrm{mean}=7.830\times10^{-5}\), \(\max=1.312\times10^{-2}\), \(\mathrm{std}=5.467\times10^{-4}\).}
  \end{subfigure}
  \hfill
  \begin{subfigure}[t]{0.32\textwidth}
    \centering
    \includegraphics[width=\textwidth]{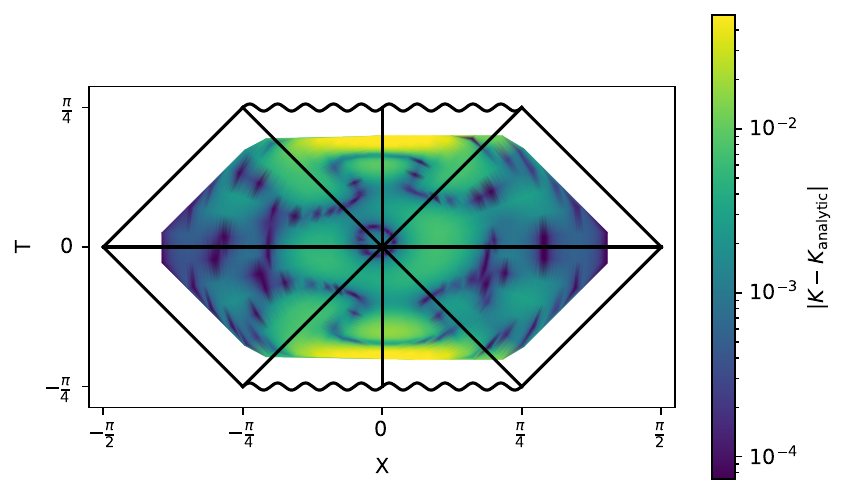}
    \caption{\(|K_{\rm pred}-K_{\rm Schw}|\) residual.\\ Statistics: \(\min=4.905\times10^{-6}\), \(\mathrm{mean}=5.050\times10^{-3}\), \(\max=1.115\times10^{-1}\), \(\mathrm{std}=1.179\times10^{-2}\).}
  \end{subfigure}
  \hfill
  \begin{subfigure}[t]{0.32\textwidth}
    \centering
    \includegraphics[width=\textwidth]{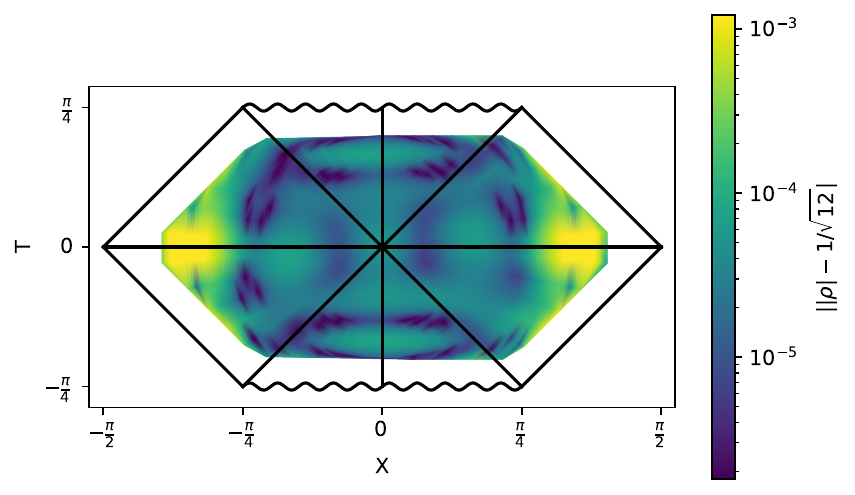}
    \caption{\(|\,|\rho|-1/\sqrt{12}\,|\) residual.\\ Statistics: \(\min=8.063\times10^{-8}\), \(\mathrm{mean}=9.389\times10^{-5}\), \(\max=2.326\times10^{-3}\), \(\mathrm{std}=2.806\times10^{-4}\).}
  \end{subfigure}
  \caption{Invariant constant residuals for a representative Schwarzschild run. The three panels compare the learned curvature invariants against the Schwarzschild speciality-index, Kretschmann, and \(\rho\)-constant targets.}
  \label{fig:schwarzschild_constant_checks}
\end{figure*}

The independent scalar checks in \cref{fig:schwarzschild_constant_checks}
visualise the deviation of each invariant of \cref{eq:schwarzschild_weyl_speciality} for a representative unsupervised trained Schwarzschild solution from the expected values of Schwarzschild.
Corroborating the low deviation from expectation shown in \cref{tab:schwarzschild_10_seed_summary}, all residuals in this figure show low values, demonstrated by the colour scales.
This again provides independent corroboration that this training process and AI architecture can indeed reproduce the Schwarzschild solution via a targeted search.

\subsection{Petrov-Type I Vacuum Search}
\label{subsec:petrov_typeI_results}

The second experiment uses the same embedded Lorentzian architecture and compactified sampling domain, but changes the target from Schwarzschild targeted search to a general type-I vacuum search. 
As described in \cref{subsubsec:petrov_repeller_network}, the Einstein residual is replaced by the homothety-invariant \(\mathcal L_{E,I}\), so the search is insensitive to overall constant metric rescalings, whilst the Schwarzschild Weyl-scale loss is removed and replaced with the speciality-index profile loss in \cref{eq:petrov_repeller_loss}. 
The vacuum type-I objective of \cref{eq:non_petrov_d_loss} therefore combines this Ricci-flatness condition with an arbitrary target speciality index profile which is representative of type I solutions, and includes two non-physical artefact losses to deter degeneracy (determinant barrier and the boundary curvature repeller). 
The \(SO(3)\) symmetry term is not used as a training constraint in this run, so the model is free to find non-symmetric solutions; the Killing residual is retained only as a diagnostic of departure from this spherically-symmetric basin.

\begin{table*}[t]
  \centering
  \begin{tabular}{lccc}
    \toprule
    Measures & Expected value & \multicolumn{2}{c}{Unsupervised Models} \\
    \cmidrule(lr){3-4}
    & & Mean & Std. \\
    \midrule
    \multicolumn{4}{l}{\textit{Loss diagnostics}} \\
    Einstein loss & \(0\) & \(9.449\times10^{-3}\) & \(5.811\times10^{-4}\) \\
    Speciality loss & \(0\) & \(7.159\times10^{-3}\) & \(2.082\times10^{-3}\) \\
    Killing loss & \(0\) & \(0.2454\) & \(6.442\times10^{-2}\) \\
    \midrule
    \multicolumn{4}{l}{\textit{Invariants}} \\
    \(\max(\det(g))\) & \(<0\) & \(-2.536\times10^{4}\) & \(1.310\times10^{4}\) \\
    \(\operatorname{Re}(\mathcal{S})\) & \(2\) & \(1.994\) & \(5.201\times10^{-3}\) \\
    \(\operatorname{Im}(\mathcal{S})\) & \(0\) & \(-5.321\times10^{-5}\) & \(3.560\times10^{-4}\) \\
    \bottomrule
  \end{tabular}
  \caption{Petrov Type-I search losses and invariants for unsupervised models, averaged over 10 random seeds.}
  \label{tab:typeI_unsupervised_summary}
\end{table*}

\cref{tab:typeI_unsupervised_summary} reports the test diagnostics for the 10 seed runs of 2000 points, showing test losses and relevant invariant values. 
The low average Einstein loss (still lower than supervised Schwarzschild benchmark) shows that these runs do find numerically Ricci-flat solutions; and that these solutions do match the arbitrary speciality index profile.
This strongly supports the idea that the AInstein architecture can be used for more general Ricci-flat vacuum solutions, with freedom to change the profile to match desired conditions.

In addition to these test losses, the Killing loss was also computed, and the reported value is much larger than in the Schwarzschild search, indicating the found solutions are not spherically symmetric (which is expected because spherical symmetry is no longer imposed as a loss in training). 
The determinant check remains strictly negative, confirming the consistent Lorentzian nature, and the speciality index is centred near the intended type-I profile value.

The full component-level metric and Ricci diagnostics for the Petrov type-I vacuum search are collected in \cref{fig:typeI_components_4x4} of \cref{app:component_plots}. These plots provide the component-level counterpart to the invariant diagnostics in \cref{tab:typeI_unsupervised_summary}: the learnt metric components are non-Schwarzschild in scale and structure, while the Ricci components remain small relative to the metric components. Taken together, the loss diagnostics, determinant check, and speciality-index profile indicate that the model has found non-degenerate, numerically Ricci-flat geometries outside the Schwarzschild/type-\(\mathrm{D}\) basin.

\subsection{Petrov-Type I Black Hole Search}
\label{subsec:blackhole_results}

The final experiment applies the black-hole objective \cref{eq:blackhole_loss} on the same embedded domain, adding the horizon anchor \cref{eq:horizon_anchor_loss} and the trapped-surface term \cref{eq:trapped_loss} to the type-I recipe of \cref{subsec:petrov_typeI_results}. 
Two variants are reported here, differing only in the trapped-surface margin and weight, $(\beta_0,\alpha_{\rm trap})=(0.05,15)$ and $(0.08,25)$ (both with $\beta_1=1$), each over $10$ random seeds, with $\alpha_{\rm trap}$ warmed up over the first $30\%$ of training. \cref{tab:blackhole_summary} collects the diagnostics, evaluated on the independent test samples of 2000 points; the interior/exterior of each fibre and its expansions $\theta_\ell,\theta_n$ are computed by finite differences of the learnt areal radius.

\begin{table*}[t]
  \centering
  \begin{tabular}{lccccc}
    \toprule
    Measure & Expected
    & \multicolumn{2}{c}{$(\beta_0,\alpha_{\rm trap})=(0.05,15)$}
    & \multicolumn{2}{c}{$(\beta_0,\alpha_{\rm trap})=(0.08,25)$} \\
    \cmidrule(lr){3-4}\cmidrule(lr){5-6}
    & & Mean & Std. & Mean & Std. \\
    \midrule
    \multicolumn{6}{l}{\textit{Loss diagnostics}} \\
    Einstein loss & \(0\)
      & \(7.046\times10^{-2}\) & \(8.330\times10^{-3}\) & \(6.326\times10^{-2}\) & \(3.919\times10^{-3}\) \\
    Killing loss & \(0\)
      & \(4.207\times10^{-2}\) & \(5.715\times10^{-3}\) & \(3.939\times10^{-2}\) & \(6.655\times10^{-3}\) \\
    Speciality loss & \(0\)
      & \(4.718\times10^{-2}\) & \(1.542\times10^{-2}\) & \(3.526\times10^{-2}\) & \(1.073\times10^{-2}\) \\
    \midrule
    \multicolumn{6}{l}{\textit{Invariants}} \\
    \(\max(\det(g))\) & \(<0\)
      & \(-18.04\) & \(2.614\) & \(-18.33\) & \(1.476\) \\
    \(\operatorname{Re}(\mathcal S)\) & \(2\)
      & \(1.993\) & \(3.973\times10^{-3}\) & \(1.992\) & \(4.399\times10^{-3}\) \\
    \(\operatorname{Im}(\mathcal S)\) & \(0\)
      & \(-3.353\times10^{-4}\) & \(4.620\times10^{-4}\) & \(-8.320\times10^{-5}\) & \(2.469\times10^{-4}\) \\
    \(\sqrt{|I|}\,r^{3}/m\) & \(\sqrt3\approx1.732\)
      & \(1.750\) & \(4.6\times10^{-3}\) & \(1.750\) & \(1.8\times10^{-3}\) \\
    \(K_{\rm hor}\) & \(0.75\)
      & \(0.763\) & \(5.2\times10^{-3}\) & \(0.766\) & \(3.1\times10^{-3}\) \\
    rel.\ Frobenius distance & \(\gg0\)
      & \(0.625\) & \(5.7\times10^{-3}\) & \(0.615\) & \(8.2\times10^{-3}\) \\
    \(\Xi_{\rm int}\) (interior median) & \(<0\)
      & \(-8.6\times10^{-2}\) & \(1.5\times10^{-3}\) & \(-1.13\times10^{-1}\) & \(2.0\times10^{-3}\) \\
    trapped fraction (interior) & \(\to1\)
      & \(0.972\) & \(5.2\times10^{-3}\) & \(0.985\) & \(3.6\times10^{-3}\) \\
    \bottomrule
  \end{tabular}
  \caption{Black-hole-search loss diagnostics and invariants for the two trapped-surface variants, averaged over 10 random seeds. Beyond the standard loss and invariant diagnostics, the quantities \(\sqrt{|I|}\,r^{3}\) and \(K_{\rm hor}\) are evaluated in the horizon band \(r\simeq2m\); rel.\ Frobenius distance is \(\|g-g_{\rm Schw}\|/\|g_{\rm Schw}\|\); \(\Xi_{\rm int}\) is the interior median of \cref{eq:trapped_scalar} and the trapped fraction is the share of interior samples with \(\Xi<0\).}
  \label{tab:blackhole_summary}
\end{table*}

Across all $20$ runs the black-hole criteria hold simultaneously and robustly. The geometry is numerically Ricci-flat, shown by the average Einstein loss matching the supervised Schwarzschild scale. %; in coordinate-invariant terms the Ricci residual is $\simeq3\%$ of the curvature scale $\sqrt K$. 
The full component-level metric and Ricci diagnostics for the black-hole search are collected in \cref{fig:blackhole_components_4x4} of \cref{app:component_plots}.
The components of the trained metric and respective Ricci tensor for a representative Schwarzschild run in the $(\beta_0,\alpha_{\rm trap})=(0.08,25)$ case  are shown in \cref{fig:blackhole_components_4x4} of \cref{app:component_plots}, which visually validate the Ricci-flat nature of the found solutions.

\cref{tab:blackhole_summary} also shows the solutions are algebraically general, $\mathcal S\simeq 1.99$, well away from the type-D value. The horizon curvatures are regular and at the Schwarzschild scale, $\sqrt{|I|}\,r^{3}\to1.75\approx\sqrt3$ and $K_{\rm hor}\to0.77\approx0.75$, so the solutions are not naked singularities; the negative max(det($g$)) confirms the solutions are consistently Lorentzian; and the metric is manifestly non-Schwarzschild, with relative Frobenius distance $\simeq0.6$ to the analytic solution. Crucially in each run, the interior is genuinely trapped: $\Xi<0$ on $97$--$99\%$ of the interior samples, with interior median $\Xi\simeq-0.1$.

Because $\Xi$ is invariant under time reversal, the loss fixes $\Xi<0$ but not the black-hole orientation. We resolve it a posteriori through the two null expansions $\theta_\ell,\theta_n$, with the convention calibrated so that the future interior $B_{\rm II}$ of the analytic Schwarzschild solution comes out future-trapped.\footnote{Interestingly, but a-posteriori unsurprisingly, the seeds divide almost evenly: half realise a genuine \textit{black hole}, with $\theta_\ell,\theta_n<0$ throughout $B_{\rm II}$, and half the time-reversed anti-trapped (white-hole) branch.} %The black-hole branch is also the cleaner vacuum, with mean Ricci MSE $8\times10^{-3}$ against $1.8\times10^{-2}$ for the anti-trapped one. 
%Our best black-hole model has Ricci MSE $6.6\times10^{-3}$, $\mathcal S=1.93$, $\sqrt{|I|}\,r^{3}=1.75$, $K_{\rm hor}=0.77$, relative distance $0.62$, and a future-trapped interior over the entirety of $B_{\rm II}$.

\begin{figure*}[t]
  \centering
  \captionsetup[subfigure]{justification=centering,singlelinecheck=true}
  \begin{subfigure}[t]{0.52\textwidth}
    \centering
    \includegraphics[width=\textwidth]{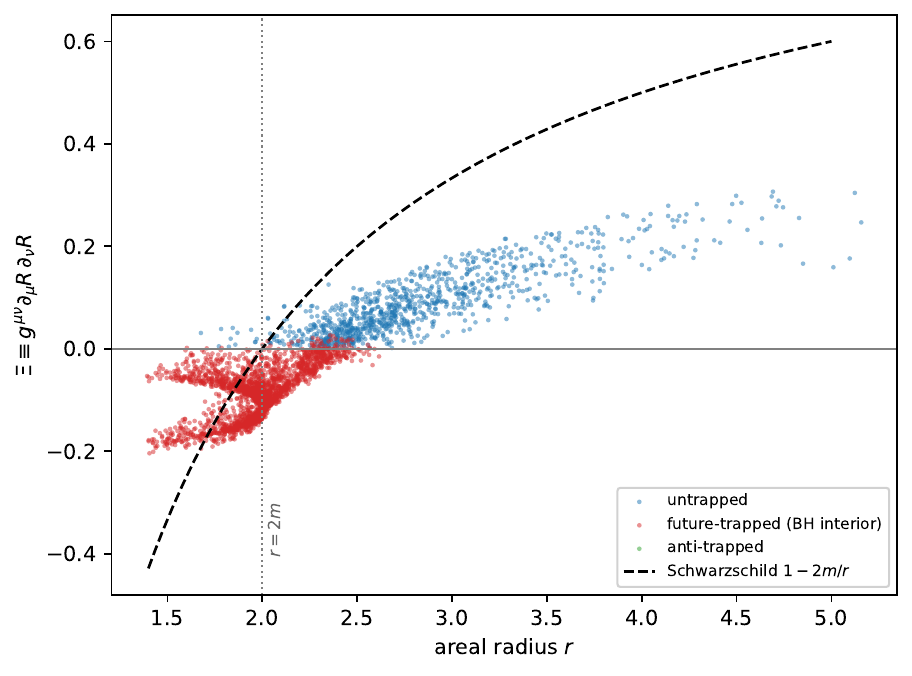}
    \caption{\(\Xi\) versus areal radius}
  \end{subfigure}
  \hfill
  \begin{subfigure}[t]{0.44\textwidth}
    \centering
    \includegraphics[width=\textwidth]{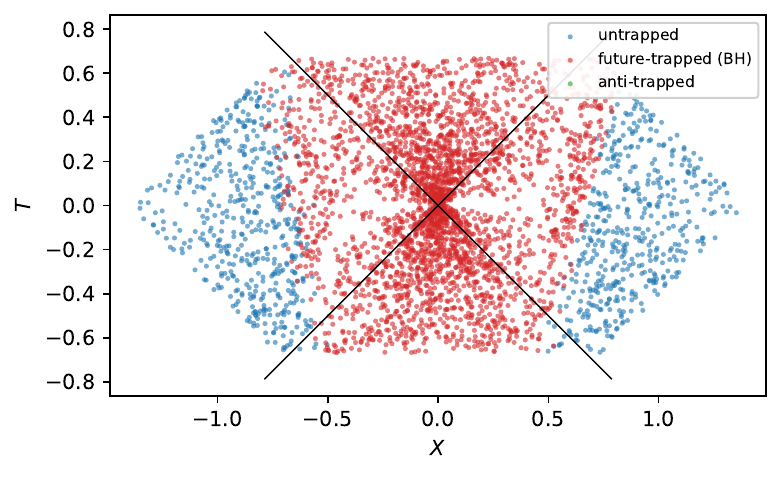}
    \caption{trapping class over the Penrose diagram}
  \end{subfigure}
  \caption{Location of the trapped region for an example learnt black-hole model. \textbf{(a)} The scalar \(\Xi\) of \cref{eq:trapped_scalar} is negative on the interior \(r<2m\), crosses zero at the horizon, and is positive outside, following the shape of the Schwarzschild \(1-2m/r\) profile (dashed) while remaining milder, as expected for a distorted type-I geometry. \textbf{(b)} Each fibre, coloured by the signs of \((\theta_\ell,\theta_n)\): the future interior \(B_{\rm II}\) is future-trapped, the exteriors are untrapped, and the null lines \(T=\pm X\) separate them, reproducing the causal structure of \cref{fig:penrose_regions}.}
  \label{fig:blackhole_trapped}
\end{figure*}

\cref{fig:blackhole_trapped} shows where the trapped region sits for this model, and reproduces the expected causal structure of \cref{fig:penrose_regions}: the trapped region coincides with the interior $r<2m$, bounded by the null horizons $T=\pm X$, with the exteriors untrapped.

Finally, as these metrics are not spherically symmetric, we confirm that $\Xi$ faithfully reports the trapping. Recomputing the expansions with the exact fibre normals -- which, owing to the small residual $\mathbb R^{2}$--$S^{2}$ cross terms of the learnt metric, do not lie exactly in the $(T,X)$ plane -- shifts $\theta_\ell,\theta_n$ by only about $5\%$, far below their interior magnitude $\simeq0.1$, and leaves the whole interior future-trapped. The classification is therefore robust against the departure from spherical symmetry.

These final results therefore validate that the AInstein architecture can be specialised to trapped surface solutions too, setting up the codebase and this methodology for general black hole search.

%%%%%%%%%%%%%%%%%%%%%%%%%%%%%%%%%%%%%%%%%%%%%%%
\section{Conclusions and Outlook}
This work upgrades AInstein from Riemannian sphere searches to Lorentzian black-hole geometries. The essential ingredients are an indefinite metric parametrisation, compactified Penrose-domain sampling, a single ambient network whose pullback handles the \(S^{2}\) topology globally, and differentiable losses built directly from curvature, invariants, non-degeneracy, and trapping diagnostics. This allows for general vacuum solution search while retaining tensorial control over the geometry being learnt.

The results show this machinery working in increasing generality. 
Local Lorentzian tests recover two-dimensional Einstein metrics with all three signs of \(\lambda\).
The Schwarzschild experiment then recovers the maximally extended vacuum black hole from geometric constraints rather than supervised metric fitting: Ricci flatness, the Weyl scale, and spherical symmetry select the Schwarzschild branch, while independent invariant checks confirm the expected type-\(\mathrm{D}\) curvature structure. 
The Petrov-profile search generalises from Schwarzschild-specific desired curvatures and uses the same framework to find non-degenerate, numerically Ricci-flat metrics in a different Petrov class, type I. 
Finally, adding a horizon curvature anchor and a trapped-surface loss leads the architecture to find and approximate algebraically general vacuum black-hole candidates with regular Schwarzschild-scale horizons and trapped interiors.

This work develops the AInstein methodology into Lorentzian solution reproduction, and then further into a flexible variational search engine for vacuum geometry.
One may specify invariant profiles and causal conditions, and the network searches the corresponding metric space directly. 
This is especially promising for black-hole problems where analytic ans\"atze are restrictive or unavailable, yet this AI methodology can provide numerical approximations. 
Natural next targets include Kaluza--Klein black strings and localized black holes, non-uniform or distorted horizons, AdS black holes with prescribed boundary data, and stationary vacuum searches where symmetry assumptions should be relaxed rather than built in from the start.

\section*{Data Availability}
The respective codebase was written in \texttt{Python}, and scripts used to generate all the results are available at the AInstein repository:\\ \url{https://github.com/xand-stapleton/ainstein}.

\section*{Acknowledgements}
The authors thank Michael Douglas, Fabian Ruehle, and Tom\'as Silva for helpful comments during the ``Mathematics and Machine Learning Program'' at Harvard University; Gianfranco Cortes and Yueqing Feng for discussions on the embedding architecture; Kymani Armstrong-Williams for comments on the Kretschmann invariant of the Schwarzschild solution; and David Berman for helpful comments on well-defined tensorial integration measures, as well as bringing us all together. 

EH is supported by São Paulo Research Foundation (FAPESP) grant 2024/18994-7. 
TSG is supported by the 2024 Max Planck-Humboldt Research Award bestowed on Geordie Williamson and Catharina Stroppel by the Max Planck Society and the Alexander von Humboldt Foundation.
AGS and TSG acknowledge support from Pierre Andurand over the course of this research. 

Computations were performed on the HPC system Viper at the Max Planck Computing and Data Facility. This research utilised computational resources of the \textit{Centro Nacional de Processamento de Alto Desempenho em São Paulo} (CENAPAD-SP). This research used Queen Mary's Apocrita HPC facility, supported by QMUL Research-IT.

%%%%%%%%%%%%%%%%%%%%%%%%%%%%%%%%%%%%%%%%%%%%%%%
\appendix

\section{Invariant Derivations for the Schwarzschild Solution}
\label{app:kretschmann_schwarzschild}

Spherical symmetry gives a preferred geometrical radius, the areal radius \(r\), defined by the area \(A\) of each two-sphere,
\begin{equation}
  A=4\pi r^{2}.
  \label{eq:areal_radius_def}
\end{equation}
In any spherically symmetric spacetime there is a corresponding Misner--Sharp mass \(M_{\rm MS}\), defined invariantly by
\begin{equation}
  1-\frac{2M_{\rm MS}}{r}
  =
  \nabla^{a}r\nabla_{a}r.
  \label{eq:misner_sharp_def}
\end{equation}
This definition depends only on the areal radius and the spacetime metric, not on a particular Schwarzschild coordinate chart. The Einstein equations imply the local mass-balance relation
\begin{equation}
  \nabla_a M_{\rm MS}=4\pi r^2\left(T_a{}^b\nabla_b r-T^{A}{}_{A}\nabla_a r\right) \, ,
  \label{eq:misner_sharp_balance}
\end{equation}
with $T^A{}_A=T^t{}_t+T^r{}_r$ being the trace over the 2D space normal to the symmetry spheres.
Therefore, in vacuum, \(T_{ab}=0\), the Misner--Sharp mass is constant:
\begin{equation}
  M_{\rm MS}=m.
  \label{eq:misner_sharp_constant}
\end{equation}

In an orthonormal frame adapted to the static Schwarzschild geometry, with \(e_{0}\) timelike, \(e_{1}\) radial, and \(e_{2},e_{3}\) tangent to the symmetry two-sphere, the independent non-zero Riemann components are
\begin{align}
  R_{0101}&=-\frac{2m}{r^{3}},
  &
  R_{0202}=R_{0303}&=\frac{m}{r^{3}},
  \nonumber\\
  R_{1212}=R_{1313}&=-\frac{m}{r^{3}},
  &
  R_{2323}&=\frac{2m}{r^{3}},
  \label{eq:schwarzschild_orthonormal_riemann}
\end{align}
up to the overall sign convention for the Riemann tensor. The Kretschmann scalar is insensitive to this overall sign, and the final Schwarzschild value agrees with the standard curvature-invariant expression \cite{Cherubini2003}. Using the usual Riemann symmetries, the contraction counts each independent sectional component four times, giving
\begin{align}
  K_{\rm Schw}
  &=
  R_{abcd}R^{abcd}
  \nonumber\\
  &=
  4\left[
    \left(\frac{2m}{r^{3}}\right)^{2}
    +\left(\frac{m}{r^{3}}\right)^{2}
    +\left(\frac{m}{r^{3}}\right)^{2}
    +\left(\frac{m}{r^{3}}\right)^{2}
    +\left(\frac{m}{r^{3}}\right)^{2}
    +\left(\frac{2m}{r^{3}}\right)^{2}
  \right]
  \nonumber\\
  &=
  \frac{48m^{2}}{r^{6}}.
  \label{eq:schw_kretschmann_geometric}
\end{align}
Equivalently, this gives the invariant check \(Kr^{6}=48m^{2}\). Substituting the Penrose-domain radius \(r=2m(1+W_{\mathcal B})\) from \cref{eq:radius_lambert} then gives the analytic target used in the implementation,
\begin{equation}
  K_{\rm Schw}
  =
  \frac{48m^{2}}{r^{6}}
  =
  \frac{3}{4m^{4}(1+W_{\mathcal B})^{6}}.
  \label{eq:analytic_kretschmann}
\end{equation}
The same orthonormal components give the cubic contraction. To see the numerical factor explicitly, raise the second antisymmetric index pair. In the ordered bivector basis
\[
  (01),(02),(03),(12),(13),(23),
\]
the non-zero diagonal entries of \(R_{ab}{}^{cd}\) are
\begin{equation}
  R_{01}{}^{01}=\frac{2m}{r^3},\quad
  R_{02}{}^{02}=R_{03}{}^{03}
  =R_{12}{}^{12}=R_{13}{}^{13}=-\frac{m}{r^3},\quad
  R_{23}{}^{23}=\frac{2m}{r^3} .
  \label{eq:riemann_bivector_eigenvalues}
\end{equation}
The signs differ from the lowered components whenever a time index is raised. Since the full Einstein summation is over ordered antisymmetric pairs, each independent bivector eigenvalue appears eight times: for example the \(01\) bivector contributes through the choices \(ab=01\) or \(10\), \(cd=01\) or \(10\), and \(ef=01\) or \(10\). The three antisymmetry signs multiply to give the same contribution in each case. Equivalently, the contraction is eight times the trace of the cube of the curvature operator on \(\Lambda^{2}\):
\begin{align}
  R_{ab}{}^{cd}R_{cd}{}^{ef}R_{ef}{}^{ab}
  &=
  8\left[
  \bigg(\frac{2m}{r^3}\bigg)^3+\bigg(-\frac{m}{r^3}\bigg)^3+\bigg(-\frac{m}{r^3}\bigg)^3+\bigg(-\frac{m}{r^3}\bigg)^3+\bigg(-\frac{m}{r^3}\bigg)^3+\bigg(\frac{2m}{r^3}\bigg)^3
  \right]
  \nonumber\\
  &=
  96\bigg(\frac{m}{r^3}\bigg)^{3}.
  \label{eq:schwarzschild_cubic_counting}
\end{align}
Thus
\begin{equation}
  J_{\rm cub}
  =
  R_{ab}{}^{cd}R_{cd}{}^{ef}R_{ef}{}^{ab}
  =
  96\frac{m^{3}}{r^{9}}.
  \label{eq:schwarzschild_cubic_invariant}
\end{equation}
Combining \cref{eq:schw_kretschmann_geometric} and \cref{eq:schwarzschild_cubic_invariant} gives
\begin{equation}
  \frac{J_{\rm cub}}{K_{\rm Schw}^{3/2}}
  =
  \frac{96m^{3}r^{-9}}{(48m^{2}r^{-6})^{3/2}}
  =
  \frac{1}{\sqrt{12}}.
  \label{eq:schwarzschild_cubic_ratio_derivation}
\end{equation}
The sign of this cubic contraction changes with the opposite Riemann-tensor convention, whereas the Kretschmann scalar does not.

The speciality-index calculation uses the complex Weyl invariants \(I\) and \(J\), not the real Riemann contraction \(J_{\rm cub}\). The relevant invariant construction is the one used by Baker and Campanelli for the speciality index, and by later Petrov-classification diagnostics \cite{BakerCampanelli2000,RosatoNakanoLousto2021}. In Newman--Penrose notation \cite{Newman:1961qr} the five Weyl scalars \(\Psi_{0},\ldots,\Psi_{4}\) are the components of the Weyl tensor in a complex null tetrad, and the scalar polynomial invariants may be written as
\begin{align}
  I
  &=
  \Psi_{0}\Psi_{4}
  -4\Psi_{1}\Psi_{3}
  +3\Psi_{2}^{2},
  \nonumber\\
  J
  &=
  \begin{vmatrix}
    \Psi_{4} & \Psi_{3} & \Psi_{2}\\
    \Psi_{3} & \Psi_{2} & \Psi_{1}\\
    \Psi_{2} & \Psi_{1} & \Psi_{0}
  \end{vmatrix}.
  \label{eq:np_weyl_invariants_appendix}
\end{align}
These are the Newman--Penrose form of the self-dual Weyl contractions in \cref{eq:weyl_petrov_invariants}, with the same normalisation. The speciality index is then \(\mathcal{S}=27J^{2}/I^{3}\), which equals unity for the algebraically special Kerr/type-\(\mathrm{D}\) limit in this convention \cite{BakerCampanelli2000}. In vacuum \(R_{abcd}=C_{abcd}\), and in the Schwarzschild principal tetrad the only non-zero Weyl scalar is
\begin{equation}
  \Psi_2=-\frac{m}{r^3},
  \label{eq:schwarzschild_psi2}
\end{equation}
which is real \cite{Cherubini2003}. All other Weyl scalars vanish in this tetrad. Substituting
\[
  \Psi_0=\Psi_1=\Psi_3=\Psi_4=0
\]
into \cref{eq:np_weyl_invariants_appendix} gives
\begin{equation}
  I=3\Psi_2^2=\frac{3m^2}{r^6},
  \qquad
  J=-\Psi_2^3=\frac{m^3}{r^9},
  \qquad
  \mathcal{S}
  =
  \frac{27J^2}{I^3}
  =
  1+0i .
  \label{eq:schwarzschild_speciality_appendix}
\end{equation}
Thus the constants in the Schwarzschild speciality-index target come directly from the standard Newman--Penrose invariant polynomials and the single Coulomb Weyl scalar of the type-\(\mathrm{D}\) Schwarzschild geometry. Comparing \cref{eq:schwarzschild_cubic_invariant} and \cref{eq:schwarzschild_speciality_appendix} gives \(J_{\rm cub}=96\,\mathrm{Re}(J)\) for the convention used here. Therefore the dimensionless scalar reported in the diagnostics is
\begin{equation}
  \rho
  :=
  -\frac{96\,\mathrm{Re}(J)}{|K|^{3/2}}
  =
  -\frac{J_{\rm cub}}{|K|^{3/2}},
  \qquad
  \rho_{\rm Schw}
  =
  -\frac{1}{\sqrt{12}},
  \qquad
  |\rho_{\rm Schw}|
  =
  \frac{1}{\sqrt{12}}.
  \label{eq:rho_schwarzschild_appendix}
\end{equation}
This last normalisation defines the diagnostic used in the numerical plots; the cited invariant results are the Schwarzschild Weyl scalar, Weyl invariants, and Kretschmann scalar. Hence the invariant targets are not additional coordinate ansatz data: they follow from the vacuum equation, spherical symmetry, and the areal-radius function \(r(T,X)\).

\section{Component-Level Metric and Ricci Visualisations}
\label{app:component_plots}

This appendix collects the full component-level validation plots for the three main four-dimensional runs. In each figure, the left block shows the learnt metric components $g_{\mu\nu}$ and the right block shows the corresponding Ricci tensor components $R_{\mu\nu}$, both arranged by tensor index over the validation samples. The shared colour bar beneath each pair fixes the component scale used across the metric and Ricci panels, while the captions report the mean absolute component sizes used in the diagnostics.

\begin{figure*}[p]
  \centering
  \captionsetup[subfigure]{justification=centering,singlelinecheck=true}

  \begin{minipage}[t]{0.47\textwidth}
  \centering
  \textbf{Metric components}\\[0.35em]

  \begin{subfigure}[t]{0.235\linewidth}
    \centering
    \includegraphics[width=\textwidth]{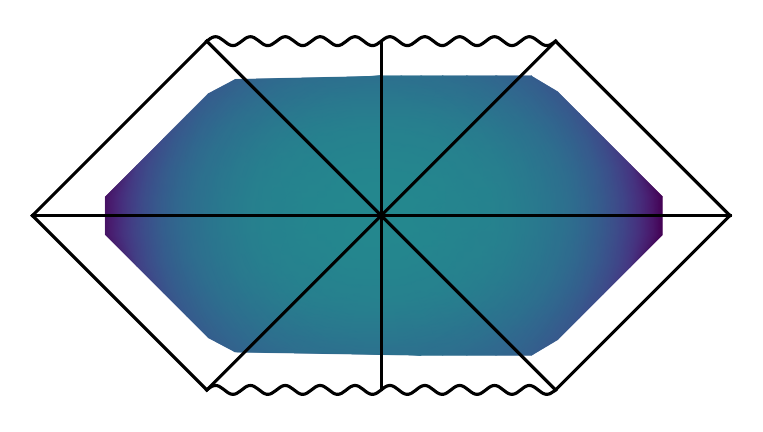}
    \caption{\(g_{00}\)}
  \end{subfigure}
  \hfill
  \begin{subfigure}[t]{0.235\linewidth}
    \centering
    \includegraphics[width=\textwidth]{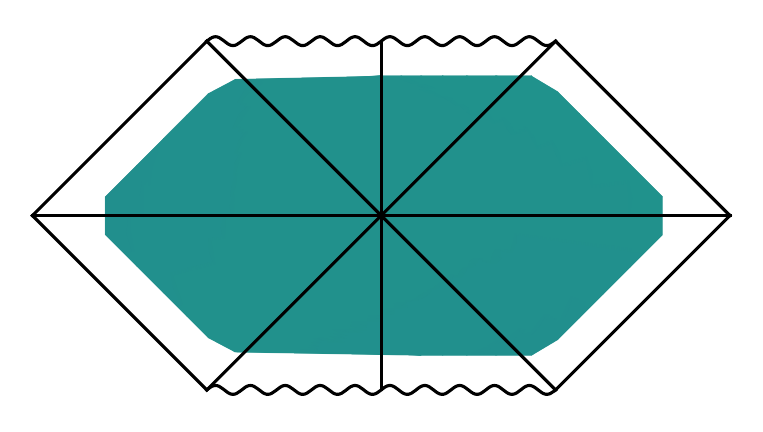}
    \caption{\(g_{01}\)}
  \end{subfigure}
  \hfill
  \begin{subfigure}[t]{0.235\linewidth}
    \centering
    \includegraphics[width=\textwidth]{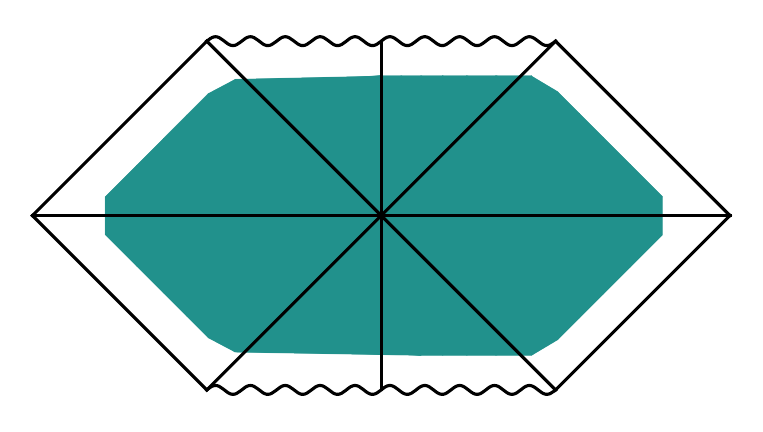}
    \caption{\(g_{02}\)}
  \end{subfigure}
  \hfill
  \begin{subfigure}[t]{0.235\linewidth}
    \centering
    \includegraphics[width=\textwidth]{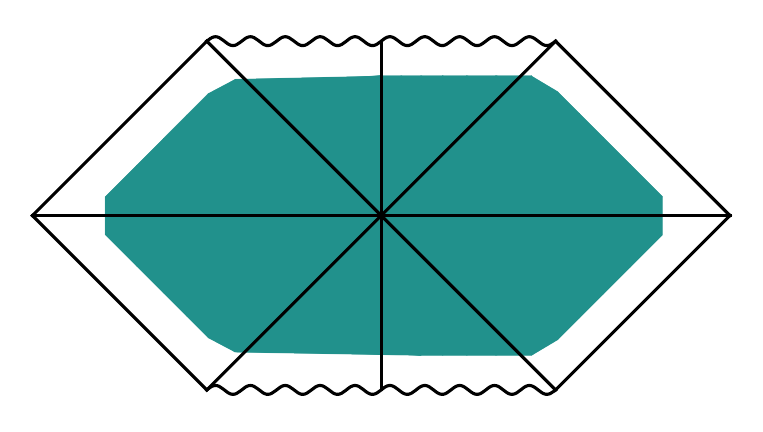}
    \caption{\(g_{03}\)}
  \end{subfigure}\\[0.25em]

  \begin{subfigure}[t]{0.235\linewidth}
    \centering
    \includegraphics[width=\textwidth]{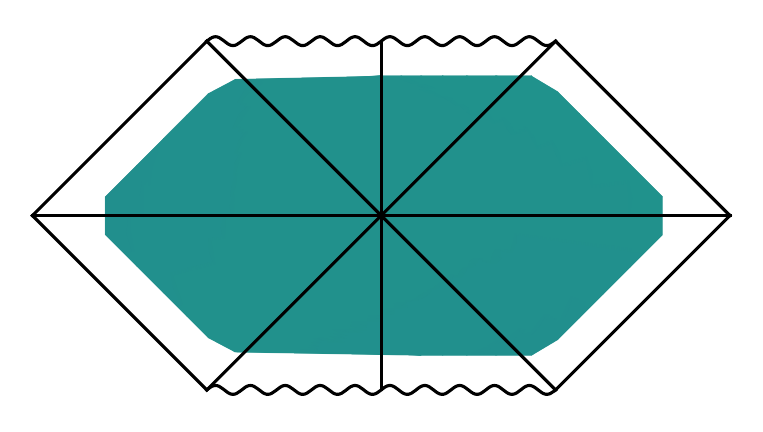}
    \caption{\(g_{10}\)}
  \end{subfigure}
  \hfill
  \begin{subfigure}[t]{0.235\linewidth}
    \centering
    \includegraphics[width=\textwidth]{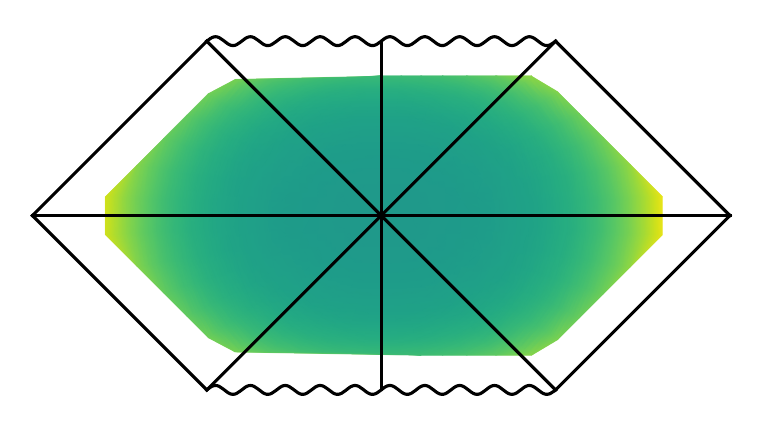}
    \caption{\(g_{11}\)}
  \end{subfigure}
  \hfill
  \begin{subfigure}[t]{0.235\linewidth}
    \centering
    \includegraphics[width=\textwidth]{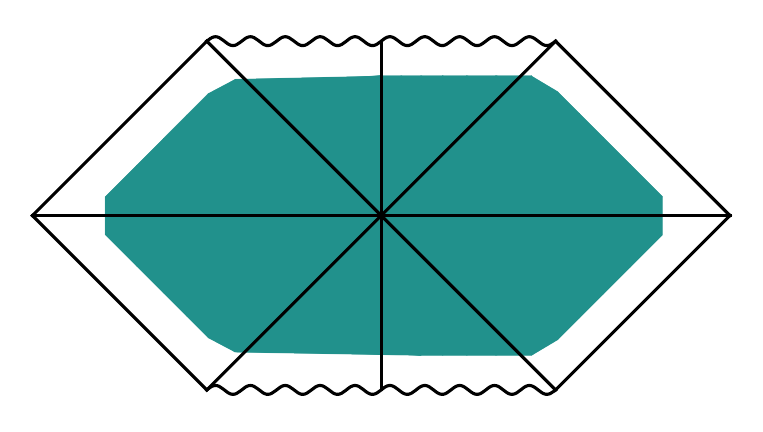}
    \caption{\(g_{12}\)}
  \end{subfigure}
  \hfill
  \begin{subfigure}[t]{0.235\linewidth}
    \centering
    \includegraphics[width=\textwidth]{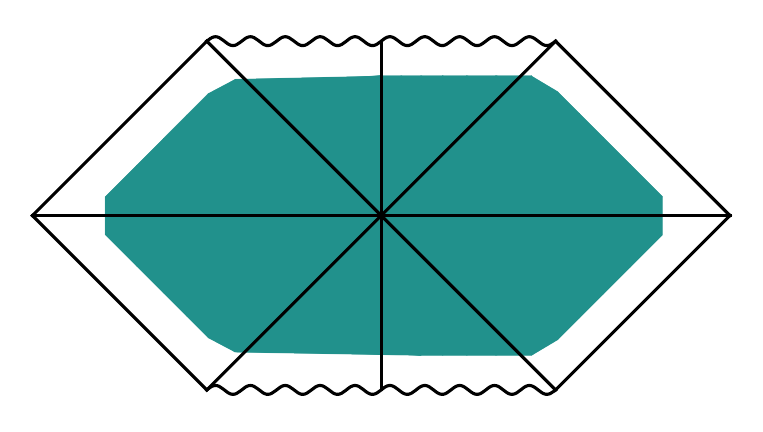}
    \caption{\(g_{13}\)}
  \end{subfigure}\\[0.25em]

  \begin{subfigure}[t]{0.235\linewidth}
    \centering
    \includegraphics[width=\textwidth]{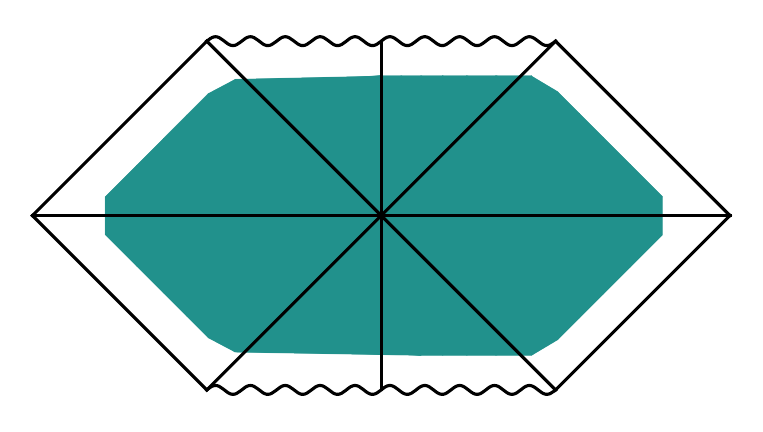}
    \caption{\(g_{20}\)}
  \end{subfigure}
  \hfill
  \begin{subfigure}[t]{0.235\linewidth}
    \centering
    \includegraphics[width=\textwidth]{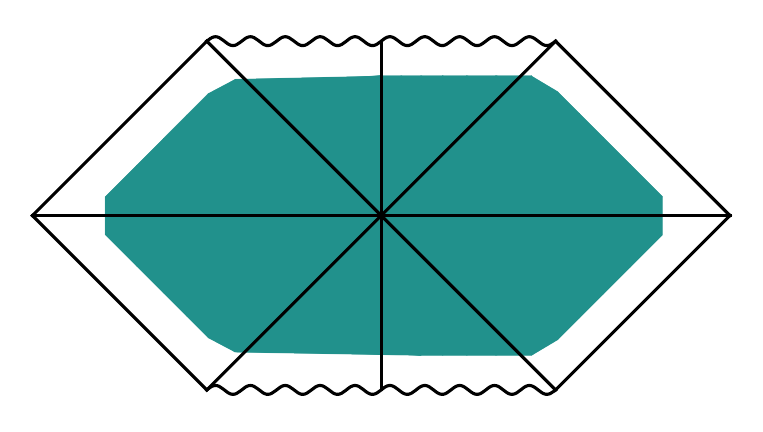}
    \caption{\(g_{21}\)}
  \end{subfigure}
  \hfill
  \begin{subfigure}[t]{0.235\linewidth}
    \centering
    \includegraphics[width=\textwidth]{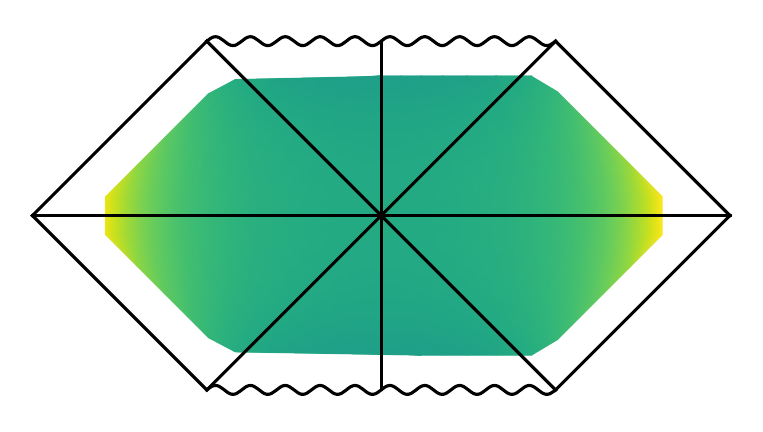}
    \caption{\(g_{22}\)}
  \end{subfigure}
  \hfill
  \begin{subfigure}[t]{0.235\linewidth}
    \centering
    \includegraphics[width=\textwidth]{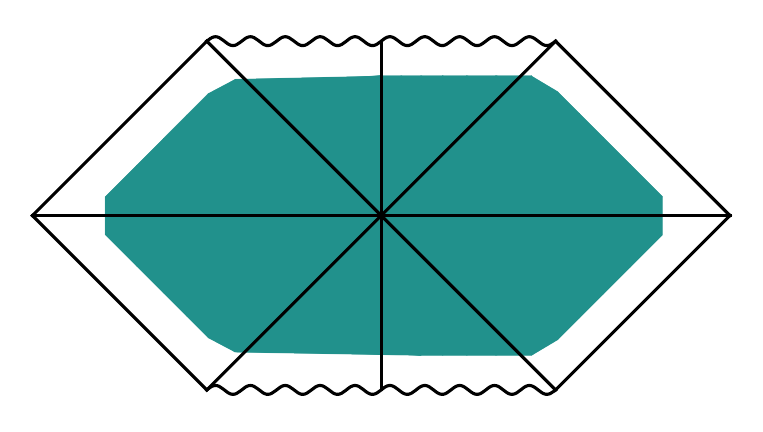}
    \caption{\(g_{23}\)}
  \end{subfigure}\\[0.25em]

  \begin{subfigure}[t]{0.235\linewidth}
    \centering
    \includegraphics[width=\textwidth]{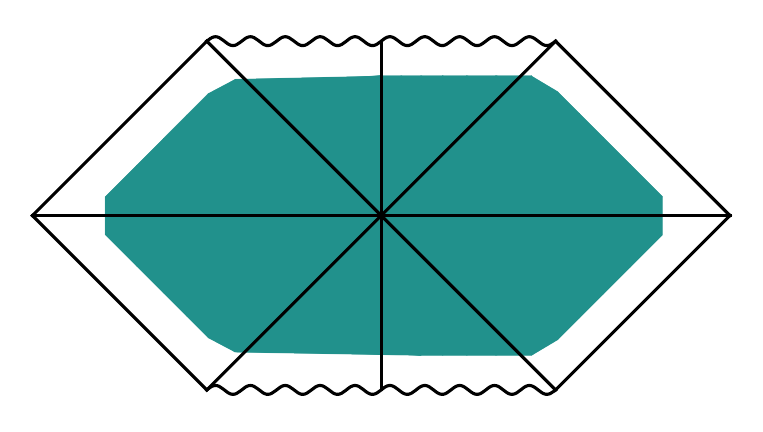}
    \caption{\(g_{30}\)}
  \end{subfigure}
  \hfill
  \begin{subfigure}[t]{0.235\linewidth}
    \centering
    \includegraphics[width=\textwidth]{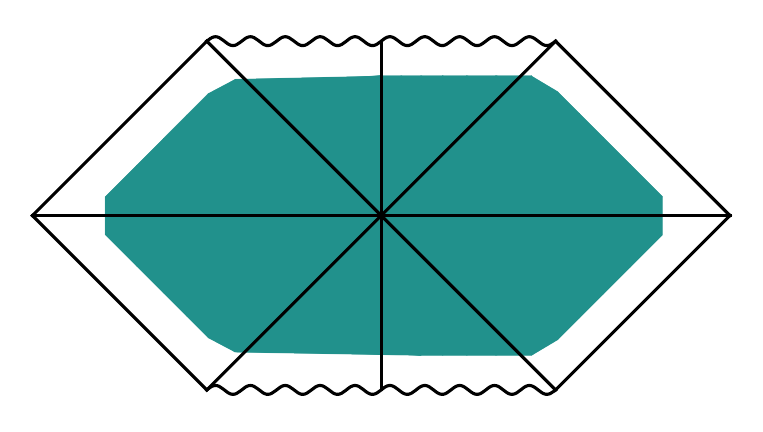}
    \caption{\(g_{31}\)}
  \end{subfigure}
  \hfill
  \begin{subfigure}[t]{0.235\linewidth}
    \centering
    \includegraphics[width=\textwidth]{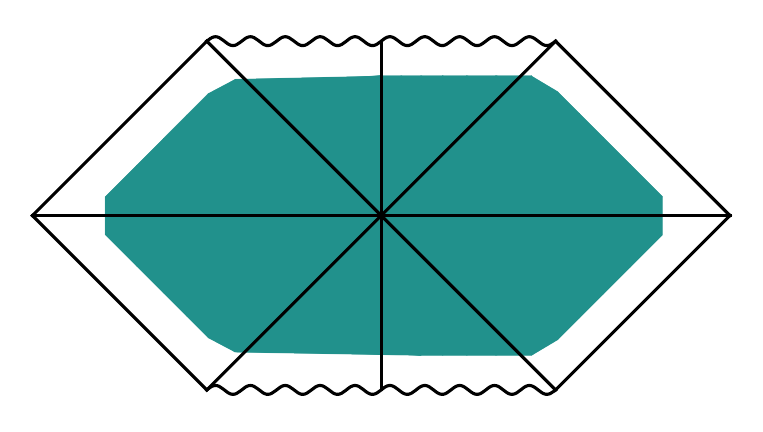}
    \caption{\(g_{32}\)}
  \end{subfigure}
  \hfill
  \begin{subfigure}[t]{0.235\linewidth}
    \centering
    \includegraphics[width=\textwidth]{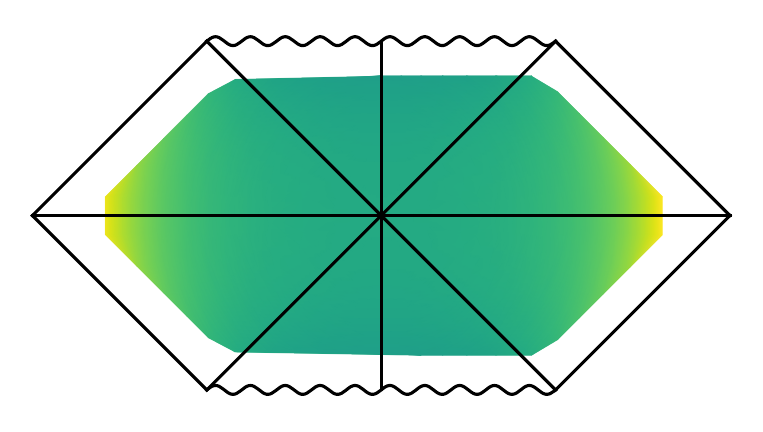}
    \caption{\(g_{33}\)}
  \end{subfigure}

  \end{minipage}%
  \hspace{0.04\textwidth}%
  \begin{minipage}[t]{0.47\textwidth}
  \centering
  \textbf{Ricci components}\\[0.35em]

  \begin{subfigure}[t]{0.235\linewidth}
    \centering
    \includegraphics[width=\textwidth]{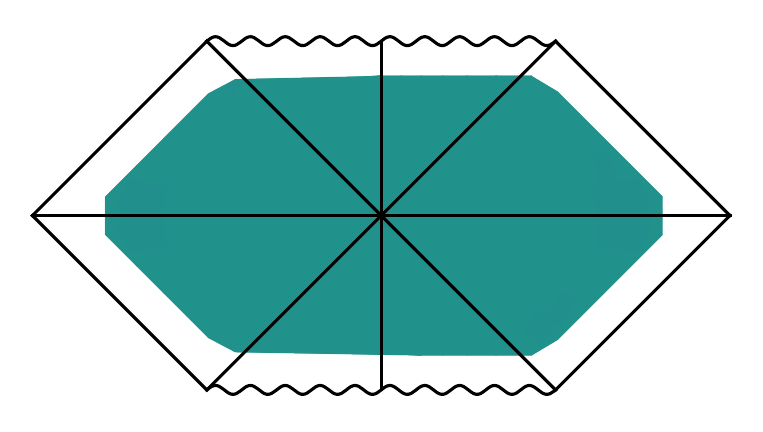}
    \caption{\(R_{00}\)}
  \end{subfigure}
  \hfill
  \begin{subfigure}[t]{0.235\linewidth}
    \centering
    \includegraphics[width=\textwidth]{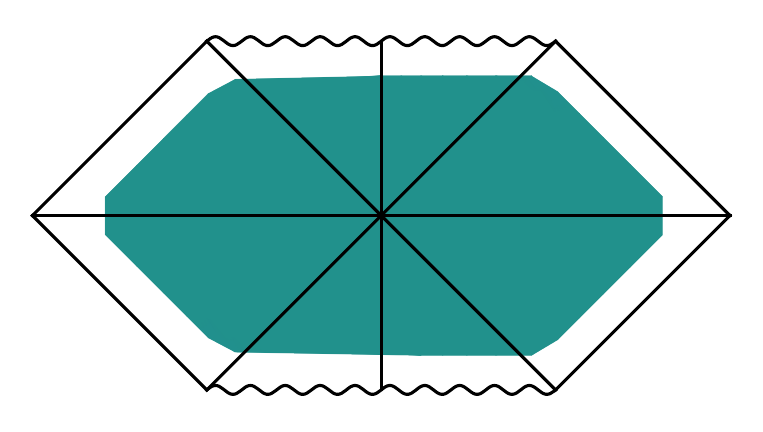}
    \caption{\(R_{01}\)}
  \end{subfigure}
  \hfill
  \begin{subfigure}[t]{0.235\linewidth}
    \centering
    \includegraphics[width=\textwidth]{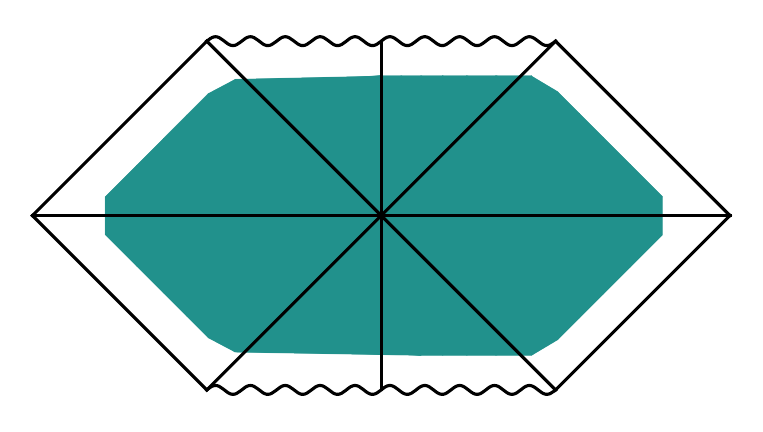}
    \caption{\(R_{02}\)}
  \end{subfigure}
  \hfill
  \begin{subfigure}[t]{0.235\linewidth}
    \centering
    \includegraphics[width=\textwidth]{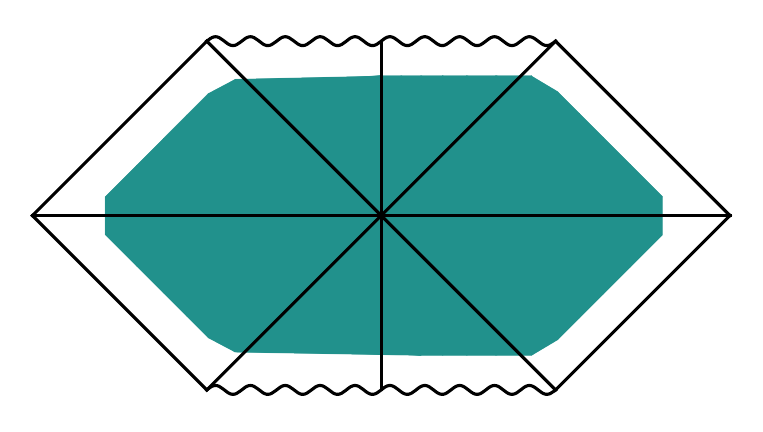}
    \caption{\(R_{03}\)}
  \end{subfigure}\\[0.25em]

  \begin{subfigure}[t]{0.235\linewidth}
    \centering
    \includegraphics[width=\textwidth]{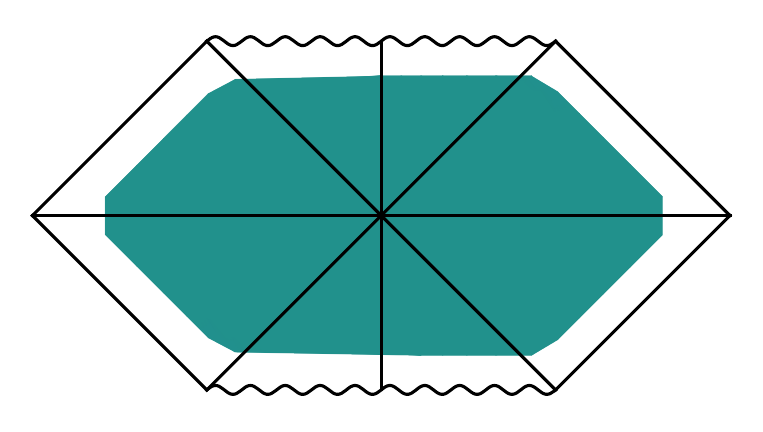}
    \caption{\(R_{10}\)}
  \end{subfigure}
  \hfill
  \begin{subfigure}[t]{0.235\linewidth}
    \centering
    \includegraphics[width=\textwidth]{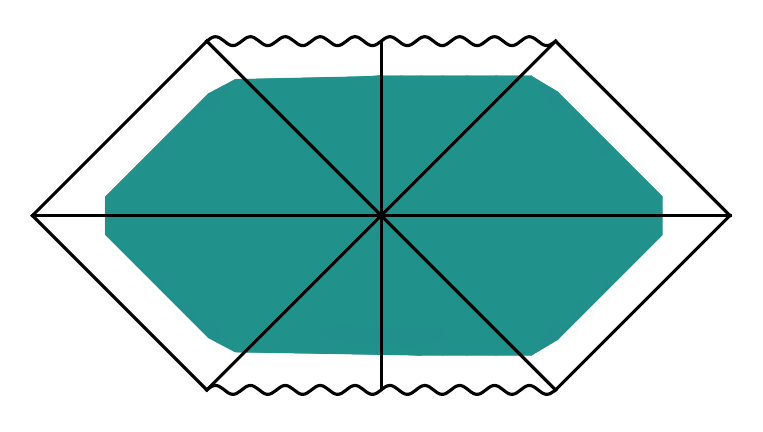}
    \caption{\(R_{11}\)}
  \end{subfigure}
  \hfill
  \begin{subfigure}[t]{0.235\linewidth}
    \centering
    \includegraphics[width=\textwidth]{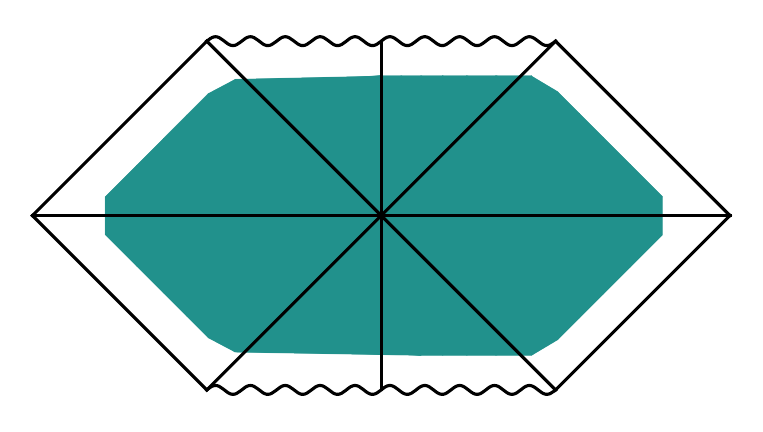}
    \caption{\(R_{12}\)}
  \end{subfigure}
  \hfill
  \begin{subfigure}[t]{0.235\linewidth}
    \centering
    \includegraphics[width=\textwidth]{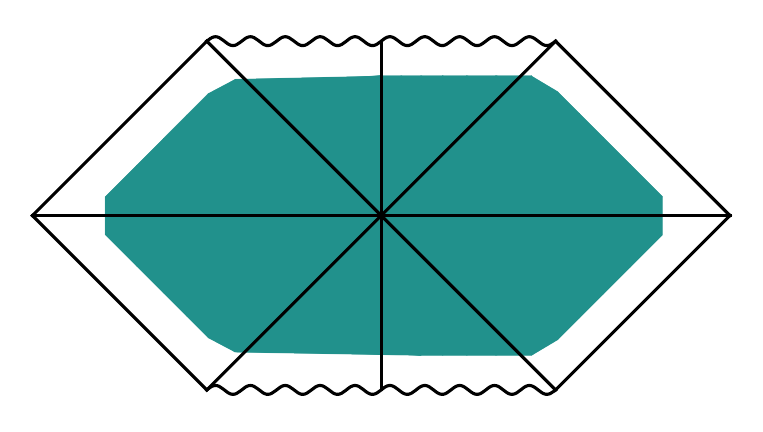}
    \caption{\(R_{13}\)}
  \end{subfigure}\\[0.25em]

  \begin{subfigure}[t]{0.235\linewidth}
    \centering
    \includegraphics[width=\textwidth]{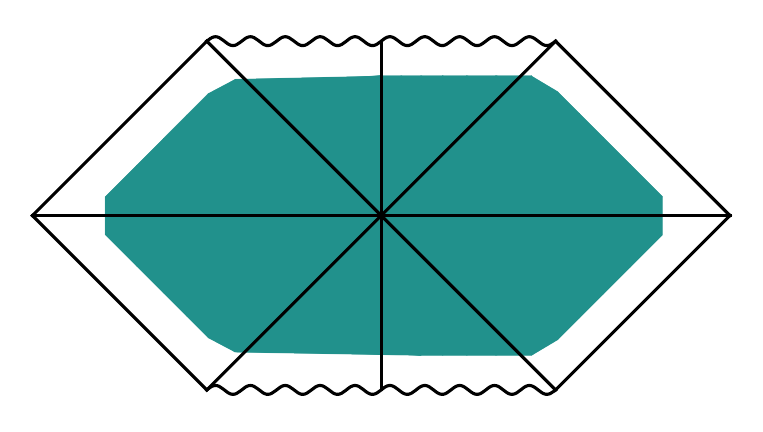}
    \caption{\(R_{20}\)}
  \end{subfigure}
  \hfill
  \begin{subfigure}[t]{0.235\linewidth}
    \centering
    \includegraphics[width=\textwidth]{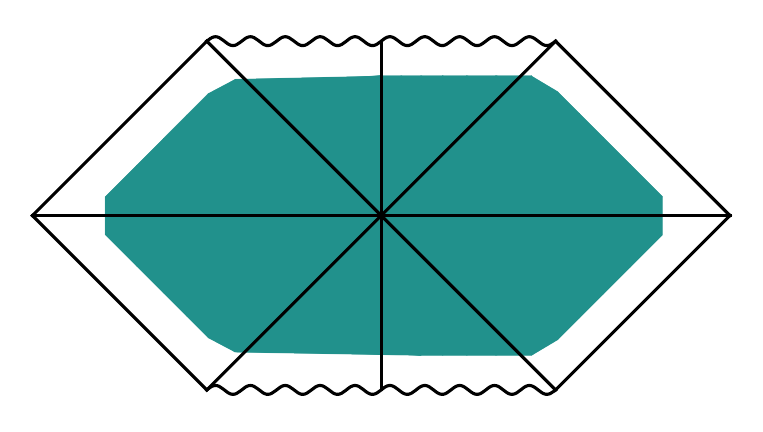}
    \caption{\(R_{21}\)}
  \end{subfigure}
  \hfill
  \begin{subfigure}[t]{0.235\linewidth}
    \centering
    \includegraphics[width=\textwidth]{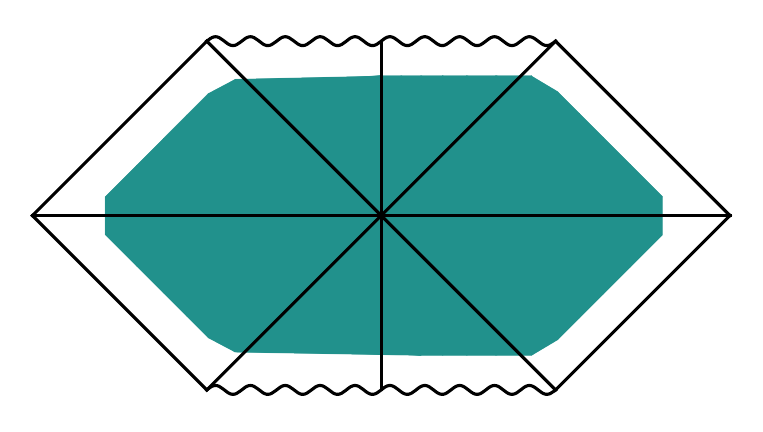}
    \caption{\(R_{22}\)}
  \end{subfigure}
  \hfill
  \begin{subfigure}[t]{0.235\linewidth}
    \centering
    \includegraphics[width=\textwidth]{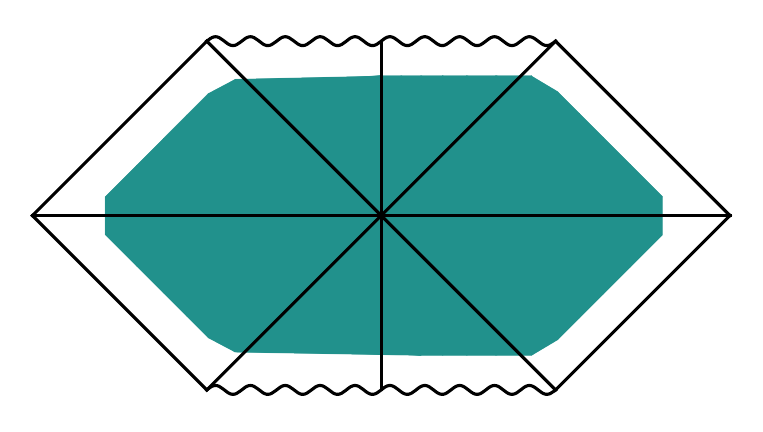}
    \caption{\(R_{23}\)}
  \end{subfigure}\\[0.25em]

  \begin{subfigure}[t]{0.235\linewidth}
    \centering
    \includegraphics[width=\textwidth]{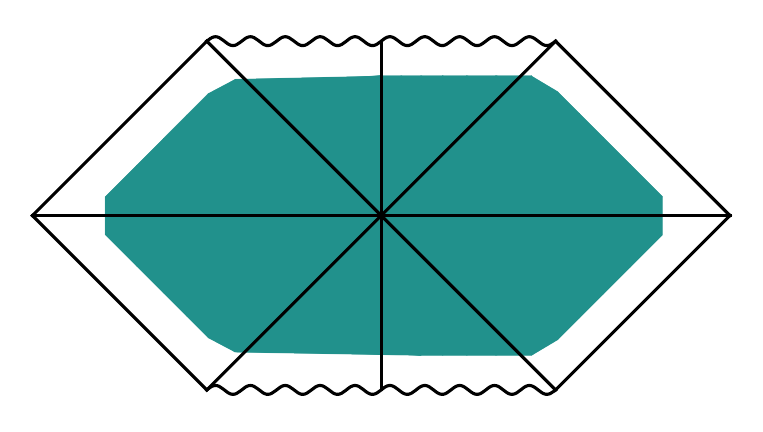}
    \caption{\(R_{30}\)}
  \end{subfigure}
  \hfill
  \begin{subfigure}[t]{0.235\linewidth}
    \centering
    \includegraphics[width=\textwidth]{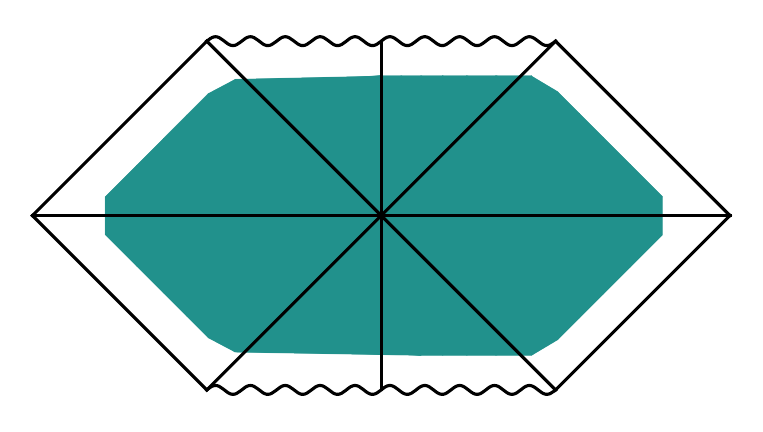}
    \caption{\(R_{31}\)}
  \end{subfigure}
  \hfill
  \begin{subfigure}[t]{0.235\linewidth}
    \centering
    \includegraphics[width=\textwidth]{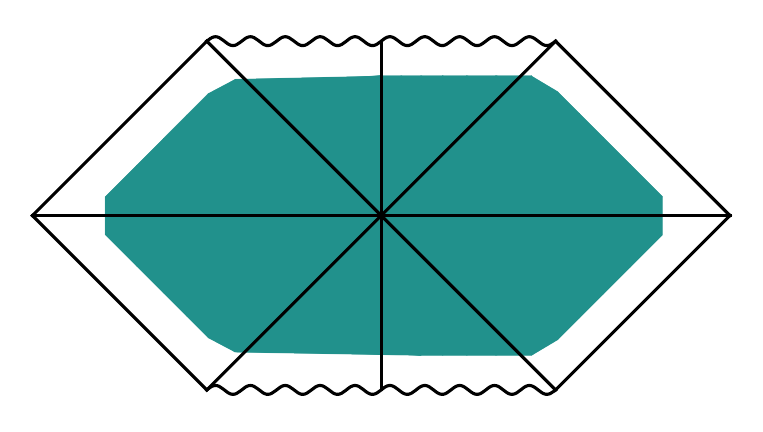}
    \caption{\(R_{32}\)}
  \end{subfigure}
  \hfill
  \begin{subfigure}[t]{0.235\linewidth}
    \centering
    \includegraphics[width=\textwidth]{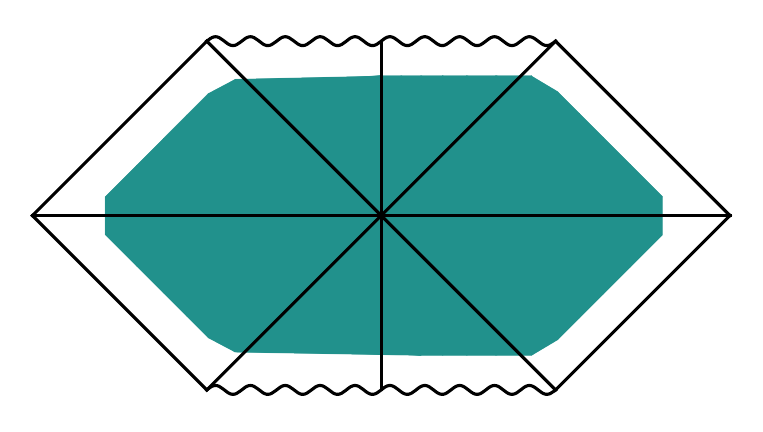}
    \caption{\(R_{33}\)}
  \end{subfigure}

  \end{minipage}

  \par\vspace{0.6em}
  \begin{subfigure}[t]{0.28\textwidth}
    \centering
    \includegraphics[width=\textwidth]{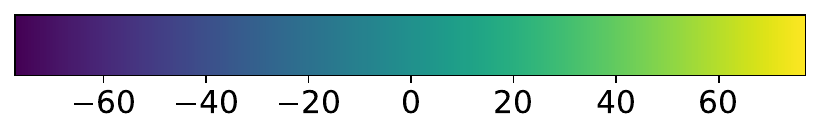}
    \caption*{Component scaling}
  \end{subfigure}

  \caption{Full component validation for the Schwarzschild run. The left block shows the learned metric components \(g_{\mu\nu}\), and the right block shows the learned Ricci tensor components \(R_{\mu\nu}\), both arranged by tensor index over the validation samples. Ricci flatness corresponds to the Ricci panels being close to zero. Over the samples and components the absolute value means are \( |\hat{g}_{ij}| = 4.316 \pm 2.374 \) and \( |\hat{R}_{ij}| = 0.015 \pm 0.012 \).}
  \label{fig:schwarzschild_components_4x4}
\end{figure*}

\begin{figure*}[p]
  \centering
  \captionsetup[subfigure]{justification=centering,singlelinecheck=true}

  \begin{minipage}[t]{0.47\textwidth}
  \centering
  \textbf{Metric components}\\[0.35em]

  \begin{subfigure}[t]{0.235\linewidth}
    \centering
    \includegraphics[width=\textwidth]{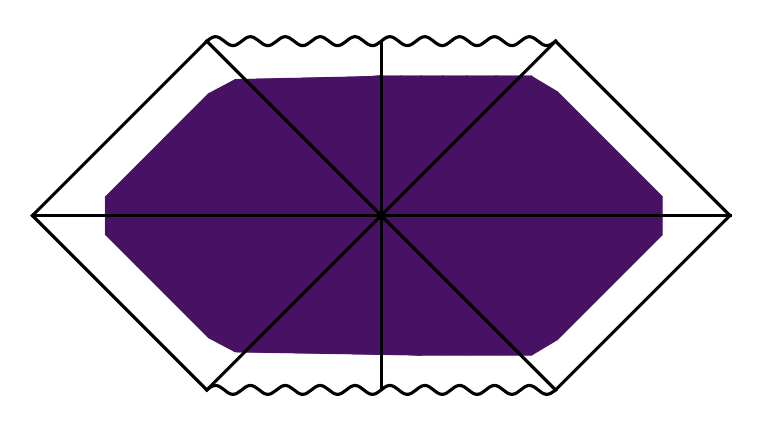}
    \caption{\(g_{00}\)}
  \end{subfigure}
  \hfill
  \begin{subfigure}[t]{0.235\linewidth}
    \centering
    \includegraphics[width=\textwidth]{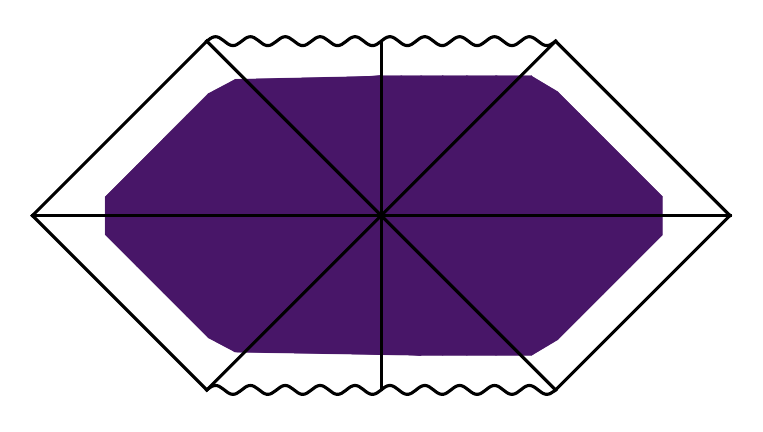}
    \caption{\(g_{01}\)}
  \end{subfigure}
  \hfill
  \begin{subfigure}[t]{0.235\linewidth}
    \centering
    \includegraphics[width=\textwidth]{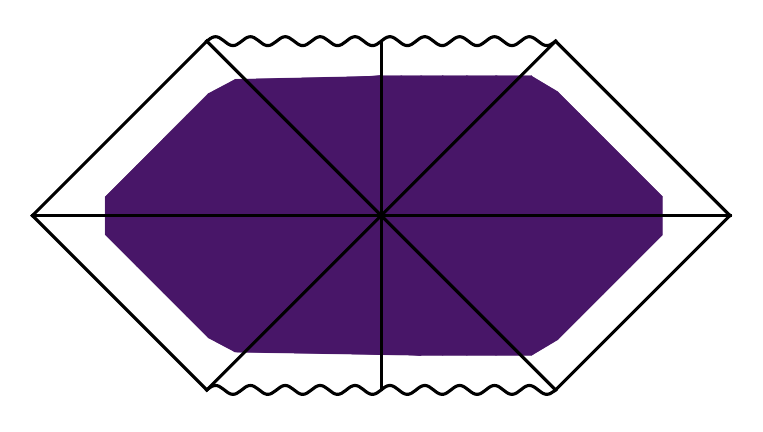}
    \caption{\(g_{02}\)}
  \end{subfigure}
  \hfill
  \begin{subfigure}[t]{0.235\linewidth}
    \centering
    \includegraphics[width=\textwidth]{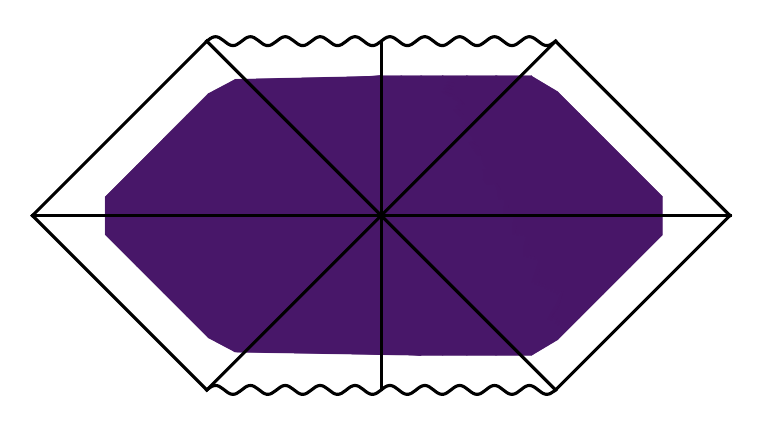}
    \caption{\(g_{03}\)}
  \end{subfigure}\\[0.25em]
  \begin{subfigure}[t]{0.235\linewidth}
    \centering
    \includegraphics[width=\textwidth]{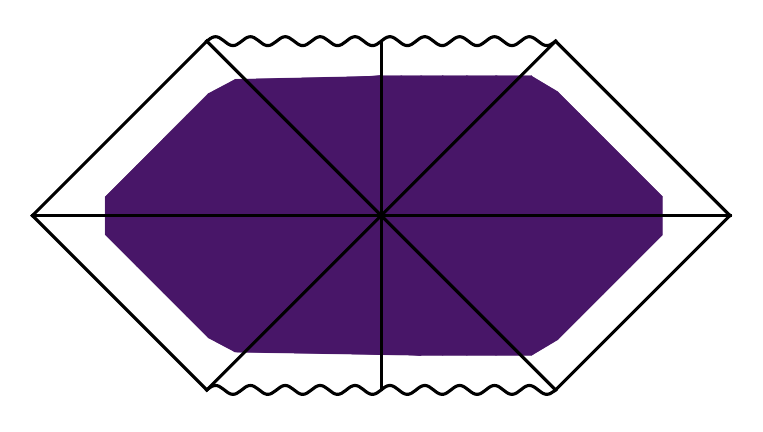}
    \caption{\(g_{10}\)}
  \end{subfigure}
  \hfill
  \begin{subfigure}[t]{0.235\linewidth}
    \centering
    \includegraphics[width=\textwidth]{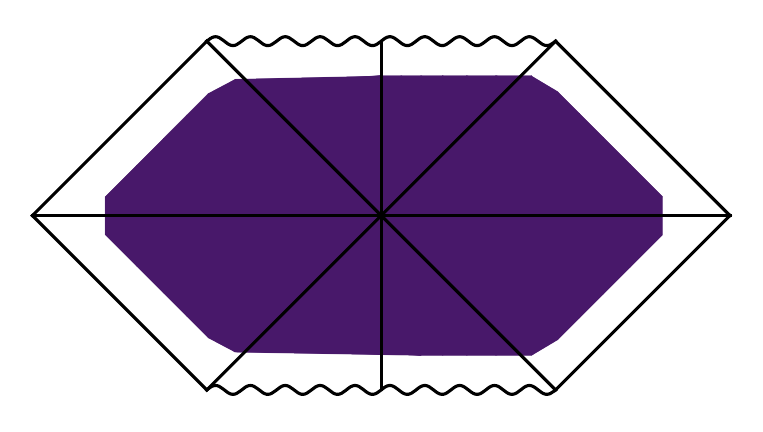}
    \caption{\(g_{11}\)}
  \end{subfigure}
  \hfill
  \begin{subfigure}[t]{0.235\linewidth}
    \centering
    \includegraphics[width=\textwidth]{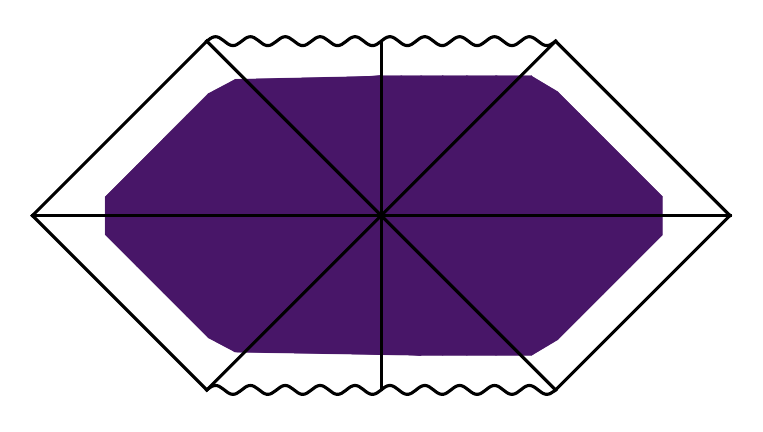}
    \caption{\(g_{12}\)}
  \end{subfigure}
  \hfill
  \begin{subfigure}[t]{0.235\linewidth}
    \centering
    \includegraphics[width=\textwidth]{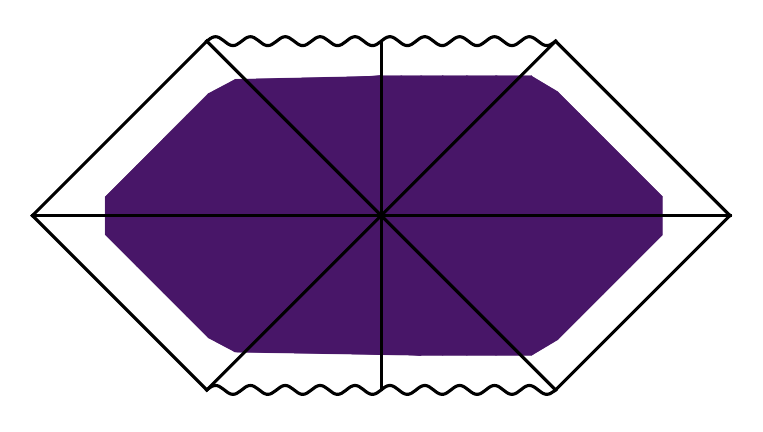}
    \caption{\(g_{13}\)}
  \end{subfigure}\\[0.25em]
  \begin{subfigure}[t]{0.235\linewidth}
    \centering
    \includegraphics[width=\textwidth]{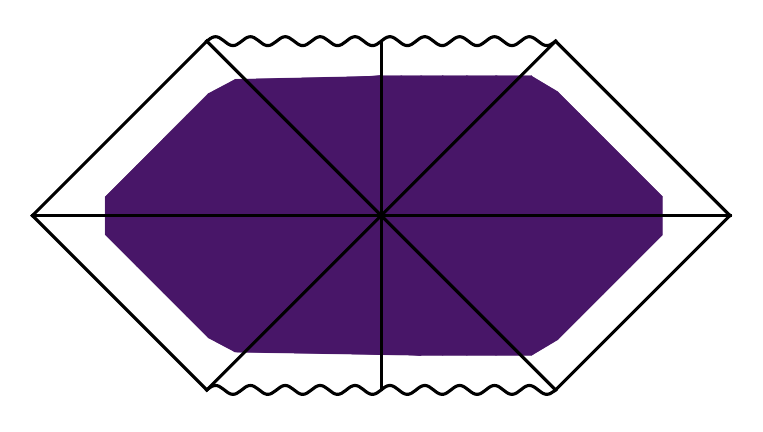}
    \caption{\(g_{20}\)}
  \end{subfigure}
  \hfill
  \begin{subfigure}[t]{0.235\linewidth}
    \centering
    \includegraphics[width=\textwidth]{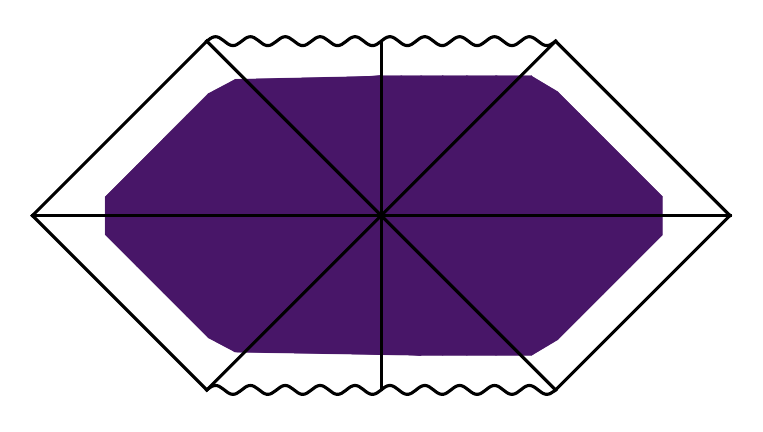}
    \caption{\(g_{21}\)}
  \end{subfigure}
  \hfill
  \begin{subfigure}[t]{0.235\linewidth}
    \centering
    \includegraphics[width=\textwidth]{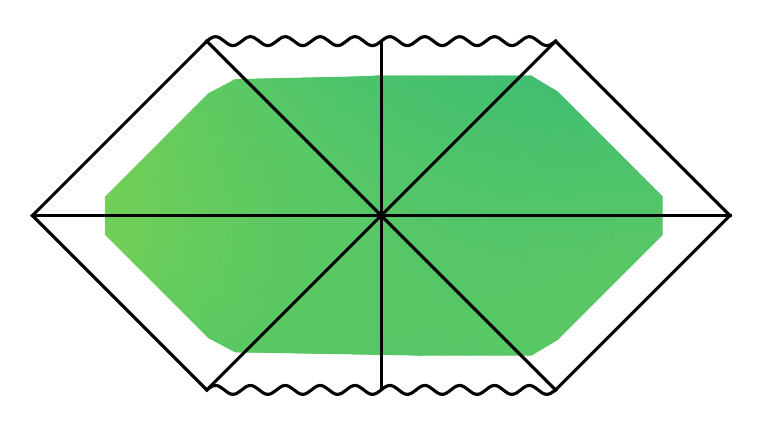}
    \caption{\(g_{22}\)}
  \end{subfigure}
  \hfill
  \begin{subfigure}[t]{0.235\linewidth}
    \centering
    \includegraphics[width=\textwidth]{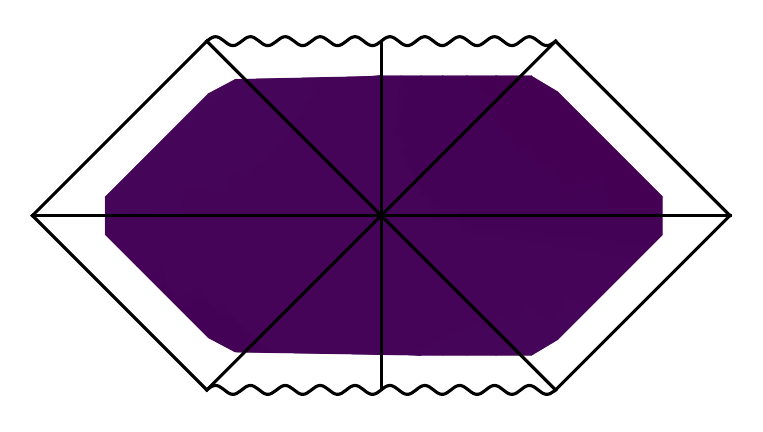}
    \caption{\(g_{23}\)}
  \end{subfigure}\\[0.25em]
  \begin{subfigure}[t]{0.235\linewidth}
    \centering
    \includegraphics[width=\textwidth]{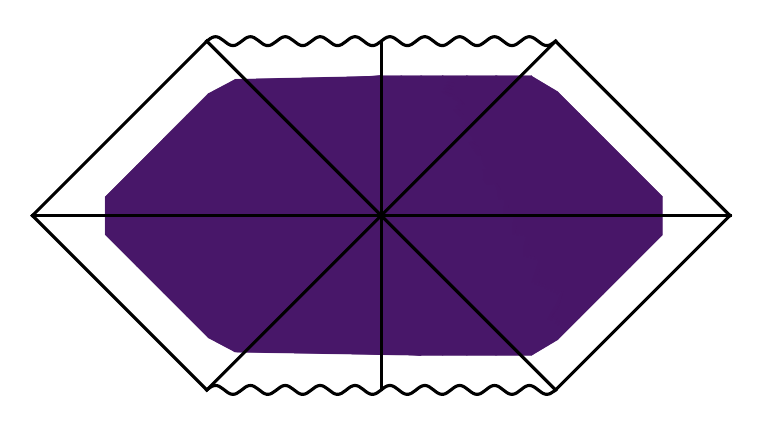}
    \caption{\(g_{30}\)}
  \end{subfigure}
  \hfill
  \begin{subfigure}[t]{0.235\linewidth}
    \centering
    \includegraphics[width=\textwidth]{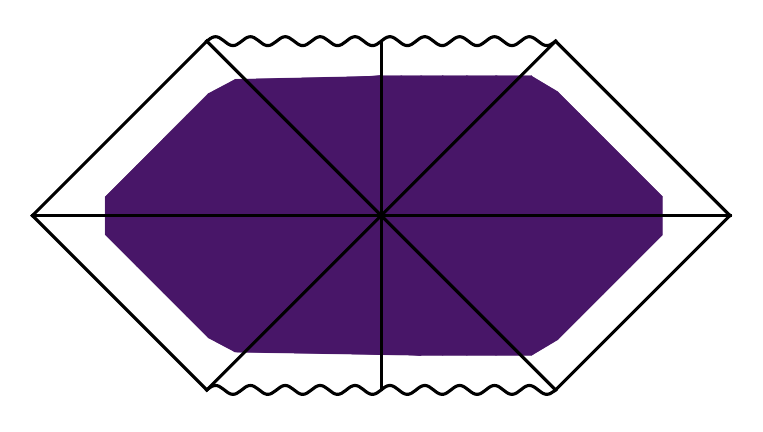}
    \caption{\(g_{31}\)}
  \end{subfigure}
  \hfill
  \begin{subfigure}[t]{0.235\linewidth}
    \centering
    \includegraphics[width=\textwidth]{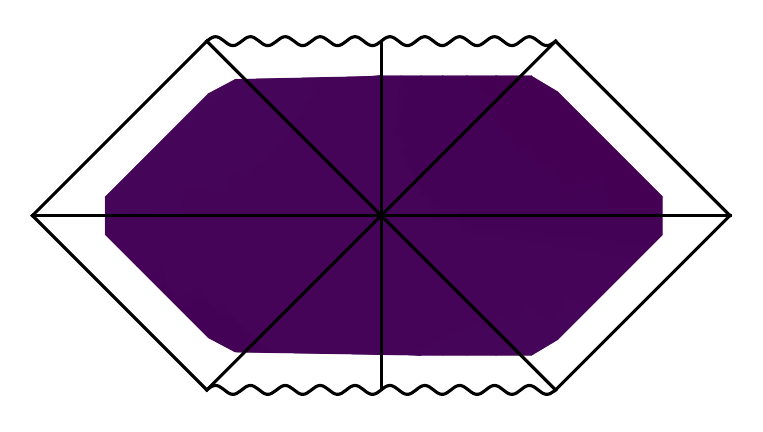}
    \caption{\(g_{32}\)}
  \end{subfigure}
  \hfill
  \begin{subfigure}[t]{0.235\linewidth}
    \centering
    \includegraphics[width=\textwidth]{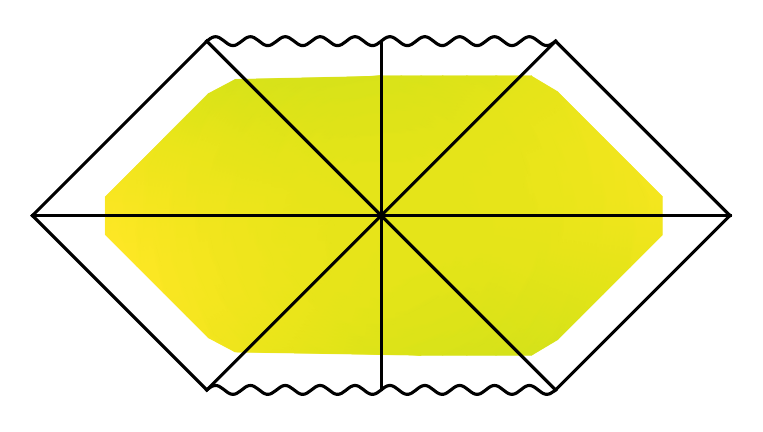}
    \caption{\(g_{33}\)}
  \end{subfigure}

  \end{minipage}%
  \hspace{0.04\textwidth}%
  \begin{minipage}[t]{0.47\textwidth}
  \centering
  \textbf{Ricci components}\\[0.35em]

  \begin{subfigure}[t]{0.235\linewidth}
    \centering
    \includegraphics[width=\textwidth]{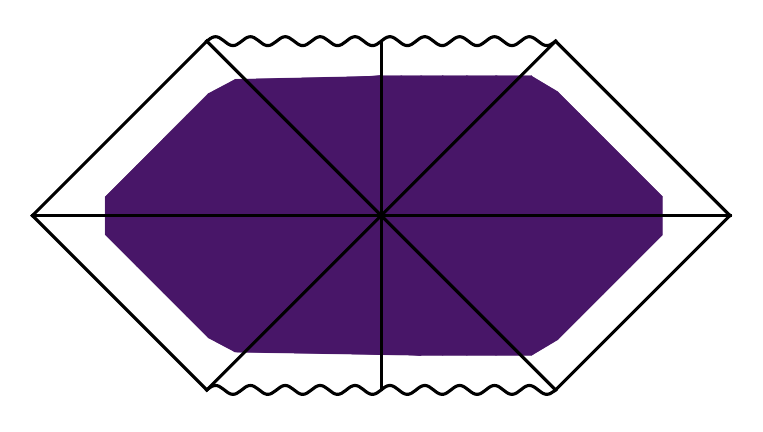}
    \caption{\(R_{00}\)}
  \end{subfigure}
  \hfill
  \begin{subfigure}[t]{0.235\linewidth}
    \centering
    \includegraphics[width=\textwidth]{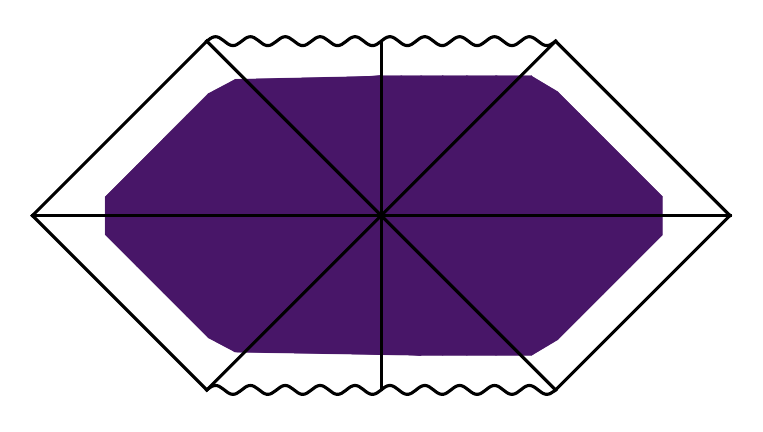}
    \caption{\(R_{01}\)}
  \end{subfigure}
  \hfill
  \begin{subfigure}[t]{0.235\linewidth}
    \centering
    \includegraphics[width=\textwidth]{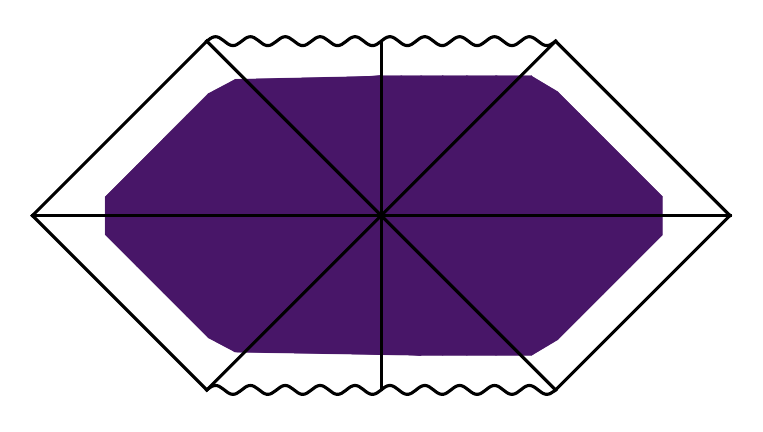}
    \caption{\(R_{02}\)}
  \end{subfigure}
  \hfill
  \begin{subfigure}[t]{0.235\linewidth}
    \centering
    \includegraphics[width=\textwidth]{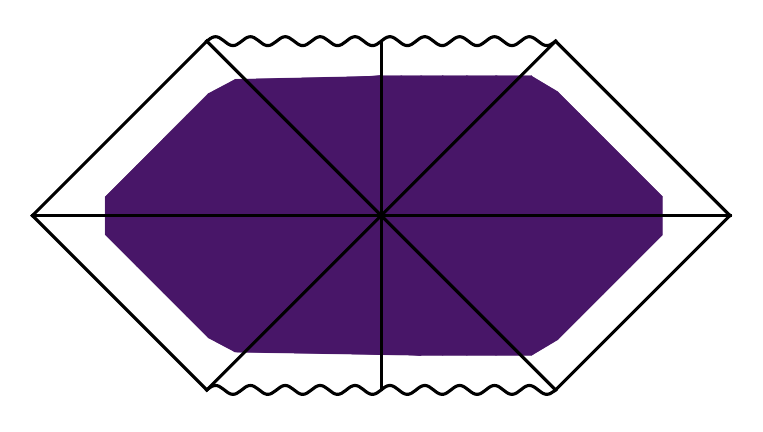}
    \caption{\(R_{03}\)}
  \end{subfigure}\\[0.25em]
  \begin{subfigure}[t]{0.235\linewidth}
    \centering
    \includegraphics[width=\textwidth]{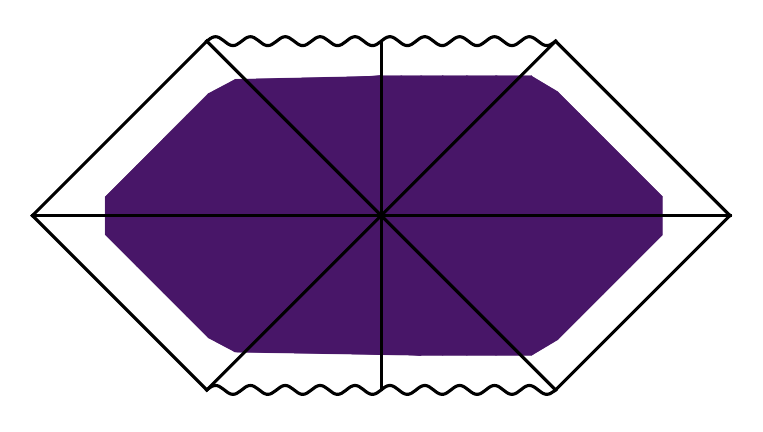}
    \caption{\(R_{10}\)}
  \end{subfigure}
  \hfill
  \begin{subfigure}[t]{0.235\linewidth}
    \centering
    \includegraphics[width=\textwidth]{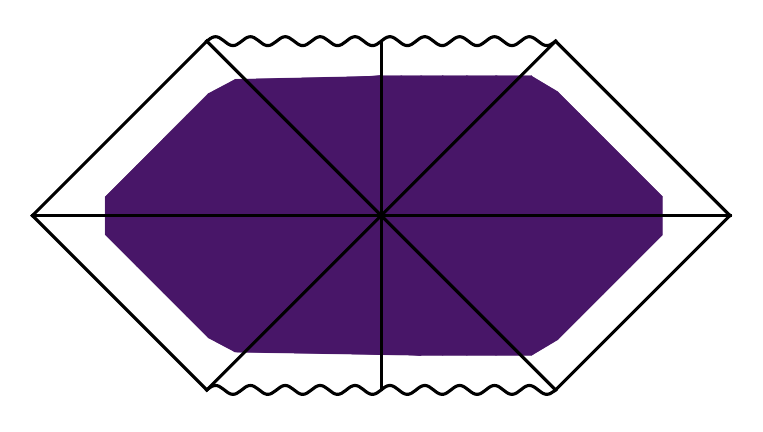}
    \caption{\(R_{11}\)}
  \end{subfigure}
  \hfill
  \begin{subfigure}[t]{0.235\linewidth}
    \centering
    \includegraphics[width=\textwidth]{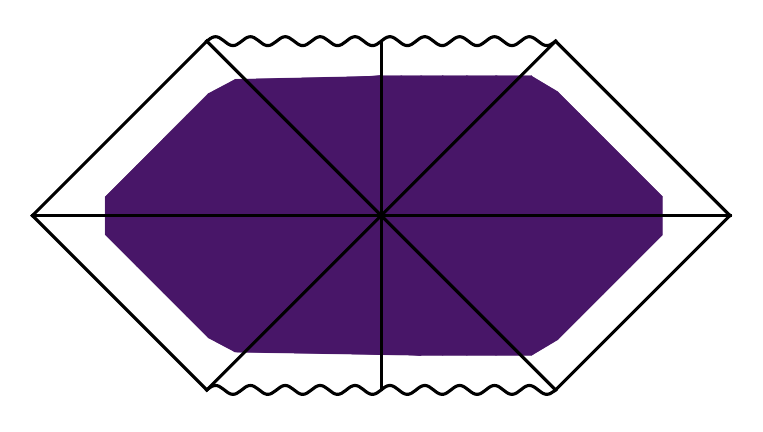}
    \caption{\(R_{12}\)}
  \end{subfigure}
  \hfill
  \begin{subfigure}[t]{0.235\linewidth}
    \centering
    \includegraphics[width=\textwidth]{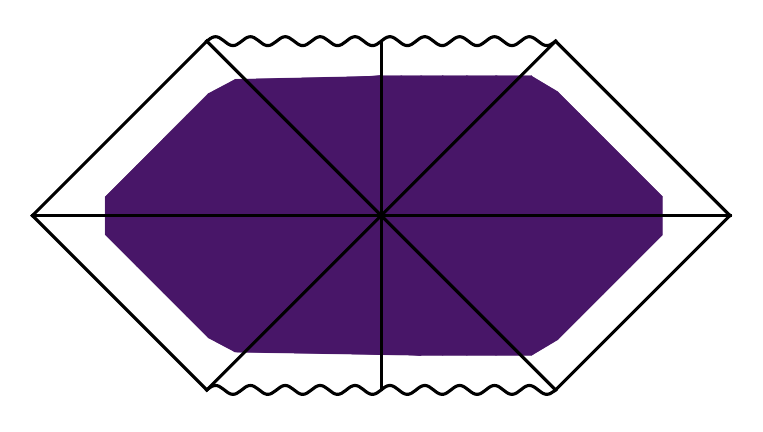}
    \caption{\(R_{13}\)}
  \end{subfigure}\\[0.25em]
  \begin{subfigure}[t]{0.235\linewidth}
    \centering
    \includegraphics[width=\textwidth]{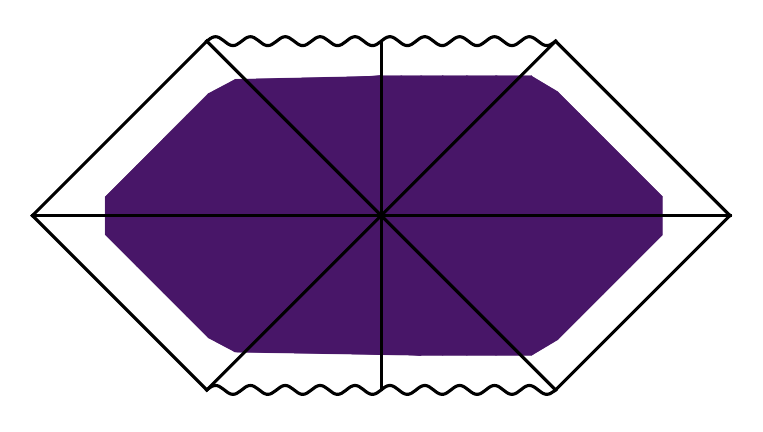}
    \caption{\(R_{20}\)}
  \end{subfigure}
  \hfill
  \begin{subfigure}[t]{0.235\linewidth}
    \centering
    \includegraphics[width=\textwidth]{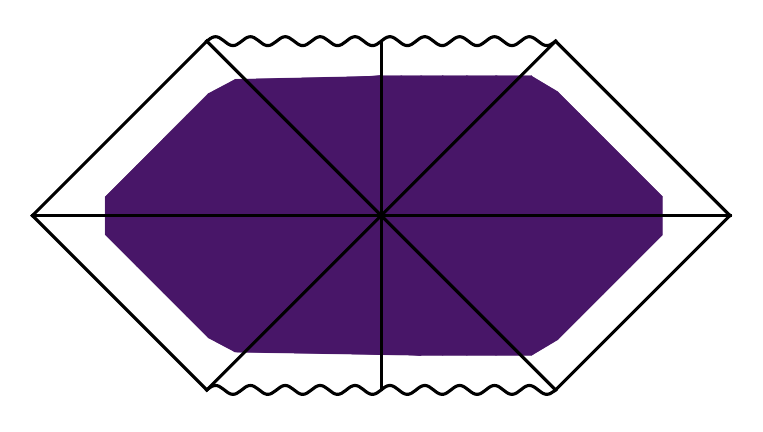}
    \caption{\(R_{21}\)}
  \end{subfigure}
  \hfill
  \begin{subfigure}[t]{0.235\linewidth}
    \centering
    \includegraphics[width=\textwidth]{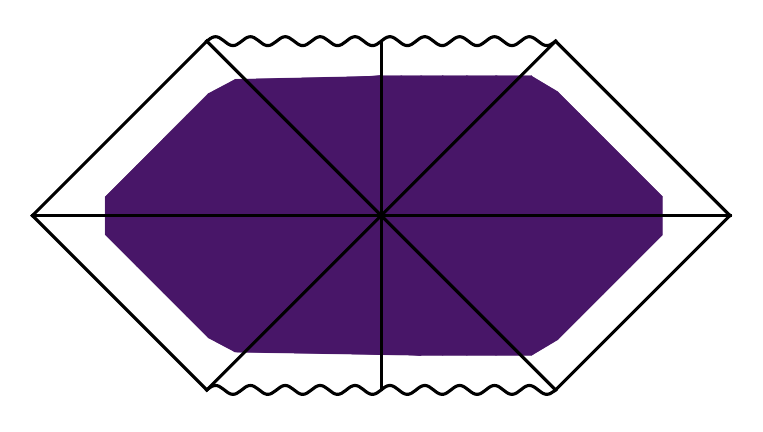}
    \caption{\(R_{22}\)}
  \end{subfigure}
  \hfill
  \begin{subfigure}[t]{0.235\linewidth}
    \centering
    \includegraphics[width=\textwidth]{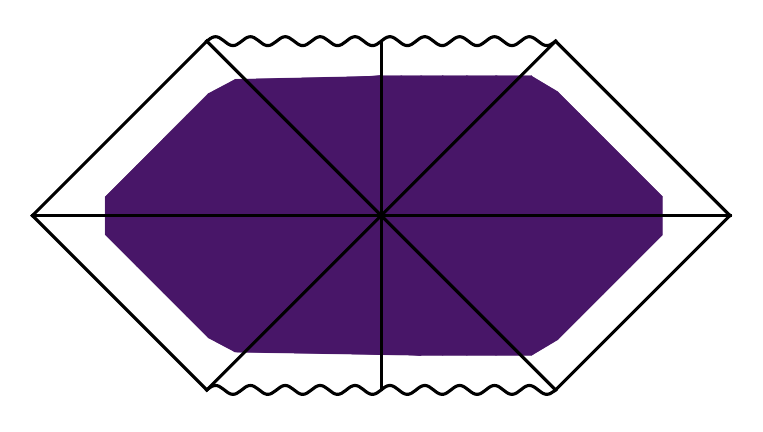}
    \caption{\(R_{23}\)}
  \end{subfigure}\\[0.25em]
  \begin{subfigure}[t]{0.235\linewidth}
    \centering
    \includegraphics[width=\textwidth]{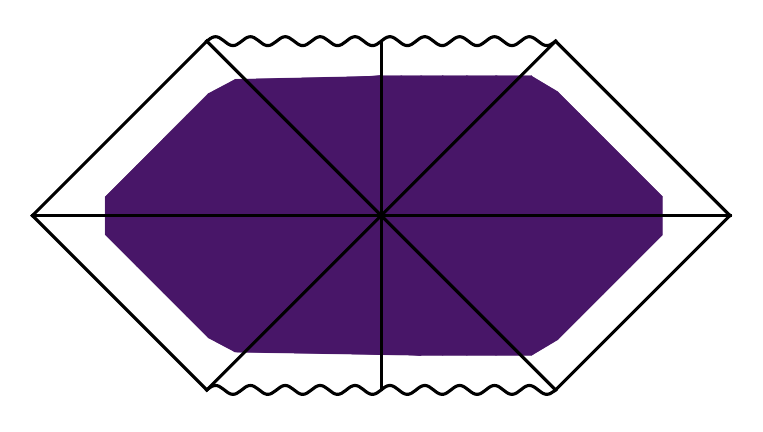}
    \caption{\(R_{30}\)}
  \end{subfigure}
  \hfill
  \begin{subfigure}[t]{0.235\linewidth}
    \centering
    \includegraphics[width=\textwidth]{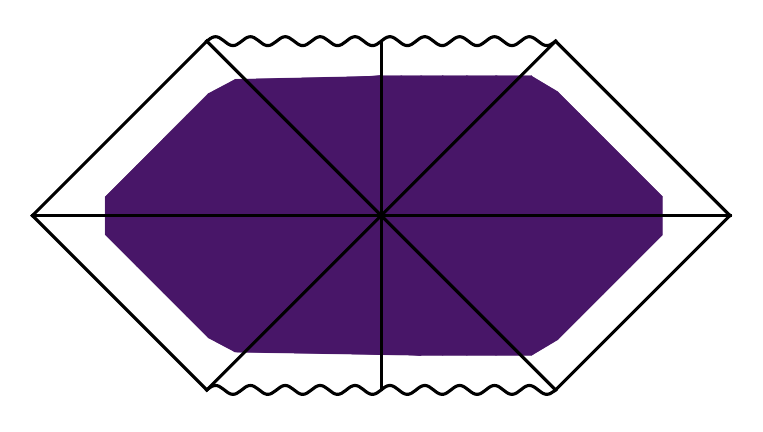}
    \caption{\(R_{31}\)}
  \end{subfigure}
  \hfill
  \begin{subfigure}[t]{0.235\linewidth}
    \centering
    \includegraphics[width=\textwidth]{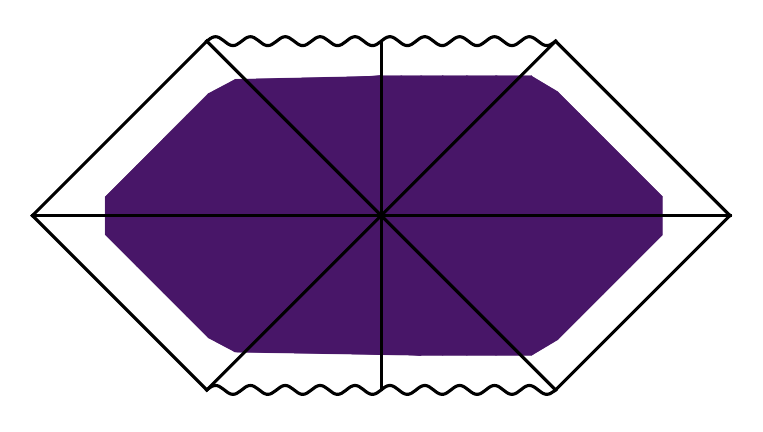}
    \caption{\(R_{32}\)}
  \end{subfigure}
  \hfill
  \begin{subfigure}[t]{0.235\linewidth}
    \centering
    \includegraphics[width=\textwidth]{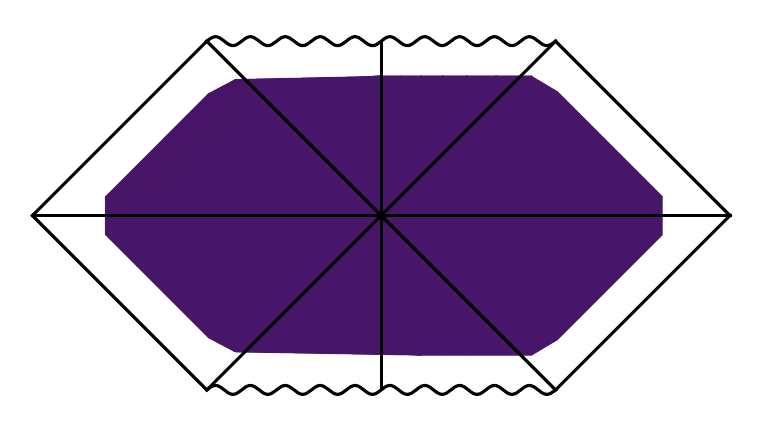}
    \caption{\(R_{33}\)}
  \end{subfigure}

  \end{minipage}

  \par\vspace{0.6em}
  \begin{subfigure}[t]{0.28\textwidth}
    \centering
    \includegraphics[width=\textwidth]{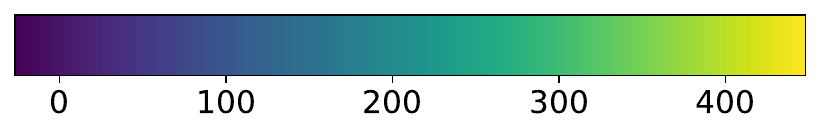}
    \caption*{Component scaling}
  \end{subfigure}

  \caption{Full component validation for the Petrov type-I search. The left block shows the learned metric components \(g_{\mu\nu}\), and the right block shows the learned Ricci tensor components \(R_{\mu\nu}\), both arranged by tensor index over the validation samples. Over the samples and components the absolute value means are $|\hat{g}_{ij}| = 48.99 \pm 2.40$ and $|\hat{R}_{ij}| = 0.038 \pm 0.006$.}
  \label{fig:typeI_components_4x4}
\end{figure*}

\begin{figure*}[p]
  \centering
  \captionsetup[subfigure]{justification=centering,singlelinecheck=true}

  \begin{minipage}[t]{0.47\textwidth}
  \centering
  \textbf{Metric components}\\[0.35em]

  \begin{subfigure}[t]{0.235\linewidth}
    \centering
    \includegraphics[width=\textwidth]{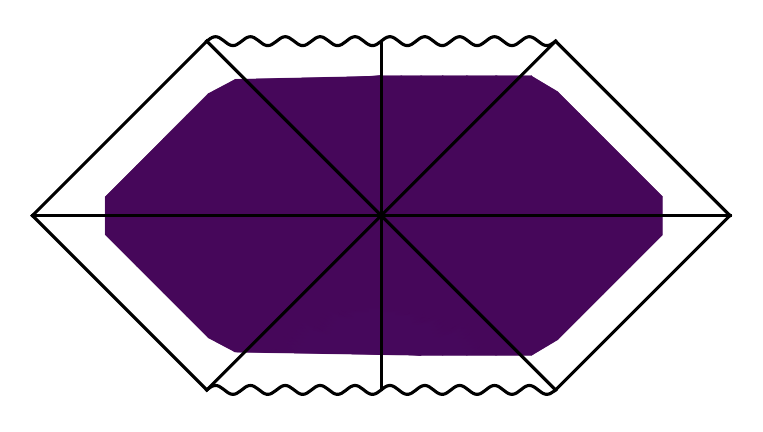}
    \caption{\(g_{00}\)}
  \end{subfigure}
  \hfill
  \begin{subfigure}[t]{0.235\linewidth}
    \centering
    \includegraphics[width=\textwidth]{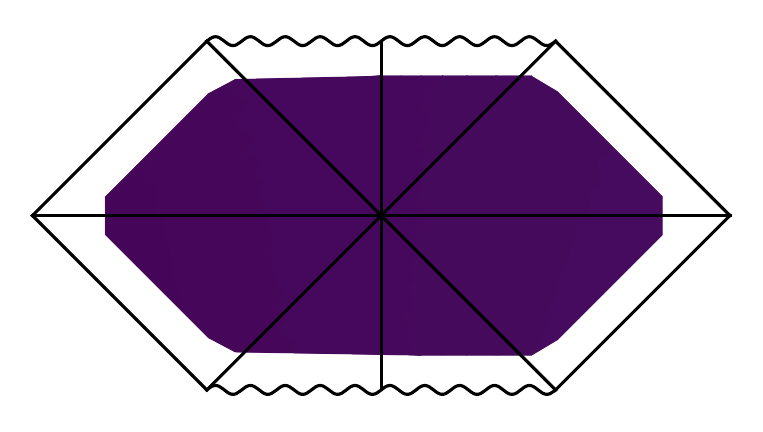}
    \caption{\(g_{01}\)}
  \end{subfigure}
  \hfill
  \begin{subfigure}[t]{0.235\linewidth}
    \centering
    \includegraphics[width=\textwidth]{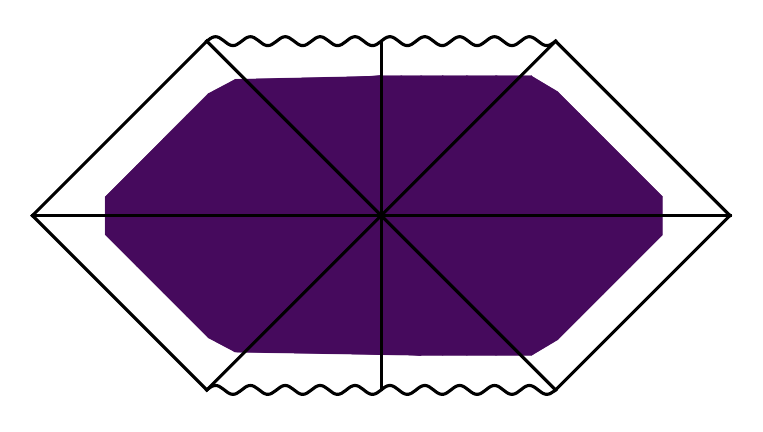}
    \caption{\(g_{02}\)}
  \end{subfigure}
  \hfill
  \begin{subfigure}[t]{0.235\linewidth}
    \centering
    \includegraphics[width=\textwidth]{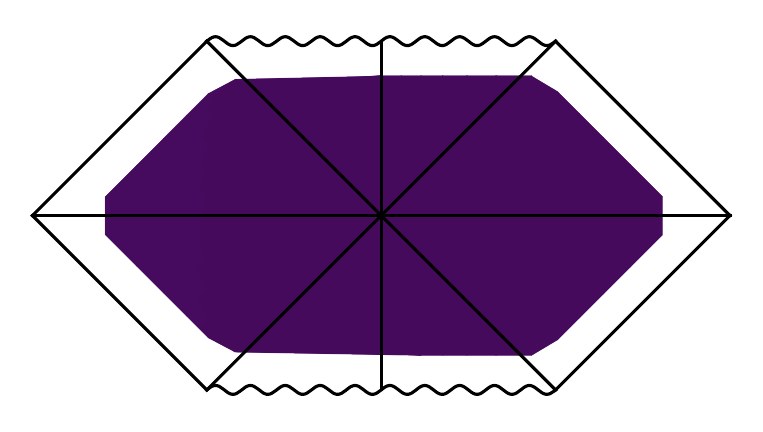}
    \caption{\(g_{03}\)}
  \end{subfigure}\\[0.25em]
  \begin{subfigure}[t]{0.235\linewidth}
    \centering
    \includegraphics[width=\textwidth]{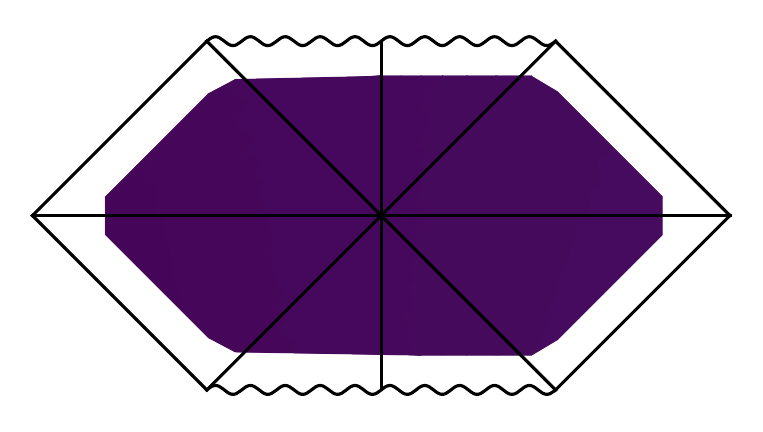}
    \caption{\(g_{10}\)}
  \end{subfigure}
  \hfill
  \begin{subfigure}[t]{0.235\linewidth}
    \centering
    \includegraphics[width=\textwidth]{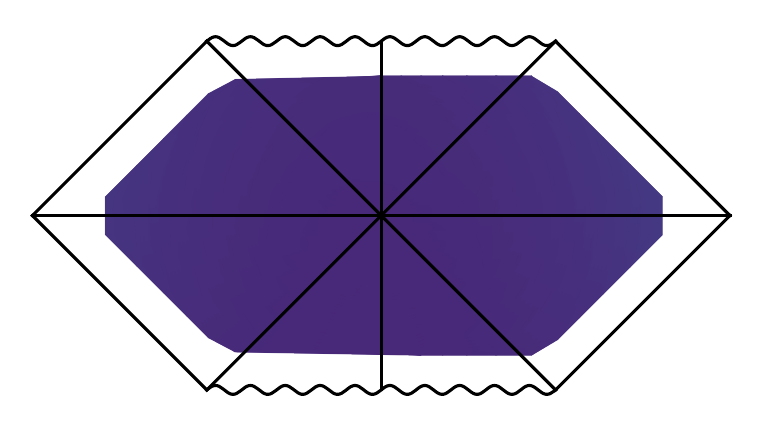}
    \caption{\(g_{11}\)}
  \end{subfigure}
  \hfill
  \begin{subfigure}[t]{0.235\linewidth}
    \centering
    \includegraphics[width=\textwidth]{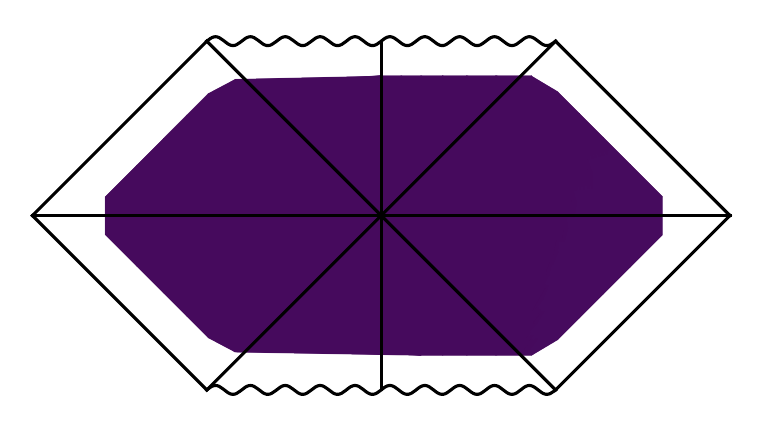}
    \caption{\(g_{12}\)}
  \end{subfigure}
  \hfill
  \begin{subfigure}[t]{0.235\linewidth}
    \centering
    \includegraphics[width=\textwidth]{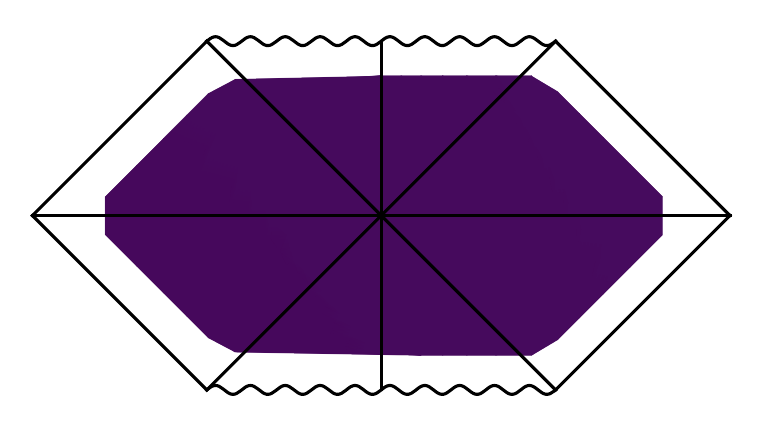}
    \caption{\(g_{13}\)}
  \end{subfigure}\\[0.25em]
  \begin{subfigure}[t]{0.235\linewidth}
    \centering
    \includegraphics[width=\textwidth]{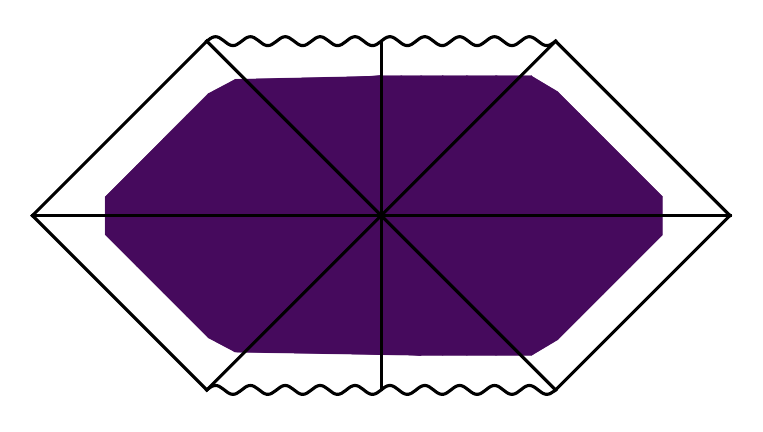}
    \caption{\(g_{20}\)}
  \end{subfigure}
  \hfill
  \begin{subfigure}[t]{0.235\linewidth}
    \centering
    \includegraphics[width=\textwidth]{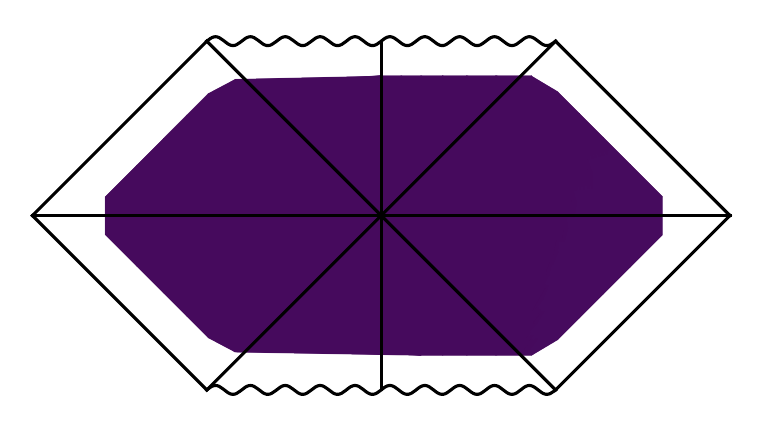}
    \caption{\(g_{21}\)}
  \end{subfigure}
  \hfill
  \begin{subfigure}[t]{0.235\linewidth}
    \centering
    \includegraphics[width=\textwidth]{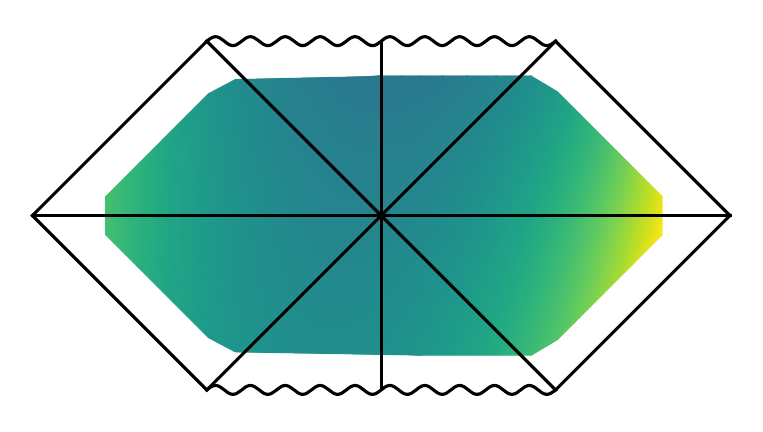}
    \caption{\(g_{22}\)}
  \end{subfigure}
  \hfill
  \begin{subfigure}[t]{0.235\linewidth}
    \centering
    \includegraphics[width=\textwidth]{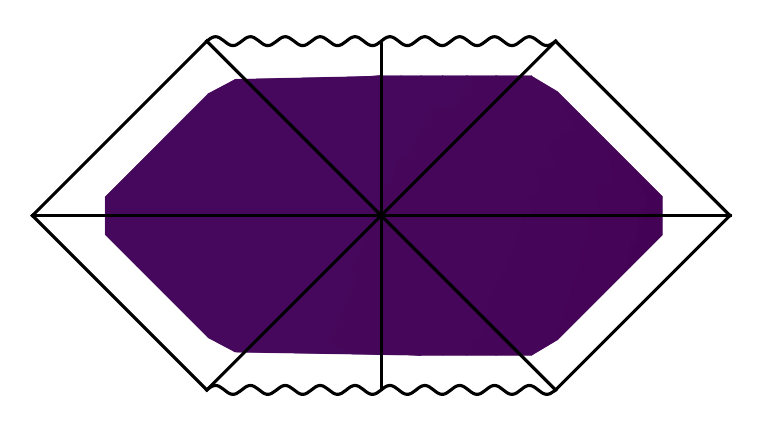}
    \caption{\(g_{23}\)}
  \end{subfigure}\\[0.25em]
  \begin{subfigure}[t]{0.235\linewidth}
    \centering
    \includegraphics[width=\textwidth]{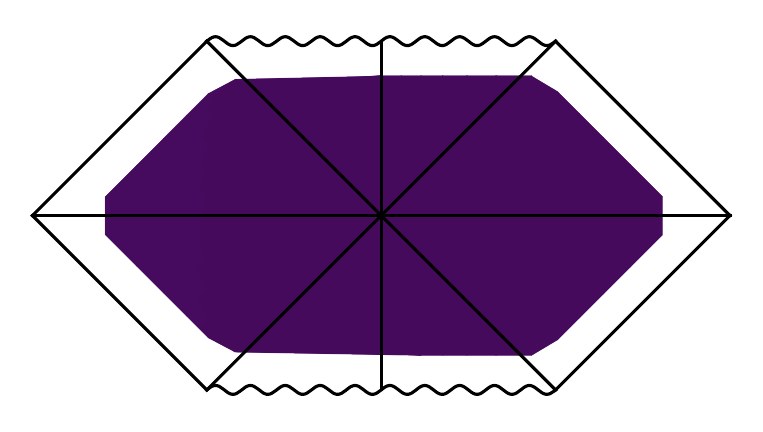}
    \caption{\(g_{30}\)}
  \end{subfigure}
  \hfill
  \begin{subfigure}[t]{0.235\linewidth}
    \centering
    \includegraphics[width=\textwidth]{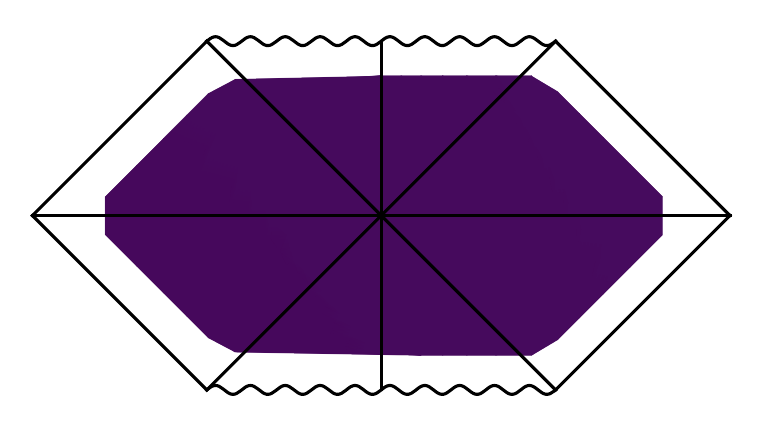}
    \caption{\(g_{31}\)}
  \end{subfigure}
  \hfill
  \begin{subfigure}[t]{0.235\linewidth}
    \centering
    \includegraphics[width=\textwidth]{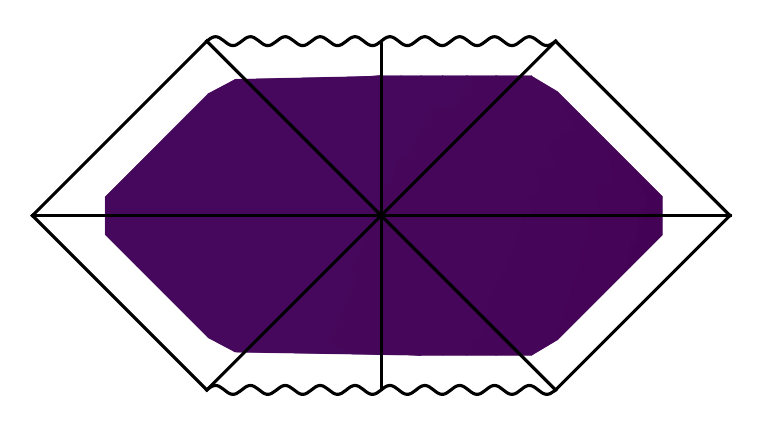}
    \caption{\(g_{32}\)}
  \end{subfigure}
  \hfill
  \begin{subfigure}[t]{0.235\linewidth}
    \centering
    \includegraphics[width=\textwidth]{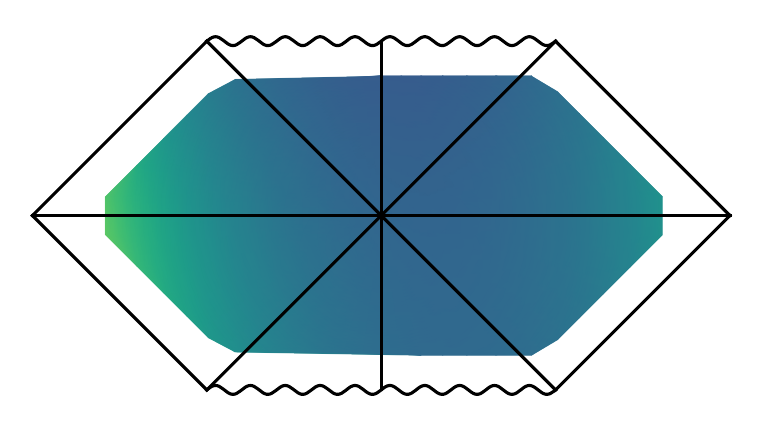}
    \caption{\(g_{33}\)}
  \end{subfigure}

  \end{minipage}%
  \hspace{0.04\textwidth}%
  \begin{minipage}[t]{0.47\textwidth}
  \centering
  \textbf{Ricci components}\\[0.35em]

  \begin{subfigure}[t]{0.235\linewidth}
    \centering
    \includegraphics[width=\textwidth]{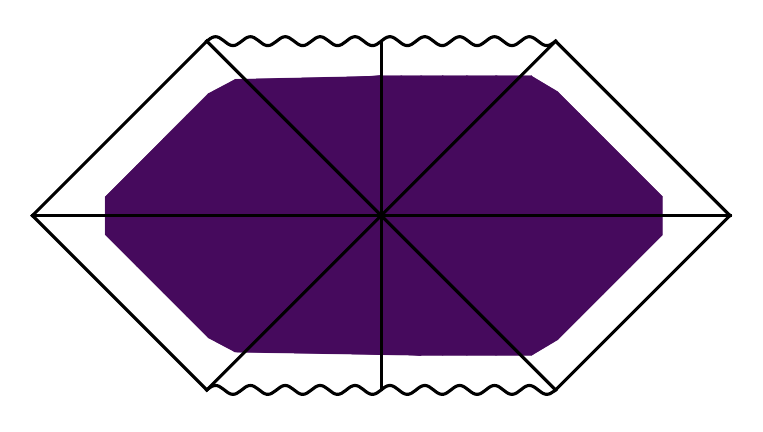}
    \caption{\(R_{00}\)}
  \end{subfigure}
  \hfill
  \begin{subfigure}[t]{0.235\linewidth}
    \centering
    \includegraphics[width=\textwidth]{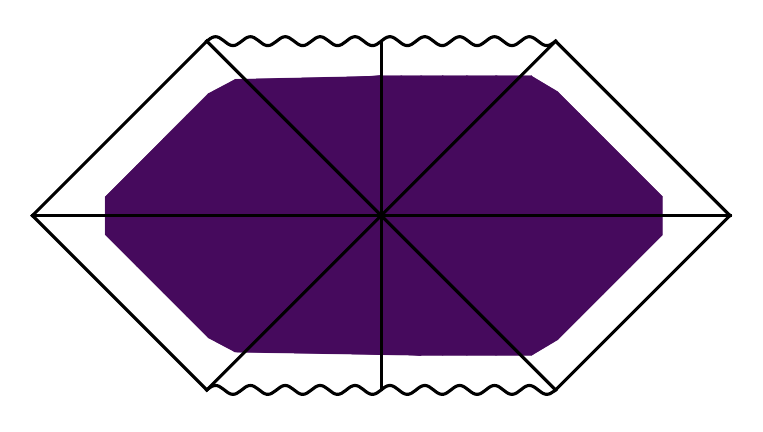}
    \caption{\(R_{01}\)}
  \end{subfigure}
  \hfill
  \begin{subfigure}[t]{0.235\linewidth}
    \centering
    \includegraphics[width=\textwidth]{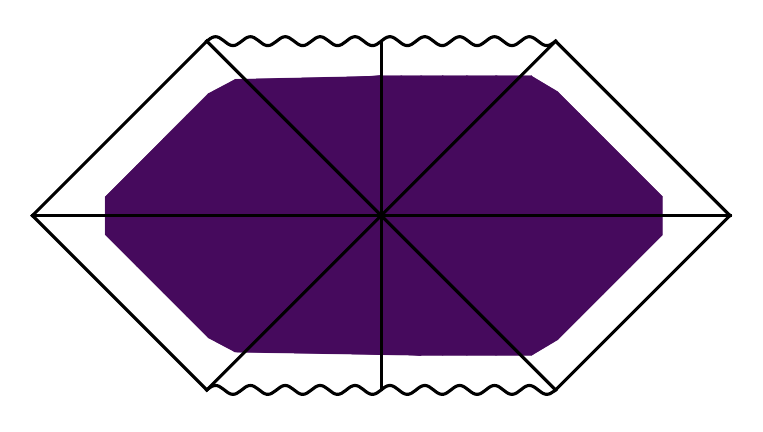}
    \caption{\(R_{02}\)}
  \end{subfigure}
  \hfill
  \begin{subfigure}[t]{0.235\linewidth}
    \centering
    \includegraphics[width=\textwidth]{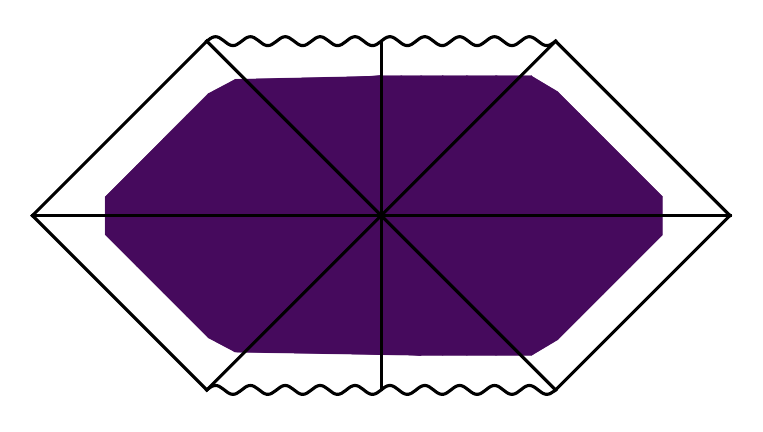}
    \caption{\(R_{03}\)}
  \end{subfigure}\\[0.25em]
  \begin{subfigure}[t]{0.235\linewidth}
    \centering
    \includegraphics[width=\textwidth]{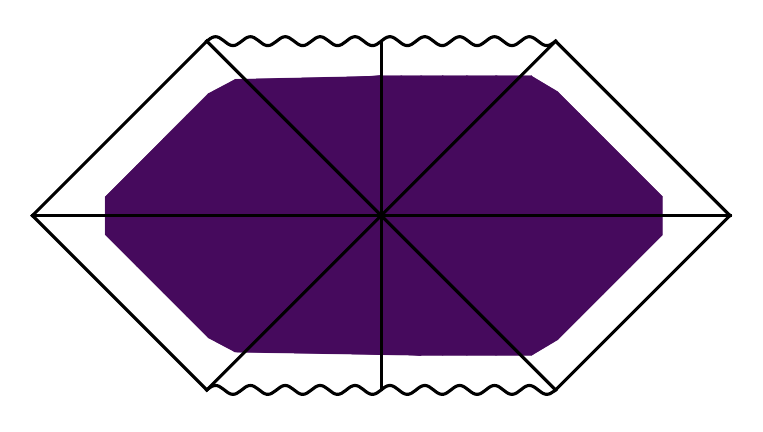}
    \caption{\(R_{10}\)}
  \end{subfigure}
  \hfill
  \begin{subfigure}[t]{0.235\linewidth}
    \centering
    \includegraphics[width=\textwidth]{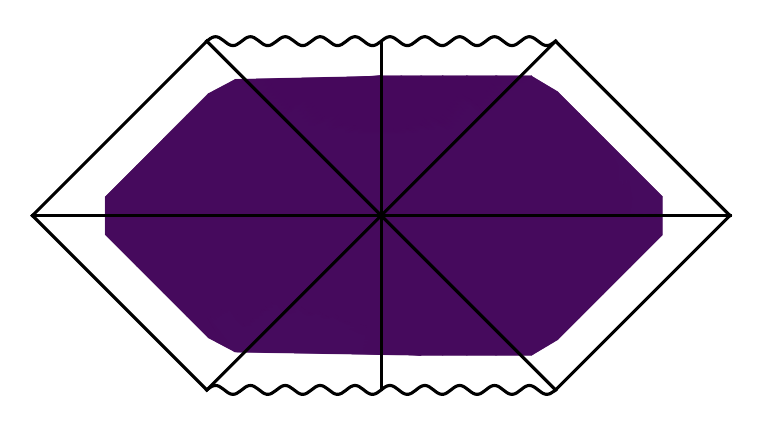}
    \caption{\(R_{11}\)}
  \end{subfigure}
  \hfill
  \begin{subfigure}[t]{0.235\linewidth}
    \centering
    \includegraphics[width=\textwidth]{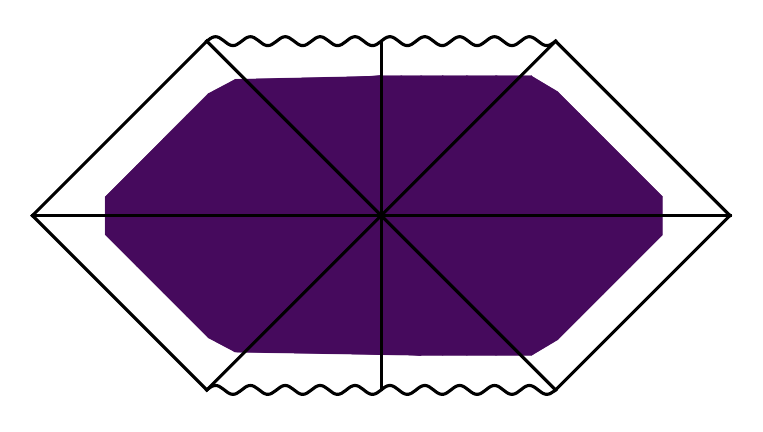}
    \caption{\(R_{12}\)}
  \end{subfigure}
  \hfill
  \begin{subfigure}[t]{0.235\linewidth}
    \centering
    \includegraphics[width=\textwidth]{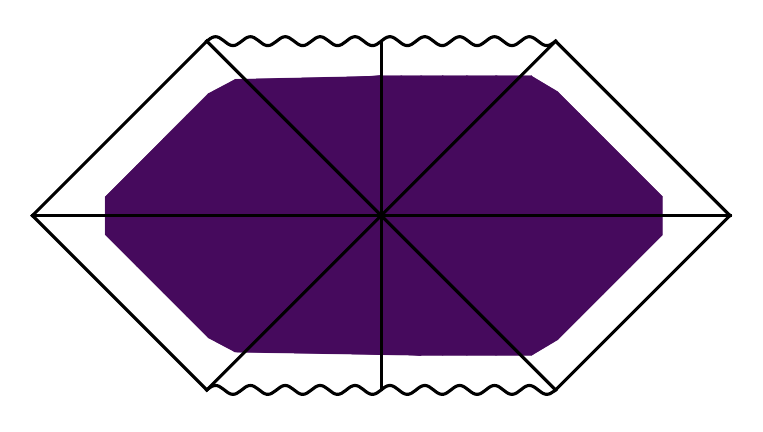}
    \caption{\(R_{13}\)}
  \end{subfigure}\\[0.25em]
  \begin{subfigure}[t]{0.235\linewidth}
    \centering
    \includegraphics[width=\textwidth]{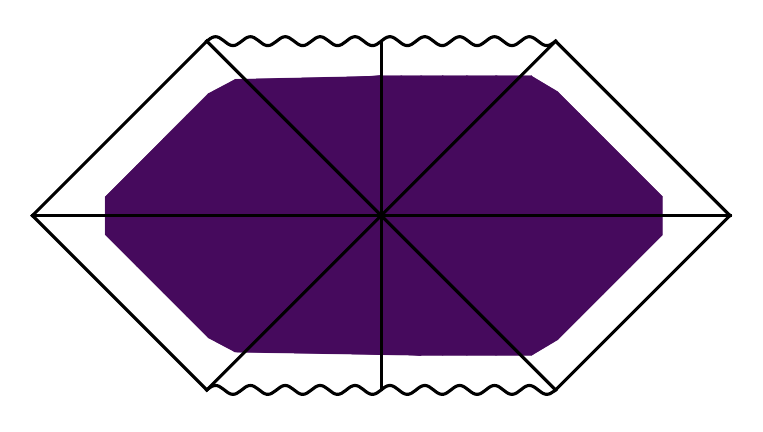}
    \caption{\(R_{20}\)}
  \end{subfigure}
  \hfill
  \begin{subfigure}[t]{0.235\linewidth}
    \centering
    \includegraphics[width=\textwidth]{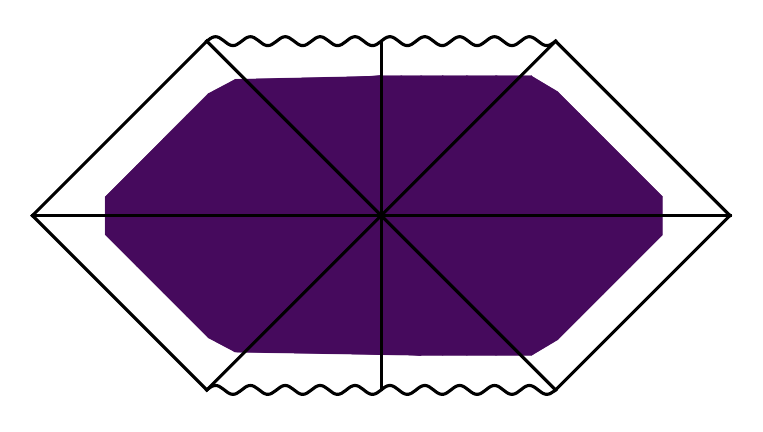}
    \caption{\(R_{21}\)}
  \end{subfigure}
  \hfill
  \begin{subfigure}[t]{0.235\linewidth}
    \centering
    \includegraphics[width=\textwidth]{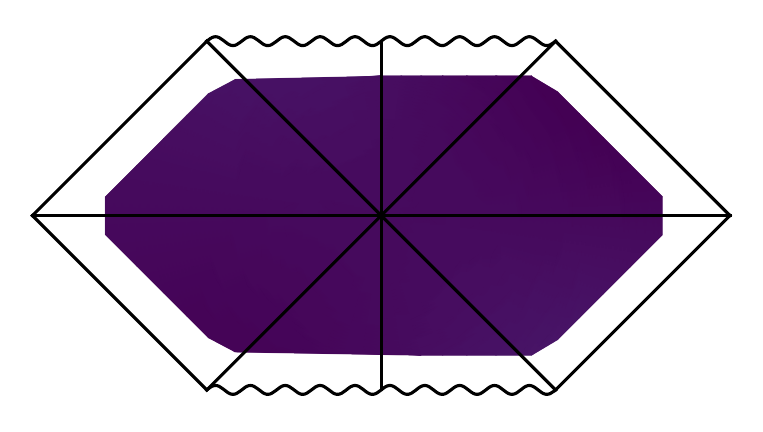}
    \caption{\(R_{22}\)}
  \end{subfigure}
  \hfill
  \begin{subfigure}[t]{0.235\linewidth}
    \centering
    \includegraphics[width=\textwidth]{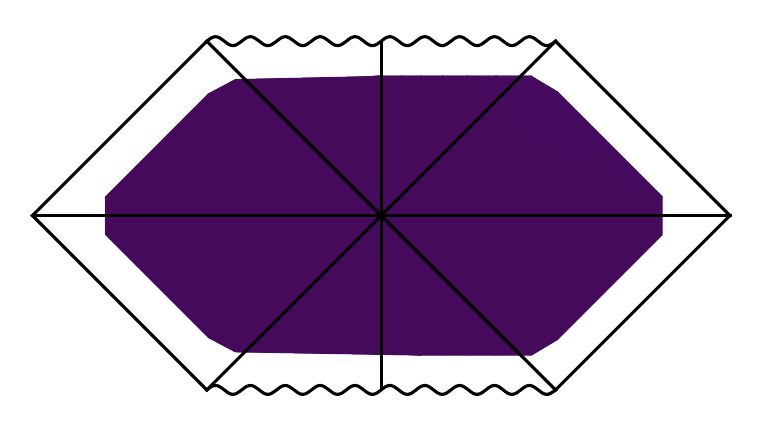}
    \caption{\(R_{23}\)}
  \end{subfigure}\\[0.25em]
  \begin{subfigure}[t]{0.235\linewidth}
    \centering
    \includegraphics[width=\textwidth]{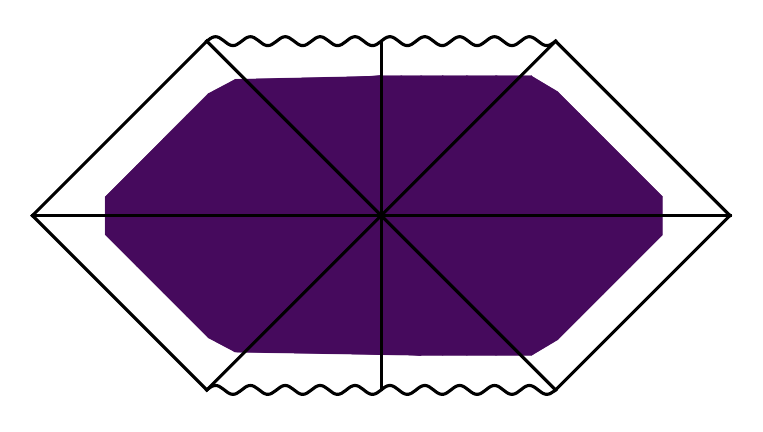}
    \caption{\(R_{30}\)}
  \end{subfigure}
  \hfill
  \begin{subfigure}[t]{0.235\linewidth}
    \centering
    \includegraphics[width=\textwidth]{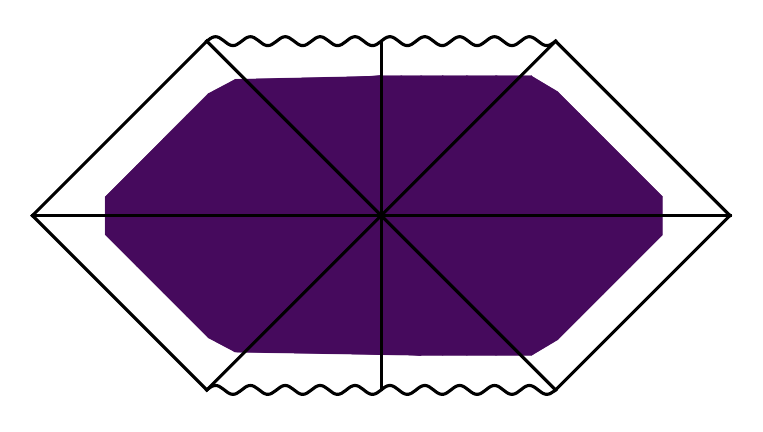}
    \caption{\(R_{31}\)}
  \end{subfigure}
  \hfill
  \begin{subfigure}[t]{0.235\linewidth}
    \centering
    \includegraphics[width=\textwidth]{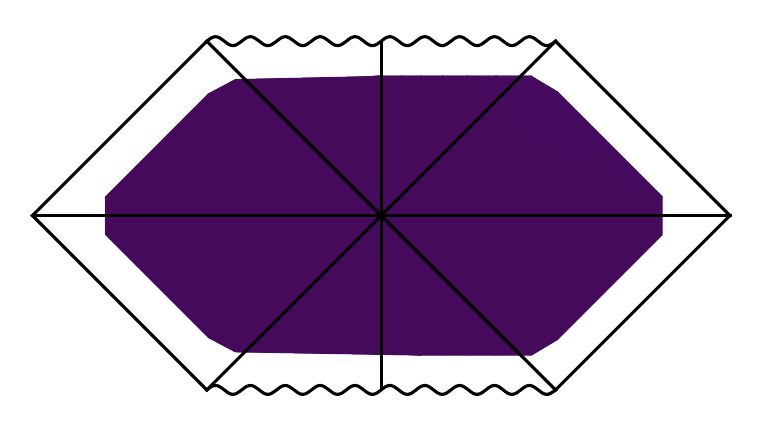}
    \caption{\(R_{32}\)}
  \end{subfigure}
  \hfill
  \begin{subfigure}[t]{0.235\linewidth}
    \centering
    \includegraphics[width=\textwidth]{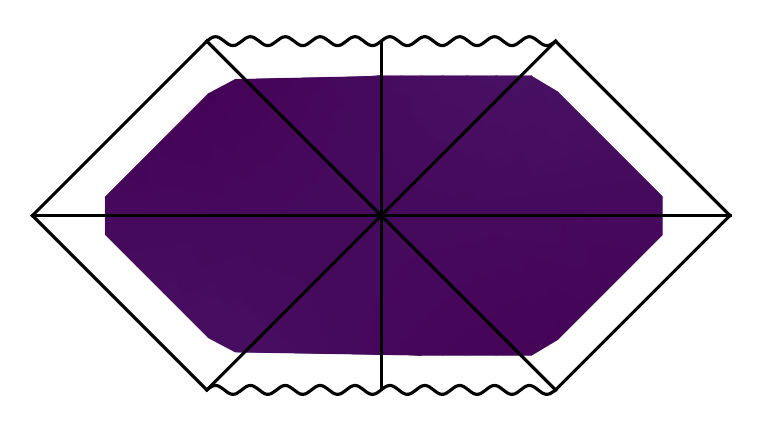}
    \caption{\(R_{33}\)}
  \end{subfigure}

  \end{minipage}

  \par\vspace{0.6em}
  \begin{subfigure}[t]{0.28\textwidth}
    \centering
    \includegraphics[width=\textwidth]{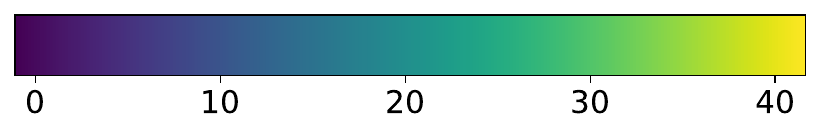}
    \caption*{Shared component scaling}
  \end{subfigure}

  \caption{Full component validation for the Petrov type-I black-hole search, with $(\beta_0, \alpha_{\text{trap}}) = (0.08, 25)$, of \cref{subsubsec:blackhole_network}. The left block shows the learned metric components \(g_{\mu\nu}\), and the right block shows the learned Ricci tensor components \(R_{\mu\nu}\), both arranged by tensor index over the validation samples. Over the 10 random seeds, the absolute component means are \( |\hat{g}_{ij}| = 1.780 \pm 0.044 \) and \( |\hat{R}_{ij}| = 0.0350 \pm 0.0007 \).}
  \label{fig:blackhole_components_4x4}
\end{figure*}

\clearpage
\section{Model Hyperparameters}

Summaries of the optimal training hyperparameters for the Schwarzschild experiment are presented in \cref{tab:best-einstein-loss}, whilst \cref{tab:type-I-vac-minimal-config} highlights the key changes for the Type-I vacuum search, and \cref{tab:type-I-bh-minimal-config} the key changes for the black hole search. An exhaustive list of parameters is presented in the GitHub repository. 

\begin{table}[h]
\centering
\small
\begin{tabular}{lll}
\hline
\textbf{Category} & \textbf{Parameter} & \textbf{Value} \\
\hline
Training & Training samples & $8{,}000$\\
Training & Epochs & $1000$ \\
Training & Batch size & $512$ \\
Training & Initial learning rate & $3 \times 10^{-4}$ \\
Training & Minimum learning rate & $3 \times 10^{-6}$ \\
\hline
Model & Hidden width & $256$ \\
Model & Hidden layers & $6$ \\
Model & Activation & GELU \\
Model & Bias & \texttt{true} \\
Model & Precision & \texttt{float64}\\
\hline
Geometry & Dimension & $4$ \\
Geometry & Overlap upper width & $0.1$ \\
Geometry & Einstein constant & $0.0$ \\
Geometry & Density power on $R^2$ & $1.0$ \\
Geometry & Density power on $S^2$ & $1.0$ \\
Geometry & Patch width on $R^2$ & $0.85$ \\
Geometry & Patch width on $S^2$ & $1.0$ \\
Geometry & Mass parameter ($m$) & $1.0$ \\
\hline
Loss weights & Einstein multiplier & $1.0$ \\
Loss weights & Weyl invariant multiplier & $1.0$ \\
Loss weights & $R^2$ determinant multiplier & $0.075$ \\
Loss weights & Speciality-index multiplier & $0.0$ \\
Loss weights & Killing-symmetry multiplier & $1.0$ \\
Loss weights & $K$-repeller multiplier & $0.0$ \\
Loss weights & Speciality-index radial-profile multiplier & $0.0$ \\
% \hline
% Loss settings & Einstein loss mode & \texttt{curvature\_normalized\_capped} \\
% Loss settings & Curvature norm epsilon & $10^{-3}$ \\
% Loss settings & Curvature norm cap & $1.0$ \\
% Loss settings & Weyl invariant norm & $0.0625$ \\
% Loss settings & Kretschmann loss mode & \texttt{weyl\_invariant} \\
% Loss settings & Determinant barrier mode & \texttt{both\_blocks} \\
% Loss settings & Use metric contraction & \texttt{true} \\
\hline
\end{tabular}
\caption{Key configuration parameters for the Schwarzschild experiment (full set are available on GitHub).}
\label{tab:best-einstein-loss}
\end{table}

\begin{table}[h]
\centering
\small
\begin{tabular}{lll}
\hline
\textbf{Category} & \textbf{Parameter} & \textbf{Value} \\
\hline
Loss weights & Einstein multiplier & $1.0$ \\
Loss weights & $R^2$ determinant multiplier & $0.075$ \\
Loss weights & Speciality-index radial profile multiplier & $1.0$ \\
Loss weights & Trapped surface multiplier & $0.0$ \\
Loss weights & Kretschmann multiplier & $0.0$ \\
Loss weights & Killing symmetry multiplier & $0.0$ \\
Loss weights & $K$ repeller multiplier & $0.0$ \\
\hline
\end{tabular}
\caption{Summary of key configuration parameter deviations of the Petrov Type-I vacuum search from the Schwarzschild experiment.}
\label{tab:type-I-vac-minimal-config}
\end{table}

\begin{table}[h]
\centering
\small
\begin{tabular}{lll}
\hline
\textbf{Category} & \textbf{Parameter} & \textbf{Value} \\
\hline
Loss weights & Einstein multiplier & $1.0$ \\
Loss weights & $R^2$ determinant multiplier & $0.075$ \\
Loss weights & Speciality-index radial profile multiplier & $1.0$ \\
Loss weights & Horizon anchor & $3.0$ \\
Loss weights & Trapped surface multiplier & $25.0$ \\
Loss weights & Kretschmann multiplier & $0.0$ \\
Loss weights & Killing symmetry multiplier & $0.0$ \\
Loss weights & $K$ repeller multiplier & $0.0$ \\
\hline
\end{tabular}
\caption{Summary of key configuration parameter deviations of the Petrov Type-I BH search from the Schwarzschild experiment.}
\label{tab:type-I-bh-minimal-config}
\end{table}
\clearpage

\bibliographystyle{apsrev4-1}
\bibliography{bib}

\end{document}